\documentclass[aps, nofootinbib, prd, superscriptaddress, showpacs, tightenlines,
notitlepage, twocolumn, floatfix]{revtex4-1}
\usepackage{graphicx}
\usepackage{xcolor}
\usepackage{amsmath,amssymb,latexsym}
\usepackage{hyperref}
\usepackage{ulem}

\newcommand\araa{Annu. Rev. Astron. Astrophys.}
\newcommand\mnras{Mon. Not. R. Astron. Soc.}
\newcommand\aj{Astron. J.}
\newcommand\apjl{Astrophys. J. Lett.}
\newcommand\apjs{Astrophys. J. Suppl.}
\newcommand\aap{Astron. Astrophys.}
\newcommand\aapr{Astron. Astrophys. Rev.}
\newcommand\pasj{Publ. Astron. Soc. Japan}
\newcommand\jcap{Journal of Cosmology and Astroparticle Physics}

\begin{document}

\title{Strongly Lensed Transient Sources: A Review}
\author{Kai Liao}\email{liaokai@whu.edu.cn}
\affiliation{School of Physics and Technology, Wuhan University, Wuhan 430072, China}
\author{Marek Biesiada}\email{marek.biesiada@ncbj.gov.pl}
\affiliation{National Centre for Nuclear Research, Pasteura 7, 02-093 Warsaw, Poland}
\author{Zong-Hong Zhu}\email{zhuzh@whu.edu.cn}
\affiliation{School of Physics and Technology, Wuhan University, Wuhan 430072, China}
\affiliation{Department of Astronomy, Beijing Normal University, Beijing 100875, China}

\begin{abstract}
The past decades have witnessed a lot of progress in gravitational lensing with two main
targets: stars and galaxies (with active galactic nuclei).
The success is partially attributed to the continuous luminescence of these sources making the detection and monitoring relatively easy.
With the running of ongoing and upcoming large facilities/surveys in various electromagnetic and gravitational-wave bands, the era of time-domain surveys would guarantee constant detection of strongly lensed explosive
transient events, for example, supernovae in all types, gamma ray bursts with afterglows in all bands, fast radio bursts and even gravitational waves.
Lensed transients have many advantages over the traditional targets in studying the Universe and magnification effect helps to understand the transients themselves at high redshifts.
In this review article, basing on the recent achievements in the literature, we summarize the methods of searching for different kinds of lensed transient signals,
the latest results on detection and their applications in fundamental physics, astrophysics and cosmology.
At the same time, we give supplementary comments as well as prospects of this emerging research direction that may help readers who are interested in entering this field.
\end{abstract}


\maketitle

\section{Introduction}
According to modern theories of gravity, primarily the General Relativity (GR), light rays travelling along (null)geodesics can be deflected due to intervening inhomogeneities in the matter distribution. In particular, a massive foreground object located between the light source and the observer can act as a deflector. This phenomenon, called gravitational lensing, is analogous to optical lensing. The history of gravitational lensing dates back to the observation of starlight deflection by the Sun during the solar eclipse in 1919 as strong evidence in favour of GR. In modern cosmology, a remarkable case of lensing is that of a distant bright quasar strongly lensed (To avoid confusion, the term ``strong lensing'' in this paper
refers to the case of a single lens case in contrast to ``weak lensing'' where a collection of massive structures distorts the shape of images in the statistical sense. Note that single deflector case can be further classified according to the lens (mass) scales as ``strong", ``milli", ``micro" lensing and so on.)  by foreground galaxy or a galaxy cluster, forming multiple magnified images of the point-like bright nucleus
while the lensed dimmer parts of the host galaxy appear in the form of stretched arcs.
Such lensed quasar systems have been attracting great attention as they are useful for efficient studies of either the nature of lens galaxies and quasars or the geometry of the Universe~\cite{2010ARA&A..48...87T}.

The reason that lensed quasars are currently the most popular targets for strong gravitational lensing research is because of the quasars themselves.
They are common in the Universe and are very bright, hence visible form large distances. Their brightness is continuous, but fortunately displays some variability.
Therefore, it is relatively easy to observe, verify, monitor and analyze quasars lensed by galaxies and galaxy clusters. There are over 200 strongly lensed quasars known so far (https://research.ast.cam.ac.uk/lensedquasars/) with more lenses being identified~\cite{2022arXiv220607714L},
though only a few of them are well-analyzed with various supplementary observations.
This number will be rapidly increasing with current and upcoming optical/infrared surveys like Legacy Survey of Space and
Time (LSST) in Vera Rubin Observatory~\cite{2010MNRAS.405.2579O}.
It has long been noticed that lensed quasars are powerful tools to address three major astrophysical issues:
studying the spatial mass distribution at kiloparsec and subkiloparsec scales, where baryons interact with dark matter shaping the galaxies as we see them;
determining the overall geometry, kinematics and components of the Universe;
and probing distant objects that are too small or too faint to be resolved or detected with current instruments~\cite{2010ARA&A..48...87T}.
A particularly striking cosmological application is time-delay cosmology~\cite{2016A&ARv..24...11T}.
Time delay between pairs of lensed images of the bright central part (active galactic nucleus - AGN), measurable by shifting the lightcurves of images, is proportional to the cosmological characteristic time scale, which is inversely proportional to the current expansion rate. The lens potential can be determined by other inputs, e.g. detailed analysis of distorted image of the dimmer parts of the host galaxy. With these two ingredients the Hubble constant $H_0$ can be measured directly and independently of the cosmic distance ladder (Historically, the method was firstly proposed by S. Refsdal with using lensed supernovae to measure time delays~\cite{1964MNRAS.128..307R}.).

In addition to quasars, galaxies have large chance to be lensed called galaxy-galaxy lensing. Since ordinary galaxies are lack of bright AGNs, lensed galaxies appear as giant arcs and more popular than lensed quasars.
There are hundreds of known galaxy-galaxy lensing systems so far.
Besides, stars in the Milky Way and M31 Andromeda galaxy were monitored as a small number of them are expected to be micro-lensed by intervening stars or compact dark objects~\cite{2007A&A...469..387T,2019NatAs...3..524N}. They are also used to find extrasolar planet based on the fact the lightcurve would be distorted by the lens when the peculiar motion causes the change of relative position~\cite{1991ApJ...374L..37M}.

While lensing of quasars, galaxies and stars continue gaining achievements, the first strongly lensed supernova ``SN Refsdal"~\cite{2015Sci...347.1123K} with multiple images opened a new window for astrophysics and cosmology and the concept of
``lensed transients" is gradually entering people's vision. In addtion,
lensed transient explosive sources like gamma ray bursts (GRBs), fast radio bursts (FRBs) and even gravitational waves (GWs) are catching attentions of
the community, thanks to the era of time-domain astronomy.
With the running of current large surveys/facilities like Gaia~\cite{2016A&A...595A...2G}, Dark Energy Survey~\cite{2017ApJS..232...15D}, Subaru Hyper Suprime-Cam Survey~\cite{2020A&A...642A.148S} in optical/infrared, CHIME~\cite{2021ApJS..257...59C} and FAST~\cite{2021arXiv211007418N} in radio, LIGO-Virgo-KAGRA network~\cite{2021arXiv211103606T} in
GW and upcoming surveys/facilities like
Vera Rubin Observatory - LSST~\cite{2010MNRAS.405.2579O}, Square Kilometre Array~\cite{2009IEEEP..97.1482D} and Einstein Telescope~\cite{2012PhRvD..86l2001R}, (more than) hundreds of thousands of transient signals will be guaranteed, a small part of
which will be lensed by galaxies, clusters and various clumped structures below galactic scales in the Universe.

Detecting and verifying these lensed transient signals are interesting and significant because they would take us to a new research field and bring more discoveries.
Firstly, while lensing of traditional sources assumes the description of geometric optics by default, lensed transient signals should sometimes take into account
wave optics  due to long wavelengths, small source size or coherent emission.
Secondly, compared to the traditional targets, lensed transients have many advantages in studying the Universe. For example, due to the well-known light curve template of
type Ia supernovae, strong lens time delay measurements may become easier and more accurate with shorter monitoring time, which benefits $H_0$ inference. This advantage becomes
much more obvious for lensed FRBs, GRBs and GWs due to their transient nature (signal durations are much smaller than time delays themselves).
Finally, studying the lensing effect may also shed light on the nature of transients themselves, especially at high redshifts where magnification effect becomes non-negligible.

The literature related to lensed transients (especially SNe and GWs) is rapidly increasing. Therefore, we find it is necessary to summarize the existing works at present time and
make prospects for the future. In this review, we focus on briefly summarizing the works especially in the last several years on detection and applications.
For more details and different points of view, we refer to the excellent previous review by Oguri~\cite{2019RPPh...82l6901O}.

The paper is organized as follows.
In Section II, we give a brief introduction of gravitational lensing theory. Methods of searching lensed signals and current observations are presented in Section III.
The applications are given in Section IV. Finally, we summarize in Section V.

\section{A primer to strong lensing}
Strong gravitational lensing occurs when the source, the deflector (the lens) and the observer are aligned. Depending on the wavelength and source size
relative to the lens scale as well as whether it is a coherent emission, this phenomenon is described by either geometric optics or wave optics formalism. We give brief introductions for both
cases in this section. We consider the standard conditions of a thin lens and weak gravitational field which are suitable for most of astrophysical cases. ``Thin lens'' means that the lens scale is much smaller
than the distances between the source, the lens and the observer while ``weak field'' means that the Newtonian potential is small enough $U/c^2\ll 1$ (Note, that ``weak field'' regime is different from the ``Newtonian approximation'', since the latter is supplemented by the condition that velocities in the system are much smaller than the speed of light. While it is true for the source, the lens -- their relative motions and internal motions of stars -- the light, by definition propagates with the speed $c$. Hence despite of the weak field approximation, gravitational lensing is inherently a relativistic phenomenon.). Readers who are familiar with gravitational lensing theory can skip
this part. We refer to the classical textbook~\cite{Schneider1992LensesBook} for more details.

\begin{figure*}
\centering
\includegraphics[width=10cm,angle=0]{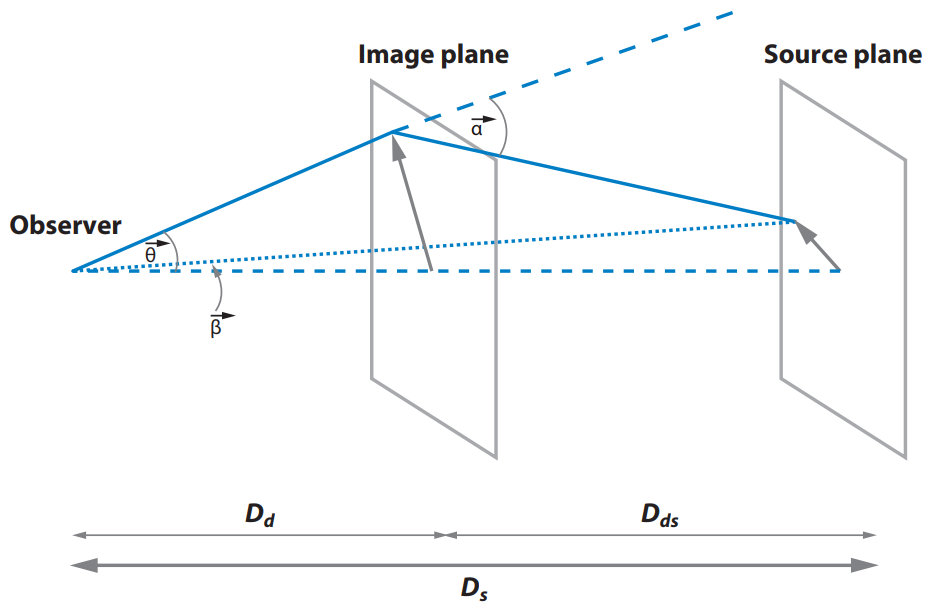}\\
\caption{Schematic picture of strong gravitational lensing system with geometric optics description (notations defined in the text), taken from Ref.~\cite{2010ARA&A..48...87T}.
}
\label{lenfig}
\end{figure*}

\subsection{Geometric optics}
For most cases of electromagnetic (EM) sources, the wavelengths of radiation are much smaller than the lens scales concerned and different parts of the source emit light independently,
thus geometric optics (Geometric optics here means that lights travel along geodesics and there is no interference between images.) is sufficient to describe lensing phenomenon.
In this case, gravitational lensing can be taken as a transformation between two-dimensional observed coordinates in the lens plane and two-dimensional
coordinates in the source plane. Given source angular position $\boldsymbol{\beta}$ (with respect to the line of sight to the center of the lens), the excess time delay relative to the case without lens is a function of the
 image position $\boldsymbol{\theta}$ in the lens plane:
\begin{equation}
t=\frac{D_dD_s(1+z_d)}{cD_{ds}}\left(\frac{1}{2}|\boldsymbol{\theta}-\boldsymbol{\beta}|^2-\psi(\boldsymbol{\theta})\right),\label{t}
\end{equation}
where $D$ stands for the angular diameter distance with subscripts $d,s$ denoting the deflector (the lens) and the source, respectively. Note that in general $D_{ds}\neq D_s-D_d$ unless one deals with Euclidean space (e.g. microlensing in the Milky Way).
The excess time delay includes two terms caused by geometry (the path length) and the ``Shapiro delay'', respectively.
The latter is attributed to the fact that light speed $c'$ in a gravitational field is smaller than $c$, thus
one can define an effective ``refractive index'' $n=c/c'$ for weak gravitational field, which is analogous to the refractive index in geometric optics.
The term $\psi$ is two-dimensional lensing potential determined by the Poisson Equation $\nabla^2\psi=2\kappa$, where $\kappa$
is the surface mass density of the lens in units of the critical density $\Sigma_c=c^2D_s/(4\pi GD_dD_{ds})$. Eq.\ref{t} actually gives a time delay surface in the lens plane.

Along this line of consideration, light travelling in gravitational field follows the Fermat's principle. Real images only occur at the extreme points of time delay surface, which
immediately leads to the lens equation from Eq.\ref{t}:
\begin{equation}
\boldsymbol{\beta}=\boldsymbol{\theta}-\boldsymbol{\alpha},
\end{equation}
where $\boldsymbol{\alpha}=\nabla\psi$ is the scaled deflection angle related to the the deflection angle $\boldsymbol{\hat{\alpha}}$ by $\boldsymbol{\alpha}=D_{ds}\boldsymbol{\hat{\alpha}}/D_s$.
We show the lensing geometry in Fig.\ref{lenfig}. Note that there might exist multiple extreme points classified in three types: minimum, maximum and a saddle point. In the particular case of perfect alignment $\boldsymbol{\beta} = 0$ the source would be seen as a ring (called the Einstein ring) around the lens. Its angular size is called the Einstein radius $\theta_E$. When the mass distribution of the lens is axisymmetric, the Einstein radius is a robust measure of the projected mass of the lens $M_d \equiv M(< \theta_E)$ according to the formula: $\theta_E^2 = \frac{4 G M_d}{c^2} \frac{D_{ds}}{D_d D_s}$.

Mathematically, the transformation between the two planes can be described by the Jacobi Matrix:
\begin{equation}
A=\frac{\partial\boldsymbol{\beta}}{\partial\boldsymbol{\theta}}=\delta_{ij}-\frac{\partial^2\psi}{\partial\boldsymbol{\theta}_i\partial\boldsymbol{\theta}_j}=\left(
                                                                                                            \begin{array}{cc}
                                                                                                              1-\kappa-\gamma_1 & -\gamma_2 \\
                                                                                                              -\gamma_2 & 1-\kappa+\gamma_1 \\
                                                                                                            \end{array}
                                                                                                          \right),\label{Jacobian}
\end{equation}
where $\gamma_1$, $\gamma_2$ and $\kappa$ depend on the second derivatives of the lensing potential. This decomposition allows us to define a pseudo-vector $\boldsymbol{\gamma}=(\gamma_1,\gamma_2)$ on the lens plane
called the \textit{shear}. While $\kappa$ determines the isotropic part of the Jacobian and is thus called the \textit{convergence}, $\gamma_1$ and $\gamma_2$ compose the trace-free part of the Jacobian
which quantifies the projection of the gravitational field (the gradient of the gravitational force) and describes the distortion of the background source.
The magnification is quantified by the inverse of the determinant of the Jacobi Matrix. For this reason, matrix $M=A^{-1}$
is called the magnification tensor. We have
\begin{equation}
\mu\equiv \mathrm{det} M=\frac{1}{\mathrm{det} A}=\frac{1}{(1-\kappa)^2-\gamma^2}.
\end{equation}
Places where $\mu=\infty$ correspond to lines on the lens plane (called critical lines) and source plane (called caustic lines), respectively.
When source passes through caustic lines, the image number changes by two and the total number should always be odd for a non-singular potential (Among the lensed images, the central one is often demagnified and outshone by the lens, thus the `observed' number may be even.) since any angular far-away source is unaffected by the lens.
Note that for realistic sources always having a finite size one in principle needs to do convolution to calculate the total magnification.

For strong lensing within the geometric optics, there
is a fundamental limit to how much information can be extracted
from the observation called the \textit{mass-sheet degeneracy}. Namely, given a lens model $\kappa(\boldsymbol{\theta})$ and a source position $\boldsymbol{\beta}$ that reproduces the image
positions and magnification ratios, it
is always possible to define a family of alternative models and source positions (quantified by the parameter $\lambda$) that
leave those observables unchanged and predict different time delays. The transformation is as follows:

\begin{equation}
\kappa_\lambda=(1-\lambda)+\lambda\kappa;\   \  \boldsymbol{\beta}_\lambda=\boldsymbol{\beta}/\lambda,
\end{equation}
such that
\begin{equation}
\Delta t_\lambda=\lambda\Delta t; \   \  \mu_\lambda=\mu/\lambda^2.
\end{equation}

\begin{figure*}
\centering
\includegraphics[width=5cm,angle=0]{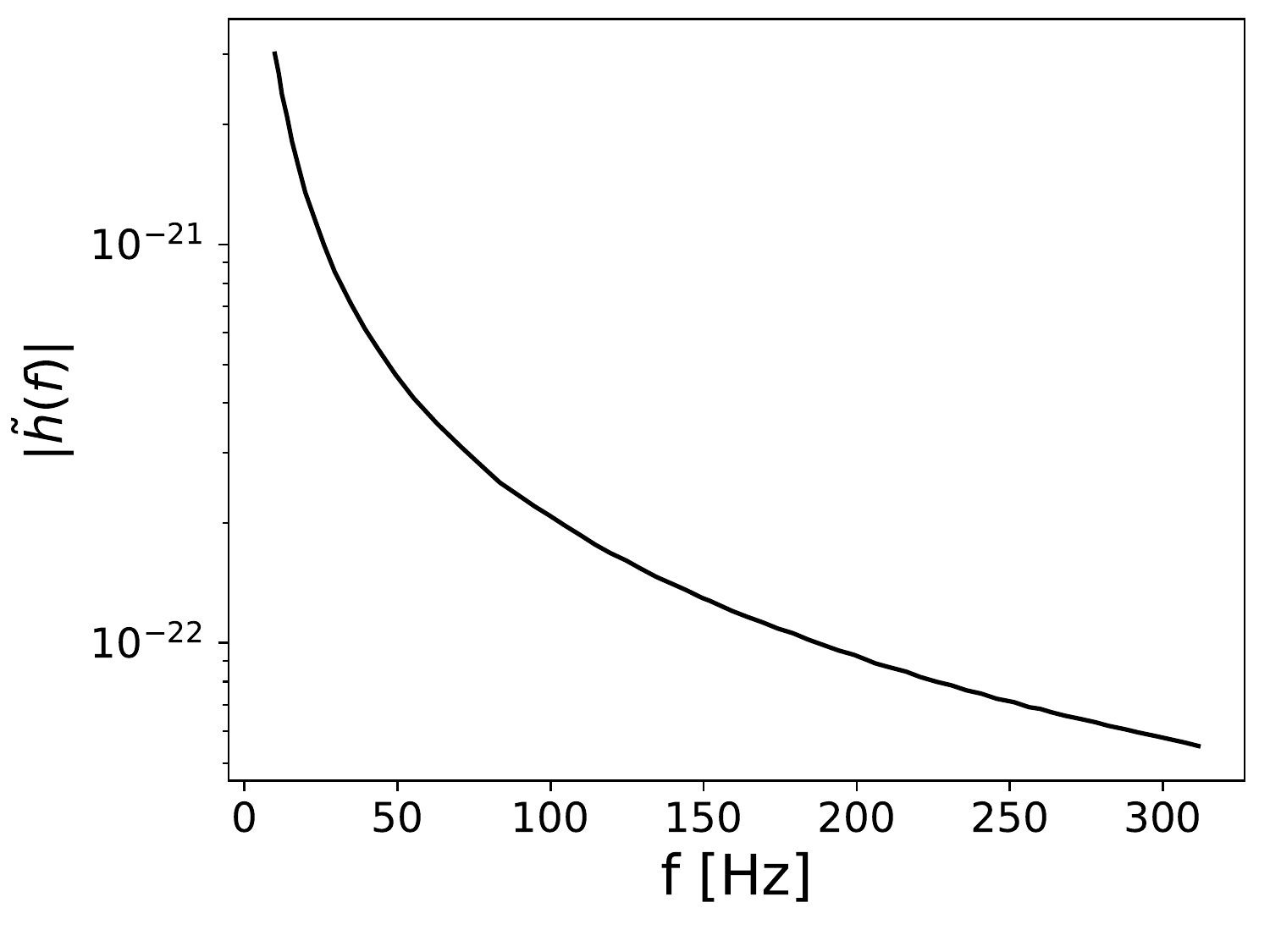}
\includegraphics[width=6cm,angle=0]{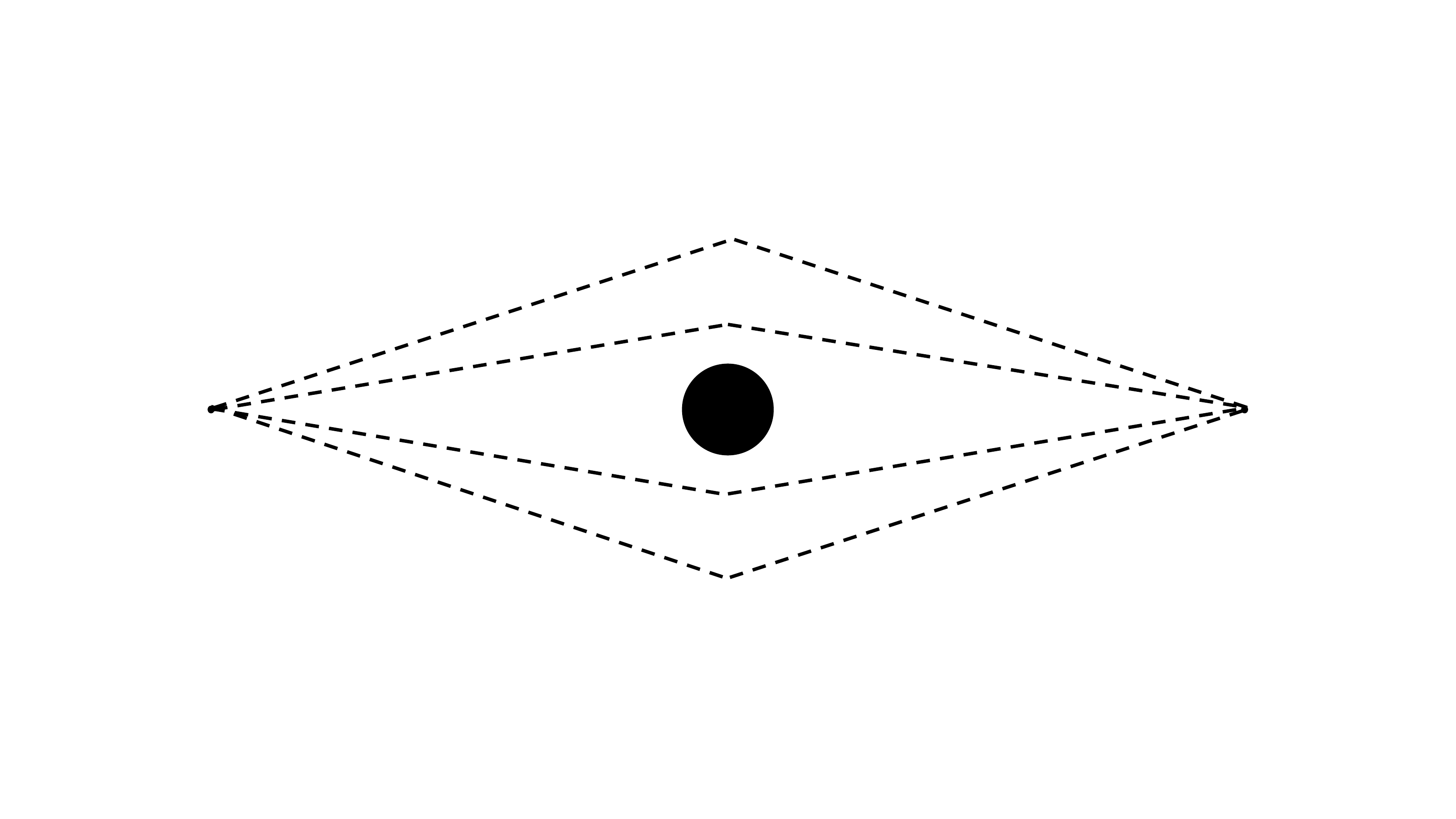}
\includegraphics[width=5cm,angle=0]{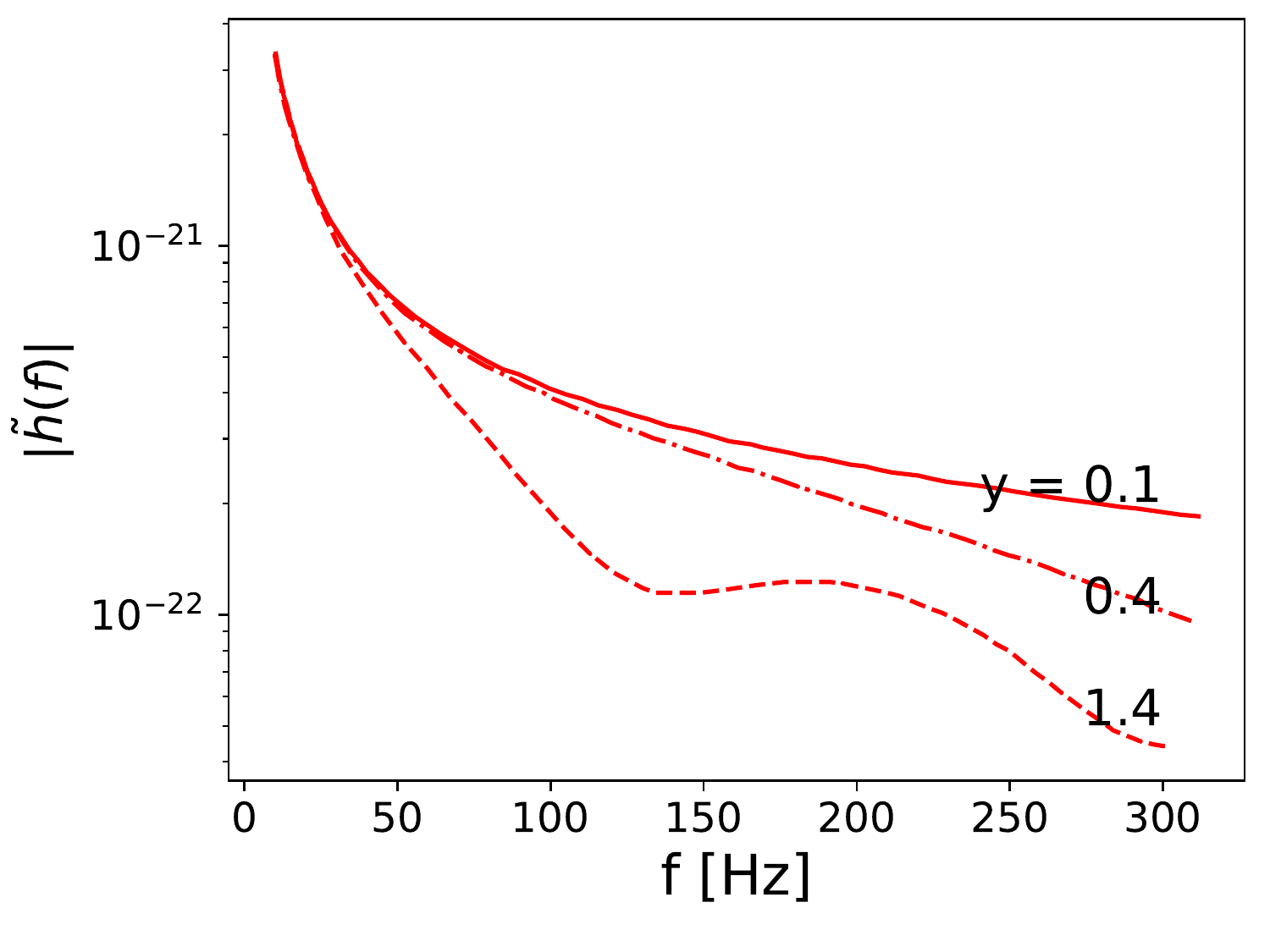}\\
\caption{An example of wave optics description: The source (left) is a binary black hole merger with redshifted masses $M_1=M_2=20M_\odot$ while the lens is a point mass with $M_d=100M_\odot$. The source position $\beta=0.1,0.4,1.4\theta_\mathrm{E}$, respectively. This set of parameters is taken from~\cite{2019PhRvL.122d1103J}. For illustration purpose, we has ignored spin-induced effect, binary-orbit eccentricity and post-Newtonian corrections. Unlike the geometric case, each path is possible for wave optics. The amplitude of lensed GW signal (right) is frequency-dependently distorted (The phase changing is not shown in this figure.).
}
\label{wavefig}
\end{figure*}

\subsection{Wave optics}
Both electromagnetic and gravitational waves can be considered as a scalar field plus its polarization. For example,
GW as the perturbation of the background spacetime:
\begin{equation}
g_{\mu \nu}=g^{(L)}_{\mu \nu}+h_{\mu \nu}
\end{equation}
can be written as
\begin{equation}
h_{\mu \nu}=\phi e_{\mu \nu}.
\end{equation}
$g^{(L)}_{\mu \nu}$ is the metric by the lens object. For a weak field and small deflection angle considered in gravitational lensing, polarization alters only at high-orders. Thus, the gravitational wave can
be approximated as a scalar field which propagates in a background spacetime following the equation:
\begin{equation}
\partial_\mu(\sqrt{-g^{(L)}}g^{(L)\mu \nu}\partial_\nu\phi)=0.
\end{equation}
The weak gravitational field can be described by the Newtonian potential $U$. For simplicity, we consider a monochromatic wave with frequency $f$,
the equation above thus becomes a Helmholtz equation which in the frequency domain reads:
\begin{equation}
(\Delta + 4\pi^2 f^2)\tilde{\phi}=16\pi^2f^2U\tilde{\phi}.
\end{equation}

Lens field changes both the intensity and phase of the wave. The impact can be described by the amplification factor
\begin{equation}
F(f)=\tilde{\phi}^L(f)/\tilde{\phi}(f),
\end{equation}
which is the ratio of lensed wave amplitude to the unlensed one. At this point, one should notice that GW detectors cannot resolve lensed signals,
in other words, only one signal exists in this situation. In the spirit of diffraction phenomenon, the detected signal is a superposition of all possible
waves on the lens plane that have different time delays corresponding to different phases:
\begin{equation}
F(f,\boldsymbol{\beta})
=\frac{D_sD_d(1+z_d)f}{cD_{ds}i}\int d^2\boldsymbol{\theta}\ exp\left[2\pi ift_d(\boldsymbol{\theta},\boldsymbol{\beta})\right],\label{amplitude}
\end{equation}
where
\begin{equation}
t_d(\boldsymbol{\theta},\boldsymbol{\beta})=\frac{D_dD_s(1+z_d)}{cD_{ds}}\left[\frac{1}{2}|\boldsymbol{\theta}-\boldsymbol{\beta}|^2-\psi(\boldsymbol{\theta})+\phi_m(\boldsymbol{\beta})\right],
\end{equation}
where $\phi_m$(y) is
chosen so that the minimum arrival time is zero.

For geometric optics limit with $f\gg t_d^{-1}$, one can evaluate Eq.\ref{amplitude} using stationary phase approximation, where only extreme points determined
by the lens equation contribute to the integral. The integral on the lens plane is reduced to the sum over these images as
\begin{equation}
F(f,\boldsymbol{\beta})=\sum_i|\mu_i|^{1/2}exp\left[2\pi ift_i-i\pi n_i\right]\label{sumextreme},
\end{equation}
where $n_i = 0, 1/2, 1$ when it
is a minimum, saddle, maximum point.
Eq.\ref{sumextreme} actually describes the interference (In geometric optics limit, ignoring the interference, magnification amplitudes are summed up directly. One should also note that GW detection is based on wave amplitude while EM observation is based on intensity. Their magnifications differ by a square.).

So far we only consider a point source. For a finite source size, $F(f)$
depends on the spatial emission distribution of the source.
The width of interference oscillations in the source
plane is on the order of $\theta_\mathrm{E}/w$, where $w = \frac{8 \pi G}{c^3} f (1+z_d) M_d(< \theta_E)$ is a useful dimentionless parameter separating the regime where the wave optics effects suppress lensing magnifications ($w <1$) and the regime where magnification pattern is pronounced and exhibits oscillatory behaviour ($w>1$).  To ensure that the interference
pattern can be observed, the source size in the angular units
$\beta_s=R_s/D_s$ should satisfy $\beta_s<\theta_\mathrm{E}/w$. This condition
is equivalent to
\begin{equation}
R_s<\frac{D_s\theta_\mathrm{E}}{w}.
\end{equation}
For more details, we refer to~\cite{2006JCAP...01..023M} by Matsunaga $\&$ Yamamoto who investigated the finite
source size effect in the context of the wave optics in gravitational lensing as well as Section 2.5 of the review by Oguri~\cite{2019RPPh...82l6901O}.

\section{Search for lensed transients}
The ``search'' here actually has two meanings. One is about whether a lensed signal can be detected, the other is about how to identify and verify such
lensed signals.
In this section, we introduce the state-of-art searching methods and the latest results for various types of lensed transients.

\subsection{Lensed SNe}
Supernovae (SNe) are violent explosions in the Universe associated with star deaths~\cite{1988ARA&A..26..295W}. They are observationally classified in different types based on
the presence or absence of certain narrow line features in the spectra and/or the shapes of their light curves~\cite{1997ARA&A..35..309F}.
For example, those without hydrogen lines and with singly ionized silicon line are classified as type Ia
who have similar peak luminosities~\cite{1992ARA&A..30..359B} and are used as standard(izable) candles to measure cosmic  distances~\cite{1998AJ....116.1009R,1999ApJ...517..565P}. On more physical basis, the classification depends
on the explosion mechanisms. Type Ia supernovae are usually the thermonuclear explosions of white dwarfs near the Chandrasekhar mass $\approx1.4M_\odot$ due to
the accretion from a companion star. Different mechanisms, for example, a double-degenerate scenario~\cite{2011A&A...528A.117P}, are summarized in~\cite{2014ARA&A..52..107M}.
Other types are produced by a core collapse of massive stars. Supernovae are bright transients lasting for months in
the observer frame. Their peak luminosities are comparable to the luminosity of the whole host galaxy.  Thus one can observe them out to high redshifts $z>1$~\cite{2018ApJ...859..101S} giving a non-negligible chance of strong gravitational lensing. Strategies of searching lensed SNe can be classified according to the images being resolved or unresolved.

In the case of resolved images, the strategy is direct: identifying multiple images. The critical factors for the success include the image
separations relative to the telescope resolution (including seeing), the flux ratios (contrast) between the
images, and the apparent magnitudes of the faintest images. Many works have been done to predict the event rates in various
surveys~\cite{2000ApJ...532..679P,2001ApJ...556L..71H,2000ApJ...531..676W,2002A&A...393...25G,2002A&A...392..757G}.
For example, Oguri and Marshall \cite{2010MNRAS.405.2579O} gave a detailed and comprehensive study for wide-field surveys based on resolved images.
For the LSST, they took the image separation $>0.^{\prime\prime}5$
with flux ratio $>0.1$ and adopted a  peak magnitude limit brighter than the telescope limit in i band
in the calculations such that the lightcurves can be well traced afterwards.
Their calculation shows that LSST is expected to find more than 100 lensed SNe, 1/3 of which are of type Ia.
Once the images are resolved, one can identify the lensing systems via spectra and the image configurations.
It is worth noting that these estimates were based on
giant elliptical galaxies (E/S0) that
are responsible for $\sim80\%$
of the total lensing probability~\cite{2010MNRAS.405.2579O}.
From the perspective of Time-Delay Cosmography based on lensed SNe, this is an advantage since elliptical lenses are easy to model.
However, cluster lensing and small-separation cases may potentially contribute considerably (see the discussion of the observed events provided below).

For the case of unresolved images, Quimby et al.~\cite{2014Sci...344..396Q} found that strongly lensed Type Ia supernovae are well separated from unlensed ones in \texttt{i}-band magnitude versus \texttt{r-i} color diagram,
and thus proposed to utilize such color-magnitude diagram to identify unresolved strong lensing candidates. Follow-up high-resolution observations can further confirm these systems and give
more detailed measurements. This approach is expected to significantly increase the number of lensed supernovae with LSST.
Meanwhile, Goldstein $\&$ Nugent~\cite{2017ApJ...834L...5G} proposed to search near the elliptical galaxies for the SNe Ia brighter than typical brightness suggested by their distances (redshifts).
Such strategy is based on the fact that strong lensing is dominated by elliptical galaxies.
They proved that this
strategy could enable rapid identifications of unresolved lensed SNe and thus increase the yield of the LSST.
Furthermore, for low-resolution imaging/small separations, one can identify multiply lensed supernovae from ellipticity.
We may be able to find lensing signature by
carefully comparing the morphology of the supernova image to
the point spread function~\cite{2018RNAAS...2..186L}. This approach does not require spectra or a good coverage of the light curves.
In addition, one might identify lensed Type Ia SNe from single combined light curves~\cite{2021ApJ...910...65B,2022MNRAS.511.1210D}. Based on the fact that the lightcurves of Type Ia SNe  are
well-known with a rise and a fall, the authors investigated whether unresolved images
can be recognized as lensed sources given only lightcurve information, and whether time delays can be extracted robustly at the same time.

The work by Wojtak et al.~\cite{2019MNRAS.487.3342W} studied both search strategies for gravitationally lensed supernovae in wide-field surveys.
They found that the detection via magnification is the only effective strategy for relatively shallow surveys.
For survey depths at the LSST capacity, both strategies yield
comparable numbers of lensed SNe. SN samples from the two methods are to a large extent
independent and combining them increases detection rates by about 50 per cent. They gave predictions for Zwicky Transient Facility (ZTF) and LSST.
Applying a hybrid method which combines searching for highly magnified or multiply imaged transients,
they predicted that LSST will detect 89 type Ia and 254 core-collapse lensed SNe per year.

Other searching strategies include monitoring plausible sites~\cite{2021MNRAS.505.1316H} or
massive clusters as cosmic telescopes~\cite{2000MNRAS.319..549S,2000A&A...363..349S,2009A&A...507...61S,2018A&A...614A.103P}.
For example, Shu et al.~\cite{2018ApJ...864...91S} proposed a new strategy of finding strongly-lensed supernovae by monitoring known galaxy-scale strong lens
systems. By targeting known strongly lensed starforming galaxies, they proved such strategy could significantly boost the detection
efficiency for lensed SNe compared to a blind search. By monitoring massive clusters,
some magnified supernovae have been discovered~\cite{2014MNRAS.440.2742N,2014ApJ...786....9P,2015ApJ...811...70R,2009A&A...507...71G,2011ApJ...742L...7A,2018ApJ...866...65R}.
However, this case belongs to weak lensing and we do not give a detailed discussion about it here.
Ryczanowski et al. ~\cite{2020MNRAS.495.1666R} discussed on building a cluster watch-list for identifying strongly lensed transients including supernovae.

There have been 4 observed strongly lensed SNe at the time of writing: PS1-afx, SN Refsdal, iPTF16geu and AT 2016jka. They will be reviewed below.
Note most of the observed lensed SNe have high total magnification which could be
partially explained by selection effect. In addition, all events have very small image separations if they belong to galaxy lensing.

\subsubsection{PS1-10afx}
PS1-10afx was a peculiar supernova discovered by the Panoramic Survey Telescope $\&$ Rapid Response System 1 (Pan-STARRS1) in 2010~\cite{2013ApJ...767..162C}.
Its redshift was measured as $z=1.388$.
It was 400 times brighter than the typical core-collapse supernova.
Therefore it was first taken as a new type of superluminous supernovae.
However, its color was found to be unusually red and thus the team conducted follow-up observations including optical and near infrared spectroscopy.
Compared to the models of all known superluminous supernovae, it is much redder (cooler) and evolved much faster.
Thus it could not be explained by superluminous supernovae which have similar high bolometric outputs.
It should be either a supermunious supernova of its own or
a normal supernova magnified by gravitational lensing.

After re-analyzing the photometric and spectroscopic data, Quimby et al.~\cite{2013ApJ...768L..20Q} proposed a possibility that PS1-10afx is a normal type Ia supernova and its unusually high luminosity is
attributed to the magnification effect by strong lensing. Note that in this case, the multiple images are unresolved due to small image separations ($<0.''4$) relative to
the telescope's resolution. Quimby et al. then showed a foreground galaxy at $z=1.117$ with the deep spectra from Keck telescope as a smoking gun of strong lensing scenario~\cite{2014Sci...344..396Q}.
The host galaxy and lens galaxy are superposed and blended in the ground-based images. The lens galaxy is faint which explains the small image
separations and time delays that are needed to be compatible with the observed properties.
Thanks to the standardizable nature of Ia SNe, the magnification factor was directly estimated. It is $\sim31$ times brighter than the standard candle.

\subsubsection{SN Refsdal}
The first lensed supernova with resolved images and time delay measurements was reported by Kelly et al~\cite{2015Sci...347.1123K}. This event is named ``SN Refsdal" to pay tribute to Refsdal who proposed
that lensed SNe can be used to infer $H_0$~\cite{1964MNRAS.128..307R}. It was serendipitously discovered by the Grism Lens-Amplified Survey from Space program~\cite{2015ApJ...812..114T} (GLASS, PI: Tommaso Treu), a follow-up near-infrared grism spectrum program of Hubble
Frontier Fields program aiming at studying distant Universe with the help of gravitational lensing magnification with six targeted massive galaxy clusters~\cite{2017ApJ...837...97L}.
SN Refsdal was located in a face-on spiral galaxy at $z=1.49$ lensed by one of the clusters: MACS J1149.6-2223 at
$z=0.54$. According to the observed light curve and spectra, it was classified as an SN 1987A-like Type II supernova~\cite{2016ApJ...831..205K}.

Interestingly, SN Refsdal along with its host was further lensed by a foreground elliptical galaxy in the cluster as a secondary lens, forming 4 images:
S1-S4 firstly seen in Oct. 2014~\cite{2015Sci...347.1123K}. The time delays among them were measured with follow-up lightcurves and spectra with relatively large uncertainties~\cite{2016ApJ...820...50R}.
According to the mass distribution modelling of the cluster via deep imaging of 100 background galaxies, a prediction was made that the other two lensed images by the cluster should exist~\cite{2015MNRAS.449L..86O,2015ApJ...800L..26S,2016ApJ...817...60T,2016MNRAS.456..356D,2016MNRAS.457.2029J,2016ApJ...822...78G}. One of the images 1.3, was predicted to have appeared $\sim17$ years before S1-S4 while
image 1.2 would appear $\sim1$ year near the center of the cluster from the appearance of S1-S4. However, there existed scatters among different modelling methods/softwares.
Luckily, follow-up monitoring observations with Hubble Space Telescope (HST) indeed observed the new image SX at the predicted position and within the time delay prediction~\cite{2016ApJ...819L...8K} (the time delay between
S1 and SX was measured to $\sim 350$d). SN Refsdal is not Type Ia and we can not get the direct magnification. According to the best-fit model, the total magnification is $\sim74$~\cite{2016ApJ...819..114K}.
SN Refsdal was used to constrain $H_0$~\cite{2018ApJ...853L..31V,2018ApJ...860...94G,2019MNRAS.482.5666W,2020ApJ...898...87G}, although the uncertainties are relatively large because of relatively poor cluster modelling and small time delays with large uncertainties between S1-S4.

\begin{figure*}
\includegraphics[width=15cm,angle=0]{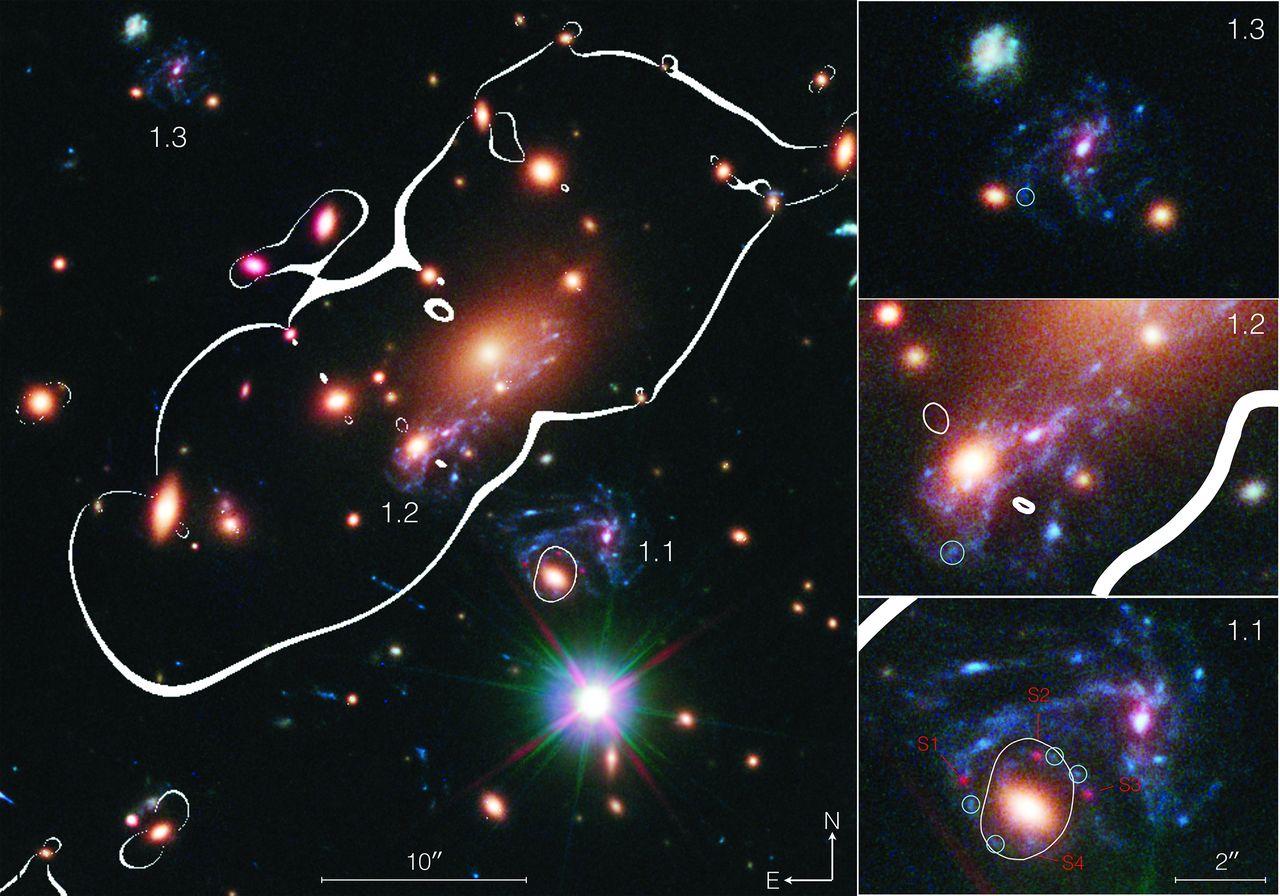}\\
\caption{Color-composite image from Kelly et al~\cite{2015Sci...347.1123K} with white critical curves.
Three images of the host galaxy formed by the cluster are marked with labels (1.1, 1.2, and 1.3) in the left panel, and each is enlarged at right.
The four images of SN Refsdal (labeled S1 to S4 in red) appear as red point sources in image 1.1 with lensed images of a bright blue knot (cyan circles).
}
\label{snrefsdal}
\end{figure*}

\subsubsection{iPTF16geu}
iPTF16geu is the first lensed Type Ia SN observed with resolved images~\cite{2017Sci...356..291G} from the intermediate Palomar Transient Factory~\cite{2013ATel.4807....1K}, a time-domain survey who searches
the sky for new transients at optical wavelengths. iPTF16geu was first detected with a statistical significance of five standard deviations on 2016 September 5.
It was first recognized near the center of galaxy SDSS J210415.89-062024.7 by a human scanner on September 11. Spectrum showed it was a normal SN Ia at $z=0.409$.
Follow-up with HST showed a Type Ia SN  lensed by a foreground galaxy at $z=0.216$ into 4 images. As in the case of PS1-10afx, the standardizable nature gives
a direct measurement of the total magnification $\sim52$. The lens galaxy is a low mass galaxy with velocity dispersion of $\sim160$ km/s leading to a relatively
small image separation $\sim0.''6$ and small time delays. The observed light curves was analyzed by Dhawan et al.~\cite{2020MNRAS.491.2639D} who gave time delays with respect to the brightest
image $-0^d.23\pm0^d.99$, $-1^d.43\pm0^d.74$ and $1^d.36\pm1^d.07$, respectively. They also considered the extinction and gave the extinction-corrected total magnification of $\mu\sim67$.

Modelling of iPTF16geu was also conducted in~\cite{2017ApJ...835L..25M,2020MNRAS.496.3270M}. It was found that while the SN image positions and host arcs were well fitted by a simple singular isothermal ellipsoid plus shear model, the flux ratios between multiple images showed obvious anomalies. These anomalies were partially or all attributed to the microlensing effect~\cite{2017ApJ...835L..25M,2020MNRAS.496.3270M,2022A&A...662A..34D}, too simplistic
assumption on the macro mass model or perturbation by substructures in the lens galaxy~\cite{2017arXiv171107919Y,2018MNRAS.478.5081F}.

\subsubsection{AT 2016jka}
In 2021, Rodney et al.~\cite{2021NatAs...5.1118R} reported a gravitationally lensed supernova by a galaxy cluster with an observable two-decade time delay: AT 2016jka (``SN Requiem"). They discovered it using data from HST program REsolved QUIEscent Magnified Galaxies (REQUIEM, HST-GO-15663, PI: Akhshik)~\cite{2021ApJ...907L...8A}, which targets massive galaxies with low specific star formation rates that have been magnified by strong lensing. The brightest target is a massive red galaxy MRG-M0138 at $z=1.95$ quadruply lensed by the foreground galaxy cluster MACS J0138.0-2155 at $z=0.338$. The authors discovered three point sources that were present in archival HST images from 18-19 July 2016 (HST-GO-14496, PI: Newman). They inferred that these are multiple images of a single astrophysical transient in MRG-M0138, most likely a type Ia supernova. According to lens modelling with HST imaging including a separate multiply-imaged [OII]-emitter, they predicted that a fourth image would appear close to the cluster core in the year 2037$\pm2$ demagnified with $\mu=0.4\pm0.2$, observation of which could provide a time delay precision of $\sim7$ days, $<1\%$ of the extraordinary 20 year baseline. A fifth image would exist but be significantly demagnified ($<0.01$) and outshone by the central bright region of the cluster. Therefore it is hard to observe even for James Webb Space Telescope.

\subsection{Lensed GRBs}
Gamma-ray bursts (GRBs) are short, intense and one-off flashes of $\sim \mathrm{MeV}$
gamma-rays with
a wide range of spectral and temporal properties~\cite{1995ARA&A..33..415F,2002ARA&A..40..137M}.
Such energetic prompt emission is followed by afterglow emissions from
X-ray to radio bands~\cite{2000ARA&A..38..379V}. GRBs were serendipitously discovered in 1967 by the Vela
satellites~\cite{1973ApJ...182L..85K}. Since then, people have been studying
their origin, energy source and progenitor. The data from the Burst and Transient Source Experiment (BATSE) on the Compton Gamma Ray
Observatory showed that GRBs are distributed isotropically which implies the cosmic origin~\cite{1992Natur.355..143M}. Besides, GRBs display a clear bimodal duration
distribution with the separation at 2s~\cite{1992Natur.355..143M}. Duration larger and smaller than 2s are classified as long and short GRBs respectively.
With more and more GRBs detected by Swift satellite~\cite{2004ApJ...611.1005G} and Fermi Gamma-ray Space Telescope~\cite{2009ApJ...697.1071A,2009ApJ...702..791M}, many efforts have been devoted to GRB studies.
Currently, we believe that the progenitors of long GRBs are massive stars based on the facts that they are associated with Type Ic core-collapse
supernovae, their hosts are young star-forming galaxies and their strong correlation with bright UV regions~\cite{2006ARA&A..44..507W}.
On the other hand, short GRBs are thought
to originate from mergers of binary compact objects like neutron stars and neutron star - black hole binary systems~\cite{2014ARA&A..52...43B}. Their hosts are always elliptical galaxies lacking the star
formation and there are no associated supernovae. GRBs and their afterglows can be detected to very high redshifts~\cite{2011ApJ...736....7C}, with a  considerable chance of being lensed.

The discussions on lensed GRBs trace back to 1990s~\cite{1986ApJ...308L..43P,1987ApJ...317L..51P,1993ApJ...402..382M,1993ApJ...406...29W,1994ApJ...435..557N,1994ApJ...435..548G}.
There were searches for GRB repeaters~\cite{1993MNRAS.265L..59Q,1996ApJ...466..757T,1998A&A...336...57G}.
No lensed signals were found in~\cite{1994ApJ...432..478N,2001PhRvL..86..580N,2009arXiv0912.3928V,2019ApJ...871..121H,2020ApJ...897..178A}.
Hirose et al.~\cite{2006ApJ...650..252H} found one candidate that might be lensed by a Population III star at redshift 20-200.

Strongly lensed GRB appears as multiple images. However, the angular separation is usually too small for current observations.
One can only resolve them via different arrival times. For point mass lens model, the time delay is roughly $\sim$ few$\times(M_d/30M_\odot)$ milliseconds, where $M_d$ is the mass of the lens.
Strategies to identify a lensed GRB could be classified in two classes: cross-correlation between signals and auto-correlation in a signal.

For the first case, time delay between signals should be much larger than the signal duration itself such that we can compare two lensed signals.
For example, macrolensing of GRBs by galaxies means a GRB
recurring with the same light curve and spectrum as the previous one, but with a time delay, a different flux and a small positional
offset~\cite{2020ApJ...897..178A}. Detection of strongly lensed GRBs by galaxies or clusters could be extremely challenging due to the poor spatial resolution,
the inefficient duty cycle and limited sky coverage. Besides, microlensing may change the signals making the confirmation hard.
Therefore, GRBs lensed by smaller-scale lenses, for example, intermediate-mass black holes or dark matter subhalos, are easier since the corresponding time delays are much smaller such that we can detect them in one search. In such case, if the lens mass/time delay is large enough such
that signals are well separated, we can use either cross-correlation or autocorrelation technique
depending on whether we take the two signals as a whole.
For the second case, the time delay is smaller than the duration such that the second signal is superimposed on the first one. Note that the time delay
should be larger than the minimum variability timescale of the intrinsic burst so that one could resolve them by correlation between the displaced
features in the overlaid images, which will generate a peak in autocorrelation of the total light curve~\cite{2018PhRvD..98l3523J}.

In addition to GRBs themselves, lensing of afterglows were discussed especially for
microlensing caused by a group of stars~\cite{1998ApJ...495..597L,2001MNRAS.325.1317K,2001ApJ...547L..97M,2001ApJ...551L..63G,2001ApJ...558..643G,2001ApJ...561..703I,2005ApJ...618..403B}.
Depending on the model of the afterglow emission, for example, a narrow emission ring~\cite{1998ApJ...495..597L}, the
light curve and polarization would be distorted according to certain patterns.
Chen et al~\cite{2022ApJ...924...49C} discussed strong lensing of GRB afterglows where the images are unresolved. They pointed out that in this case the X-ray afterglows are likely to
contain several X-ray flares of similar width in the linear scale and similar spectrum while the optical afterglow lightcurve shows
re-brightening signatures.

Among the $\sim10^4$ observed GRBs, there have been a handful of claimed lensing detections. However, some of them are not statistically robust. Currently, there are a few reported lensed GRB signals with high confidence levels.

\subsubsection{GRB 950830}
GRB 950830 was the first lensed GRB uncovered by Paynter et al.~\cite{2021NatAs...5..560P}.
The team initially wanted to find some lensed signals by intermediate-mass black holes.
They analysed the BATSE dataset containing $\sim2700$ bursts including both
long and short GRBs. In their search, they required the time delay between the burst and its echo being $<240$s. The minimum detectable time delays are
$\sim 1$s and $\sim 40$ms for long bursts and short burst, respectively, which are limited
by the $\gamma$-ray pulse widths. The time delay range corresponds to a lens mass range of $10^2-10^7M_\odot$.

Auto-correlation algorithm was used to find preliminary candidates. They utilised the four available broadband energy channels of BATSE independently.
 For each candidate, they
employed Bayesian model selection to determine the Bayesian odds comparing the lensing hypothesis to non-lensing hypothesis.
At the same time, two parameters: time delay and flux ratio are well-constrained, being independent of photon energy according to the equivalence principle .
They discovered one statistically significant lensed candidate: GRB 950830 (BATSE trigger 3770) which is a short burst. The echo has similar light curves
in all four channels. The time delay plus flux ratio yields the (redshifted) mass of the deflector $(1+z_d)M_d=5.5^{+1.7}_{-0.9}\times10^4M_\odot$ (90$\%$ credibility) assuming a point mass lens model.
This value corresponds to an intermediate-mass black hole. 

\begin{figure*}
\includegraphics[width=10cm,angle=0]{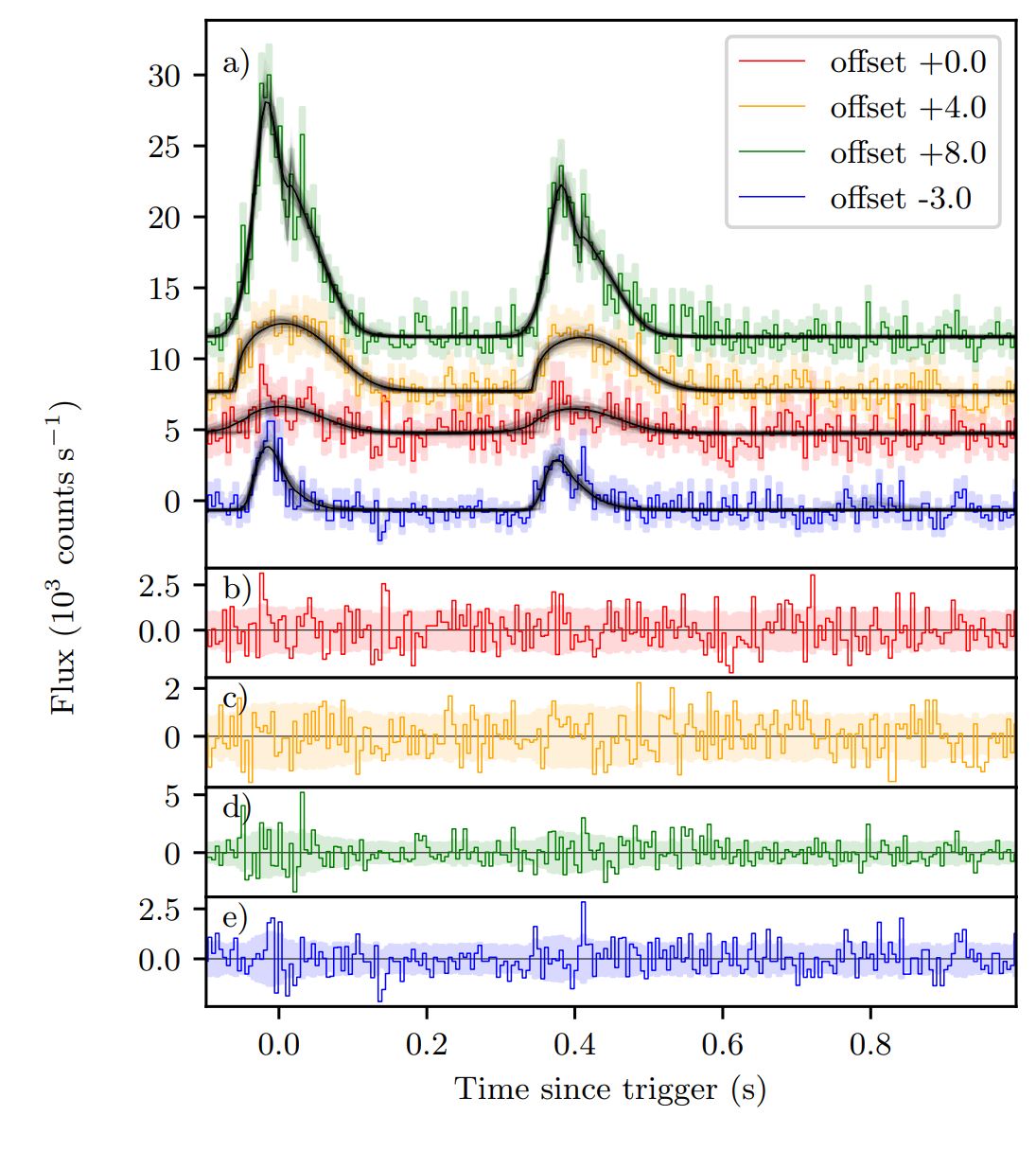}\\
\caption{GRB 950830 from Paynter et al.~\cite{2021NatAs...5..560P}. a) The light curve
is the pre-binned 5ms tte BFITS data. Each colour indicates a different energy channel, red: 20 - 60 keV, yellow:
60 - 110 keV, green: 110 - 320 keV, blue: 320 - 2, 000 keV. The coloured shaded regions are the 1-$\sigma$ statistical errors
of the gamma-ray count data. The solid black curves are the posterior predictive curves. b-e) The lower panels are the difference between the true light curve
and the posterior predictive curve.
}
\label{grb950830}
\end{figure*}

\subsubsection{GRB 200716C}
GRB 200716C was observed as a possible long GRB in the first place by multiple satellites.
The Fermi GBM team reported the detection on 16 Jul 2020~\cite{2020GCN.28135....1V}. The GBM light curve consists of two separated pulses with
a total duration $T_{90,\mathrm{GBM}}\sim5.3s$ in the 50-300keV band. The SWIFT/BAT signal has duration $T_{90,\mathrm{BAT}}\sim5.44s$~\cite{2020GCN.28136....1B}. HXMT/HE detected
GRB 200716C in a routine search which has a duration of $T_{90,\mathrm{HXMT}}\sim2.16s$~\cite{2020GCN.28145....1X}. The duration $>$ 2s and thus it belongs to
long GRB in principle.

Wang et al~\cite{2021ApJ...918L..34W} searched a large GRB database including 3099 Fermi/GBM events, 1297 Swift/BAT events and 311 HXMT/HE events and found that
GRB 200716C was a promising candidate detected simultaneously by all these instruments. According to both auto-correlation and Bayesian inference analysis,
they found that two pulses have highly correlated light curves in all energy channels and their temporal evolutions of the spectra were quite similar.
They also found that both pulses as well as the whole burst belong to short GRBs in the Amati diagram. The flux ratios are constant in all channels.
If the event is indeed lensed, the time delay is $\approx$1.9s. Time delay together with flux ratio leads to the redshifted lens mass $4.25_{-1.36}^{+2.46}\times 10^5M_\odot$.
Considering the uncertainty of lens redshift $<z_s$, the lens could be an intermediate-mass black hole.
At the same time, Yang et al~\cite{2021ApJ...921L..29Y} independently gave a similar analysis and conclusion. They presented a Bayesian analysis identifying gravitational lensing in
both temporal and spectral properties. Their inferred lens mass was a little smaller than Wang et al. One reason may be that Wang et al. included the BGO and HXMT
data in the analysis.

\subsubsection{GRB 081126A and GRB 090717A }
Lin et al.~\citep{2022ApJ...931....4L} gave a systematic search for millilensing among 3000 GRBs observed by the Fermi GBM up to 2021 April. Eventually they found 4 interesting candidates by performing auto-correlation test, hardness test and time-integrated/resolved spectrum test. According to their analysis, GRB 081126A and GRB 090717A were ranked as the first class candidates based on their excellent performance both in temporal and spectrum analysis. If it is true, they corresponds to lens mass scale of $\sim10^6M_\odot$. The other two: GRB 081122A and GRB 110517B were ranked as the suspected candidates, mainly because their two emission episodes show clear deviations in part of the time-resolved spectrum or in
the time-integrated spectrum.

\subsubsection{GRB 210812A}
Veres et al.~\cite{2021ApJ...921L..30V} presented Fermi-GBM observations of GRB 210812A showing two emission episodes, separated by 33.3s with flux ratio of about 4.5. An exhaustive temporal and spectral analysis
showed that the two emission episodes have the same pulse and spectral shape, which poses challenges
to GRB models. They reported multiple lines of evidence for a gravitational lens origin. Assuming a point mass lens, the mass of the lensing object is about 1 million solar masses.

\subsection{Lensed FRBs}
Fast radio bursts (FRBs)~\cite{2019ARA&A..57..417C,2019A&ARv..27....4P} are bright radio transients with extremely short durations $\sim \mathrm{ms}$ and cosmic origin.
Since the first discoveries in 2007~\cite{2007Sci...318..777L} and 2013~\cite{2013Sci...341...53T}, hundreds of FRBs have been detected at frequencies from several hundred MHz to several GHz by radio telescopes such as Parkes~\cite{2016MNRAS.460L..30C},
UTMOST~\cite{2016MNRAS.458..718C}, ASKAP~\cite{2018Natur.562..386S}, CHIME~\cite{2021ApJS..257...59C} and FAST~\cite{2021arXiv211007418N}. The catalog can be searched in Transient Name Server (https://www.wis-tns.org/).
Photons with smaller frequencies propagate slower in plasma and arrive later, thus
FRB searches are based on forming a large number of time series corresponding to different amounts of dispersion over
a wide range. The dispersion measure ($\mathrm{DM}_e$ (To distinguish from the abbreviation of dark matter, we use the subscript ``e" standing for electron.)) is measured by the time delay between the highest and lowest frequencies of the observation and
determined by the integrated electron number density along the light of sight quoted in units of $\mathrm{cm^{-3}pc}$.
The excess (relative to the contribution of the Milky Way) of $\mathrm{DM}_e$ implies a cosmological origin. FRB 121102 that has been precisely localized to a dwarf galaxy at
$z=0.193$ further reinforces the extragalactic origin of these events~\cite{2017ApJ...834L...7T}. While the majority shows single bursts, a small fraction $\sim20$ shows repeated emission~\cite{2016Natur.531..202S,2021arXiv211007418N}, separated
by a wide range of time scales from seconds to years, implying they may not be catastrophic events. It is possible, however, that all FRBs be repeat since some
bursts may be too faint or only one burst was captured. Although the origin of FRB is still puzzling, FRBs can be used to study the history of the Universe.
According to the detection rate and the field of view of radio telescopes, FRB event rate should be $\sim10^4 \mathrm{sky^{-1}day^{-1}}$.
Besides, one may detect them at very high redshifts~\cite{2018ApJ...867L..21Z}.
Therefore, detecting lensed FRBs is very promising.

Lensed FRB signals are similar to the case of GRBs. Since the signal duration scales of FRBs are much smaller than GRBs', they are sensitive to lensing by much smaller mass objects.
For point mass lens, time delay typically depends on mass $\sim$ few$\times(M_d/30M_\odot)$ milliseconds.
Motivated by the opportunity that FRB lensing can be used to constrain compact dark matter~\cite{2016PhRvL.117i1301M} (see next section),
Liao et al.~\cite{2020ApJ...896L..11L} analyzed 110 FRBs and found no lensed signals. Their search was based on the
fact that compact dark matter objects could act as lenses and cause splitting of FRBs. They proposed to use dynamic spectra to
identify any lensed signals. Using cross-correlation method in dynamic spectra, Zhou et al.~\cite{2022MNRAS.511.1141Z,2022ApJ...928..124Z} searched the latest $\sim600$ FRB events and found no lensed FRBs either.
Furthermore, FRBs were found to have much smaller substructure and lensing signature can be identified even in an unsplit signal.
For example, Sammons et al~\cite{2020ApJ...900..122S} discussed the case with the resolution of temporal structures of $\sim10\mu$s.
Leung et al~\cite{2022arXiv220406001L} used a novel interferometric
method to search for lensed FRBs with a time resolution of 1.25ns corresponding to very tiny lenses. They found no lenses in 172 bursts.

Compared to strong lensing of GRBs by galaxies and clusters, strongly lensed FRBs, especially repeating FRBs, are promising to detect with high spatial resolution of radio telescopes.
In this review, we give a comment on how to identify macro-lensed repeating FRB signals from radio data.
In other words, what are the signs of macro-lensed repeating FRB signals?
For a series of bursts, if one can identify any two pairs of bursts that have the same time delay, flux ratio and
$\mathrm{DM}_e$ difference, it could be a candidate.
However, due to limited observing strategies, one may not get any complete lensed pairs.
The most probable way to identify lensed repeating FRBs would be through the inconsistent source positions. Unlike unlensed
repeating FRBs whose multiple bursts have a common location, lensed repeating FRBs by elliptical galaxies would imply 2-4 location centers
with typical separations $\sim$1 arcsecond, corresponding to multiple images. Er $\&$ Mao~\cite{2022MNRAS.516.2218E} studied the effects of plasma on the magnification and time delay of strongly lensed FRBs.
The radio signals of lensed FRBs will be deflected by plasma in lens galaxies in addition to gravity~\cite{2014MNRAS.437.2180E}. Such deflections by both gravity and plasma will cause frequency dependent time delays, which show distinguishing behaviours either between the multiple images, or from the dispersion relation. Such information provides not only a potential way to search for lensed FRBs, but also constraints on the mass and plasma distributions in the lens galaxy.
Connor and Ravi~\cite{2022arXiv220614310C} recently studied the rates of FRB lensing and the applications. Xiao et al.~\cite{2022arXiv220613534X} calculated the lensing rates and proposed to detecting cosmic strings with lensed fast radio bursts.

\subsection{Lensed GWs}
Einstein's General Relativity is currently the most successful theory of gravity.
In GR, dynamical perturbation of spacetime can appear as gravitational wave~\cite{2009LRR....12....2S}.
In 2016, LIGO and Virgo Collaboration~\cite{2016PhRvL.116f1102A} claimed the first direct detection of GW signal:
GW150914 from the merger of two black holes with masses $\sim$30$M_\odot$, which
occurred  $410\;\mathrm{Mpc}$ away. After this breakthrough, $\sim$90 GW events have been detected with the upgrade of the detector network~\cite{2021arXiv211103606T}.
Detections of GWs help
to understand the astrophysical processes in compact star mergers and mark the advent of the era of gravitational wave astronomy.
Among these GW events, the merger of binary neutron stars system (GW170817) with a wide range of electromagnetic (EM) counterparts~\cite{2017PhRvL.119p1101A} is important since it
opened the multi-messenger astronomy.
In addition, mergers of black hole-neutron star binaries are expected to have EM counterparts, though no EM emission has been detected in two such events detected so far~\cite{2021ApJ...915L...5A}.
At present, the observed gravitational waves
all occurred at relatively low redshifts (smaller than $z\sim0.1$).
The third-generation ground-based gravitational wave detectors,
such as the European Einstein telescope~\cite{2012PhRvD..86l2001R} and
American Cosmic Explorer~\cite{2017CQGra..34d4001A} will be able to explore
gravitational wave bursts with redshifts of z$>$2, which
would make it possible to use gravitational wave bursts as cosmic probes to study the Universe.

Naturally, one could discuss the gravitational lensing effects of GWs since
they propagate along geodesic lines like EM waves~\cite{1996PhRvL..77.2875W}.
From the observational perspective, GWs and EM waves are substantially different. For common EM sources of astrophysical relevance,
the wave lengths are much smaller than the scales of the lenses. Different parts of the source emit independent EM waves, hence the emission is incoherent. Therefore,
geometric optics is sufficient for the description.
For GW sources of common concern, the wavelengths are much larger than those of EM waves, sometimes comparable to the scales of
the lenses that we consider. Furthermore, the emission is coherent. Therefore, one should note that sometimes wave optics is more appropriate for GW lensing~\cite{1998PhRvL..80.1138N,2003ApJ...595.1039T,2020PhRvD.101f4011H}.

Verification of strongly lensed GWs can happen according to three approaches.
Firstly, for a single signal, coming from the standard siren (chirping binary), whenever the redshift information is acquired, for example from its EM counterparts or assuming a prior on the chirp mass, one can compare the distances expected from
a cosmological model (or by matching standard candles in $z$) and directly from the GW signal. If the GW is magnified by lensing, it shows a smaller distance. In turn, the magnification effect can make neutron stars/black holes look heavier~\cite{2018MNRAS.475.3823S,2022arXiv220412978B}.

Secondly, if one has two GW signals in hand (suspected of coming from lensed images), consistency test can identify whether they are from the same source. The signature of
strongly lensed GW signals would be that they differ only by amplitude having the same duration, frequency drift, rate of change of the amplitude (i.e. the chirp), and come from the same location strip on the sky~\cite{2021arXiv211205932W}. The amplitude scale of
the signal could also be affected by the detector orientation factor changing
between the arrivals of lensed signals due to rotation of the Earth, but this could be
accounted for once the time delay is known. Haris et al.~\cite{2018arXiv180707062H} gave a pipeline to identify lensed GW signals, where they suggested the lensed signals
would correspond to the same set of inferred parameters except luminosity distance biased by magnification. For the given pair of signals,
they computed the odds ratio between two hypotheses: lensed and unlensed cases. Then they obtained Bayes factors to assess the lensing probability.

Thirdly, for a single signal (corresponding to the specific lensed image), the waveform
would be slightly distorted by lensing~\cite{2021ApJ...923L...1J}.
A lensed template can be used to distinguish lensing signal from unlensed one. There are 3 mechanisms. Firstly, the
Morse phase in strong lensing for the saddle point image changes the waveform even in the geometric-optics limit~\cite{2017arXiv170204724D,2021PhRvD.103j4055W}.
In other words, type II images cannot be fully matched by unlensed templates. Secondly, for binary neutron stars, the distortion will appear at higher orders if we know
the equation of state of neutron stars. Finally, if the lens scale is comparable to GW wavelengths, the diffraction by the lens object would imprint the waveform~\cite{1998PhRvL..80.1138N,2003ApJ...595.1039T,2014PhRvD..90f2003C,2020PhRvD.102l4076G}. Moreover, due to unresolved nature of images in GW lensing only the last arriving signal would have the intrinsic chirp pattern. Earlier signals interfere with each other, which distorts the waveform, primarily in the inspiral phases well before the coalescence.

Efficient detection strategies would help to increase the number of registered lensed events. For example, a targeted sub-threshold search for strongly-lensed GWs ~\cite{2019arXiv190406020L} (since multiple signals increase the signal-to-noise ratio),
or searches combined with optical surveys~\cite{2022arXiv220412977S,2022arXiv220408732W}. Janquart et al.~\cite{2022arXiv220511499J} suggested that a lens model be important for GW lensing searches.

Many works have calculated the rates of GW events lensed  by galaxies or clusters.
Lensed GWs are expected to be detected even by the ongoing detectors~\cite{2018PhRvD..97b3012N,2021ApJ...921..154W}.  For
the third-generation detectors, there will be $\sim100$ lensed GWs per year~\cite{2013JCAP...10..022P,2015JCAP...12..006D,2018MNRAS.476.2220L,2022MNRAS.509.3772Y}, most of which
are binary black hole mergers. For space detectors like LISA, there will be several such lensed events per year~\cite{2010PhRvL.105y1101S}. The probability becomes higher if wave-optics effects are considered~\citep{2022arXiv220602803C}. It was found that there can be a few tens to a few hundreds of lensed gravitational wave events per year with the beat pattern observed by DECIGO and B-DECIGO~\cite{2021PhRvD.103d4005H}.

Hannuksela et al.~\cite{2019ApJ...874L...2H} searched for signatures of GW lensing detected by Advanced LIGO and Virgo during the first two observational runs.
They looked for all three types: 1) strong lensing magnification in individual signals; 2) multiple images due to
strong lensing by galaxies; 3) diffraction signals by point-mass lenses. After a detailed analysis, they found no compelling evidence of any
of these signatures.

LVC team~\cite{2021ApJ...923...14A} searched for GW lensing in O3a data. Several pairs of signals showed similar parameters. However, taking into account population priors,
selection effects and the prior odds against lensing, they claimed no compelling evidence was found either.
On the other hand, Diego et al~\cite{2021PhRvD.104j3529D} reanalyzed the data and claimed a higher probability for the evidence.

Dai et al.~\cite{2020arXiv200712709D} searched for lensed GWs including Morse phase information in O2. The Morse phase information would help to constrain considerably viable lenses and identify lensing systems. They found a candidate consisting of GW170104, GW170814 and a sub-threshold GWC170620
from a BBH merger. A similar analysis was done in~\cite{2021ApJ...908...97L}.

Kim et al.~\cite{2022arXiv220608234K} presented the result of a deep learning-based search for beating patterns. They examined the binary black hole events in the first and second gravitational-wave transient catalogs. The search identifies beating patterns induced by lenses with masses between $10^3-10^5M_\odot$. They found no significant evidence of beating patterns from the evaluated binary black hole events.

\section{Applications of lensed transients}

The significance of studying lensed transients is due to the following reasons. Firstly, such detections can further verify and test the General Relativity and deepen our
understandings on the nature of gravity. Secondly, the signal-to-noise ratio can be improved through the lens amplification effect thus enabling us to detect transients at higher redshifts.
Thirdly, such a study would provide new templates for the GW detectors, enrich the template database for matched filtering technique and consequently enlarge the resulting catalogs of events. Fourthly, it could play a key role in solving specific problems in astrophysics and cosmology.
Finally, the strong lensing is a source of systematic error that needs consideration. In the following, we present the applications of strongly lensed transients under the headings of: fundamental physics, astrophysics and cosmography.

\subsection{Fundamental physics}

\subsubsection{Wave nature of GWs}

The nature of light had once been fiercely debated as either particles or waves.
Young's double-slit experiment, Kirchoff-Fresnel and Fraunhoffer's diffraction supported
the wave nature, while Einstein's photoelectric effect and Compton scattering revealed its particle nature. Finally, the consensus was
reached within wave-particle duality. Naturally, one may conjecture that GWs should have the same properties.
Recent detections of GWs have confirmed the classical nature of this phenomenon. Nevertheless,
to gain a better understanding of the physical nature of GW, in particular to
better investigate its wave nature within or beyond GR, it is essential to observe more features related to the wave-nature of registered signals.

In diffraction experiments, a slit or an obstacle is supposed to exist.
For GWs from distant Universe, the lens object plays the role of diffraction obstacle.
For the signals from mergers of binary stars, the waveform would be distorted by the lens and
the impact is frequency-dependent~\cite{1998PhRvL..80.1138N,2003ApJ...595.1039T,2018PhRvD..98j4029D,2018PhRvD..98j3022C}.
Liao et al~\cite{2019ApJ...875..139L} proposed an observational strategy by monitoring the amplitude modulation of
lensed sources of continuous gravitational signals in microlensing case. In other words, they wanted to measure the spatial diffraction/interference fringes.
In the case of monochromatic light, a characteristic fringe pattern is visible on the screen. In astrophysical
scales, the ``screen'' has astronomical scales and is inaccessible directly. This corresponds to
a fixed $\beta$ parameter (see Section II). However, in the microlensing case \cite{pacz1991}, $\beta$ changes in time (due to relative
motions) and one can explore the ``screen'' by watching the manifestation of fringes.
Discussions on continuous GWs lensed by galactic supermassive black hole can be found in~\cite{2022arXiv220500022B} (for a review of strong lensing of continuous GWs -- see \cite{universe7120502}).

Oguri $\&$ Takahashi~\cite{2022arXiv220400814O} discussed how small-scale density perturbations on the Fresnel scale affect amplitudes and phases
of gravitational waves that are magnified by gravitational lensing in the geometric optics regime.
The interference between GWs from lensed images might generate beat patterns, which would bear important information about the mass of the lens and the time delay distance~\cite{2020PhRvD.101f4011H,2021arXiv211210773B,2009PhRvD..80d4009I,2021PhRvD.104b3503C}.
Lensing in hierarchical triples in galactic nuclei with space-borne detectors were discussed in~\cite{2020PhRvD.101h3031D,2021PhRvD.104j3011Y} where they claimed stellar-mass BBHs have a high probability to merge in the vicinity of a supermassive black hole and this approach offers new testing grounds for strong gravity.

\subsubsection{Propagation of gravitational and electromagnetic signals}

Within the framework of GR, GWs propagate with the speed of light $c \equiv c_{\gamma}$, i.e. along null geodesics. This may no longer be the case in modified gravity theories.
The consistency between the registered GW signals and theoretical templates built within the GR places very tight constraints on the nature of gravity in the strong field limit.
In some alternative theories of gravity, however, significant deviations could exist in the weak field limit. For example, the
propagation speed of GWs could be different: $c_\mathrm{GW}\neq c_{\gamma}$. This case is essentially equivalent to the graviton having a non-zero mass. The absence of detectable dispersion in the GW signals places a tight limit on the graviton mass. LIGO/Virgo collaboration~\cite{PhysRevD.100.104036} derived the bound of $m_g \leq 4.7 \times 10^{-23} \; eV/c^2$ for the graviton mass based on the analysis of events registered during the first and second observing run contained in the GWTC-1 catalog.
Probing the difference between $c_\mathrm{GW}$ and $c_{\gamma}$
can be a test of GR against alternative gravity theories. Measuring the arrival time of gravitational waves and
electromagnetic counterparts is an obvious way of measuring the relative speeds. However, it is reliable only if the intrinsic
time-lag between emission of the photons and gravitational waves is well understood. Attempting to circumvent this limitation, Collet $\&$ Bacon~\cite{2017PhRvL.118i1101C} and Fan et al.~\cite{2017PhRvL.118i1102F} proposed to test
the GW speed from  GW and EM signals strongly lensed by galaxies. Conceptually, this method does not rely on any specific theory of massive gravitons
or modified gravity. Its differential setting, i.e., measuring the difference between time delays in GW and EM domains, makes it robust
against lens modelling details and against internal time lag between GW and EM emissions.

Even though assuming EM and GW intrinsically emit at the same time, the arrival time differences between GW and EM signals may exist depending on the gravitational lensing described by either geometric-optics or
wave-optics~\cite{2017ApJ...835..103T,2019arXiv191107435M}. Note, however, that this viewpoint is controversial~\cite{2020ApJ...896...46S,2020PhRvD.102b3531E}. Similar discussions can be also found in~\cite{2021JCAP...06..050T}.

If gravitons have mass, the dispersion relation and speed of gravitational waves will be affected in a frequency-dependent manner, which would leave traces in the diffraction pattern if the waves are lensed. Chung $\&$ Li~\cite{2021PhRvD.104l4060C} propose to use lensing of GWs as a novel probe of graviton mass.

Finke et al~\cite{2021PhRvD.104h4057F} proposed to probe the GW propagation in modified gravity theories with strongly lensed coalescing binaries basing on the fact that the luminosity distance inferred from standard sirens in modified gravity would be different from that obtained from electromagnetic standard candles.

The speed of light $c$ is commonly considered as a fundamental constant of Nature. However, there have been ideas that it might have been an effective one allowing it to vary in time at cosmological scales. Such varying speed of light theories could be an alternative to inflation \cite{Albrecht99,Barrow99} and even explain the scale-invariant spectrum of Gaussian
fluctuations in the cosmic microwave background data \cite{Magueijo03}.  Although $c$ has been measured with very high precision,
however, most of these measurements were carried out on
Earth or in our close cosmic surroundings. Precise measurements of the speed of light using extragalactic objects
seem to be still missing.
Cao et al.~\cite{2018ApJ...867...50C} proposed a new test of the speed of light over cosmological distances by using the combination of strongly lensed and unlensed SN Ia. The transient nature of SN Ia was a crucial element of this idea.

\subsubsection{GW Polarization}

GWs are tensor perturbations, thus their polarization is worth testing.
Hou et al~\cite{2019PhRvD.100f4028H} discussed the rotation of polarization plane in gravitational lensing of GWs and found it is negligible in most cases due to the small deflection angles.
While GR predicts that GWs have only two polarization modes, some theories of gravity can predict up to six polarization modes.
To extract all polarization modes, there should be at least as many detectors as polarization modes. Therefore, a detector network allows us to study the nature of GW polarizations. Currently, this approach is limited by the number of independent detectors.
Since strongly lensed GWs have multiple signals arriving at the Earth which rotates, they naturally ``increase'' the (effective) detector number~\cite{2021PhRvD.103b4038G}.

\subsubsection{Lorentz Invariance Violation}

In some quantum gravity (QG) theories, the picture of a space-time foam at short distances leads to Lorentz Invariance Violation (LIV) \cite{2022PrPNP.12503948A,2013LRR....16....5A,2001PhRvD..64c6005A,1998Natur.395Q.525A}.
Many QG theories predict that LIV could be revealed at high energies close to the Planck energy scale.
In particular LIV can manifest itself by the energy dependent modification of standard relativistic dispersion relation. In such case, photon velocity in the vaccum
depends on its energy such that high-energy photons have different velocity compared to low-energy photons. By measuring the arrival
time difference of the photons originating from the same astrophysical source in different energy bands, one can give a constraint
on the energy scale for LIV: $E_\mathrm{QG}$.

GRBs were proposed to test LIV since
they are highly energetic explosions from cosmological distances \cite{1998Natur.395Q.525A}. One can observe the photons from GRB in different energy channels and register time of arrival differences.
If such differences are to serve as a test of LIV, one has to ignore or make assumptions regarding the intrinsic delays for different energy
channels~\cite{2020ApJ...890..169P,2021JCAP...05..029A}. On the other hand, the emission mechanism of GRB has not been fully
understood, especially concerning long GRBs. There is no reason to think that the low and high
energy photons should be emitted simultaneously at the
source, and while detecting distinct signals at different
energy channels, we have no idea which one was sent
first.

Biesiada $\&$ Pi{\'o}rkowska~\cite{2009MNRAS.396..946B} proposed to use gravitational lensing time delays to test LIV in a way, which is fully independent of intrinsic time delays.
In classical GR, the time delays are achromatic (i.e. do not depend of photon's energy). Therefore, a possibility opens up to study time delays induced by LIV. Namely, one may reveal extra delays due to distorted dispersion relation in LIV theories by comparing time delays between signals registered in different energy channels from lensed GRB images, for example, optical-gamma-ray or low energy and TeV high energy photons. Such test would be free from the intrinsic time lags -- they cancel out in the differential setting of this method. In the original paper \cite{2009MNRAS.396..946B} the case of galaxy lensing was considered, while the lensing by point-mass lenses may also be important as the intermediate mass black holes can also serve as lenses at cosmological distances. More discussions on this method can be found in \cite{2022Univ....8..321P}.

As already mentioned, it is quite hard to detect strong lensing of gamma-ray bursts by galaxies. Existing GRB data did not reveal
any case of a GRB lensed by a galaxy. Recently, several possible lensed GRBs were claimed to be identified. Light curves in different energy channels
were similar and used to measure time delays. Time delays and reasonable assumptions regarding distances suggested that the lenses could be intermediate mass black holes modelled by a point-mass.
Lan et al. (in preparation) for the first time applied lensing time delay difference method to test LIV. By directly fitting the time delay data of GRBs 950830 and 200716C, they obtained
the limits on LIV energy scale: $E_\mathrm{QG,1}\geq3.2\times10^9 GeV$ and $E_\mathrm{QG,1}\geq6.3\times10^9 GeV$, respectively.

Similar approach to test fundamental physics can be found in~\cite{2017PhRvD..95f3512B} where the authors gave a general discussion on how gravitational lensing time delays in multi-messenger signals, for example, massless/massive photons, gravitons, neutrinos and axions, originating from the same source, can be used to test corresponding theories or models.

\subsection{Astrophysics}

\subsubsection{Dark matter (sub)structure}
Cold dark matter (CDM) hierarchical model has successfully explained the formation of the large-scale structure with voids and filaments, as well as its composition up to galaxy levels. Actually, the simulations trace the formation of compact halos built from collision-less DM. Baryonic matter, i.e. the gas falls into the potential well of a halo forming the galaxy composed of stars and gas as we observe. This scenario works well up to big galaxies like the Milky Way.
However, it suffers the problem of predicting too much power on smaller scales.
According to simulations,  part of dark matter halos seeding big galaxies should be in the form of subhalos
surviving from tidal stripping process. These numerous sub-halos might be appearing as satellite dwarf galaxies~\cite{2008Natur.454..735D} accompanying big galaxies or globular clusters inhabiting halos. However,
we have only observed much fewer such satellites around the Milky Way or in the nearby galaxies~\cite{1999ApJ...522...82K,2015ApJ...813..109D} (see the recent detections in~\cite{2022arXiv220912422C} and the references therein).
This is the essence of the puzzle called ``Missing Satellite Problem''.
Various mechanisms were proposed to solve this mismatch. Detecting substructure in distant galaxies is important for understanding the nature of dark matter.

Strong lensing provides a possibility to probe dark matter substructure at cosmic distances~\cite{1998MNRAS.295..587M}. Strong lensing theory (see Sec.II) predicts the sequence of image magnifications, given the macro lens model (some analytic relations can be found in~\cite{2008MNRAS.389..398C}). Measurements, which are not in agreement with these predictions are called flux ratio anomalies.
Lensed quasars are used to detect substructure with the observed flux ratio anomalies: while the smooth lens model
can fit the image positions, flux ratio could be impacted by substructure. However, one should be careful of microlensing effect~\cite{2002ApJ...580..685S}, which is not always easy or straightforward. Usually, appropriate bands corresponding to a large source size are chosen such that microlensing by stars would not impact.

In addition, Keeton $\&$ Moustakas~\cite{2009ApJ...699.1720K} proposed that time delay anomalies can be used to detect substructure. The perturbation on time delay is typically smaller than 1 day. To identify such time delay anomalies, one needs a very precise time delay measurement. However, in lensed quasar systems, time delays measured with light curves typically have uncertainties larger than 1 day~\cite{2015ApJ...800...11L}, limiting the test.
Moreover, Tie $\&$ Kochanek~\cite{2018MNRAS.473...80T} pointed out that time delay measurements may be biased by days due to
microlensing effect based on differential magnification of the finite accretion disk. Therefore, lensed quasars can not be used to test substructure by this method.

Liao et al.~\cite{2018ApJ...867...69L} proposed that lensed GW signals from coalescing binaries can have a very precise time delay measurement due to their transient nature. In addition, GWs with long wavelengths should be free of microlensing impact.
By conducting a detailed simulation, they demonstrated that the lensed GWs could be an excellent tool to detect dark matter substructure. This advantage works for other lensed transients like GRBs, FRBs as well.
Furthermore, Pearson et al.~\cite{2021PhRvD.103f3017P} showed
that the timing precision of lensed repeating FRBs could be used to search for long wavelength GWs, for example, sourced by supermassive black hole binaries because the GW source near the FRB host galaxy impacts the timing of burst arrival.

Another intriguing possibility of using strongly lensed GWs to study the properties of DM has been pointed out by Cao et al. ~\cite{2022A&A...659L...5C,2021MNRAS.502L..16C}. Namely, the DM sector of particle physics, even though interacts very weakly (if at all) with charged baryonic matter could be expected to possess its own rich phenomenology of self-interactions. Indeed, the self-interacting DM has been invoked as a cure for the so called core-cusp problem of CDM scenario. Namely, the non-interacting CDM simulations predict steep density profile of the halo with a cusp at the center. On the other hand, there is a plenty of observational evidence that in reality the central density profile is much shallower, exhibiting a core. The solution could be possible self-interaction of DM manifested as an effective viscosity of DM fluid. The core-cusp problem at the scale of dwarf galaxies could be solved with self-interaction cross sections per mass of the order $\sigma_\mathrm{SI}/m_\mathrm{DM} \sim 0.5 - 10\;\mathrm{cm^2/g = 0.9 - 18\; barn/GeV}$, while the same problem at the scale of cluster halo profiles favours weaker self-scattering: $\sigma_\mathrm{SI}/m_\mathrm{DM} \sim 0.2 - 1.\;\mathrm{cm}^2/g = 0.36 - 1.8\; \mathrm{barn/GeV}$. It has been long known that GWs travel through a perfect fluid unaffected. However, dissipative fluid characterized by a non-zero shear viscosity attenuates the GW amplitude
$h_{\alpha, visc} = h_{\alpha} e^{-\beta D}$, where $D$ is the comoving distance and damping parameter $\beta$ is related with DM self-interaction $\sigma_\mathrm{SI}/m_\mathrm{DM} = 6.3 \pi G \langle v \rangle / (c^3 \beta) $. GW attenuation leads to the mismatch between the true luminosity distance and that inferred from standard sirens. It is not easy to test this on real data. In the papers~\cite{2022A&A...659L...5C,2021MNRAS.502L..16C} it has been proposed that strongly lensed signals from transient sources can be used for this purpose. Simulations demonstrated that already with ten strongly lensed transients one would be able to constrain the $\beta$ parameter with the precision of $\Delta( \sigma_\mathrm{SI}/m_\mathrm{DM}) \sim 10^{-4} \;\mathrm{cm^2/g }$ allowing one to differentiate between different scenarios solving the core-cusp problem at dwarf galaxy and cluster scales.

\subsubsection{Compact dark matter, primordial black holes}

One may conjecture that dark matter (or at least a part of it) consists of compact objects (COs), such as
the massive compact halo objects, primordial black holes (PBHs), axion miniclusters and compact mini halos.
Among these candidates PBHs ~\cite{2021JPhG...48d3001G} within the mass range $10-100M_\odot$ are currently been paid much attention,
partially due to the binary black holes detected by
LIGO which fall mostly within this range of masses and purely astrophysical (stellar evolutionary) scenarios leading to such massive binary BHs are not straightforward. Traditional probes of COs are mainly via microlensing of stars~\cite{2007A&A...469..387T} which constrains the population of low mass ($\leq10M_\odot$) COs in nearby galaxies
since the Einstein ring crossing time scales of these lenses correspond to relatively short monitoring time of a survey~\cite{2019NatAs...3..524N}.
Lensed transients are recently being used in constraining COs with the increasing of data, as outlined below.

1) Lensed SNe Ia: The distribution, particularly the skewness of high redshift type Ia SN brightnesses relative to the low redshift
sample can be used to constrain COs~\cite{2007PhRvL..98g1302M}. The latest results give constraints on compact
objects: less than $35.2\%$ (Joint Lightcurve Analysis) or $37.2\%$ (Union 2.1) of the total matter content in the Universe, at $95\%$ confidence-level. The results are valid for masses larger than 0.01$M_\odot$, limited by the size of SN relative to the lens Einstein radius~\cite{2018PhRvL.121n1101Z}.
In addition to statistical effect, we can make a conjecture here that COs could directly leave imprint on a single SN Ia event.
Since we have known the multi-band light-curve templates of SNe Ia, we may probe COs in a single event with distorted light-curves.
This process is similar to SN microlensing analysis~\cite{2018ApJ...855...22G} but with a simple point-mass lens model. One needs to assume a SN
model like W7 or N100 to calculate the microlensing effects.

2) Lensed GRBs: COs as lenses could produce echoes in GRB signals for lens masses $>10M_\odot$. The trailing image
could be overlapped with the previous one in the lightcurve if the lens mass is not sufficiently large. With auto-correlation analysis, one can identify such signals.
The latest results by Ji et al.~\cite{2018PhRvD..98l3523J} showed that currently existing data from Fermi/GBM and Swift/BAT are noise-limited and can not give
a stringent constraint. They used dedicated simulations to capture the relevant phenomenology of the GRB prompt emission and
calculated the signal-to-noise ratio required to detect GRB lensing events as a function of flux ratio and time delay. They found future observatories
having better sensitivity will be able to probe down to the $f_\mathrm{DM}\sim1\%$  level across the [$10M_\odot,1000M_\odot$] mass range. As we discussed in the previous section, the possibly detected lensed GRBs were also used to identify intermediate-mass black holes~\cite{2021NatAs...5..560P,2021ApJ...918L..34W,2021ApJ...921L..29Y} .

3) Lensed FRBs: FRBs have much finer temporal resolution than GRBs.
They were also proposed to constrain compact dark matter~\cite{2016PhRvL.117i1301M}. It was suggested that time delays caused by COs in the
lens mass range $10-100M_\odot$ should be comparable to the FRB widths themselves. Splitting FRB signals by the lens would provide evidence of compact dark matter. On the contrary, a null search can exclude some
regions in the parameter space. Liao et al~\cite{2020ApJ...896L..11L} gave the first such constraint using 110 FRB data. The constraint is comparable to that from wide binaries which could be perturbed by large mass COs~\cite{2009MNRAS.396L..11Q}.
They also made an improved forecast for CHIME and emphasized the importance of using dynamic spectra to identify lensed signals.
Latest results by Zhou et al~\cite{2022MNRAS.511.1141Z,2022ApJ...928..124Z} used 593 FRBs consisting of the first CHIME catalog to constrain COs.
Leung et al~\cite{2022arXiv220406001L} used a novel interferometric
method to search for lensed FRBs with a time resolution of 1.25 ns corresponding to very tiny lenses. They found no lenses after analysing 172 bursts.
Using FRB lensing to probe small-scale structure was also discussed in~\cite{2020PhRvD.102b3016L}.

4) Lensed GWs: COs as the lenses can distort the waveforms of GWs. Thus the distorted waveforms can be used to probe COs in turn. This viewpoint was proposed in~\cite{2019PhRvL.122d1103J}.
Compared to EM waves, GWs have longer wavelengths and are coherent. For ground-based detectors,
when the lens has mass $1-10^5M_\odot$, the diffraction occurs since the chirping wavelengths are comparable to the scale of the lens.
GW signals would be an effective tool to probe the nature of COs,  especially for the third-generation detectors like the Einstein Telescope. Liao et al~\cite{2020MNRAS.495.2002L} predicted that
for a null search of the fringes, one-year observation of Einstein Telescope can constrain the CO density fraction down to $\sim10^{-2}-10^{-5}$ in the mass range
10-100$M_\odot$. Similar work can be found in~\cite{2022MNRAS.509.1358U}.
Probing the nature of dark matter via gravitational waves lensed by small dark matter halos can be found in~\cite{2022PhRvD.106b3018G}.

\subsubsection{Early phase of SNe}

Strong lensing with multiple images provides the possibility to completely observe the trailing images from the beginning.
This opportunity was noted in Oguri et al.~\cite{2003ApJ...583..584O}.
Early observations of SN light curves are critical in constraining the properties of SN progenitors.
The shock breakout process lasts only seconds to a fraction of an hour. In usual observational campaigns, it is easy
to miss out such phase. In the case of gravitational lensing, one can always predict the occurrence of the trailing image.
Suwa 2018~\cite{2018MNRAS.474.2612S} found it possible to predict the time and position of the fourth image of quadruply lensed SNe,
with combined information from the three previous images.
Foxley-Marrable et al.~\cite{2020MNRAS.495.4622F} studied in detail the issue of observing the earliest moments of supernovae using strong lenses.
Assuming a plausible discovery strategy, they predicted that LSST and ZTF will discover $\sim110$ and $\sim1$ systems
where the final image has not appeared yet. Systems will be identified $11.7^{+29.8}_{-9.3}$ days before the final image, leaving
sufficient time for the follow-up monitoring. They also gave a detailed study on Type IIP and Ia SNe.
They concluded that if all Type Ia SNe evolved from the double-degenerate channel, the lack of early blue flux in 10(50)
trailing images would rule out more than $27\%$ ($19\%$) of the population having 1$M_\odot$ main sequence companions at $95\%$ confidence.

On the other hand, type Ia SN microlensing is expected to have an achromatic phase lasting for weeks before the peak~\cite{2018ApJ...855...22G}. The reason is that because for SNe Ia in the early phase, the ratio of intensities in any two bands is roughly constant over all velocity of the shell, resulting in the same impact for different bands and thus microlensing-free color curve. The HOLISMOKES team studied microlensing effect with four SN Ia models. They found different durations of the achromatic phase for different SN Ia models, where the N100 model in this specific case has the shortest duration and the merger model the longest duration~\cite{2021A&A...646A.110H}. In addition, the color curves show differences during the first few days. Therefore, the observed lensing systems may be used to support/exclude SN Ia models via lensed (early)color curves.

\subsubsection{Detection at high redshifts}
Supernovae are associated with the death of stars and therefore their rates as a function of galaxy type or redshift help understand the cosmic history
of star formation. Supernovae
are produced only when masses of progenitor stars fall in a
particular range, from which we can obtain information on
the stellar initial mass function. Lensing magnification effect make higher-redshift events detectable~\cite{2003ApJ...583..584O}.
For example, the latest paper by Rydberg et al.~\cite{2020MNRAS.491.2447R} discussed possibility of detecting strongly lensed supernovae at $z\sim5-7$ with LSST.
They calculated the detection rates at such high redshifts.
Clusters can be cosmic telescopes to detect high-redshift SNe~\cite{2013MNRAS.435L..33P,2016A&A...594A..54P,2018ARep...62..917P,2019PASJ...71...60W}.
The same idea works for GWs~\cite{2019ApJ...881..157B}.
With the increasing of redshift, the chance of strong lensing becomes larger. Besides, weak lensing becomes more important.
Weak lensing plus strong lensing effect would bias the measured flux.
The impact of lensing on the distributions of black holes was discussed in
~\citep{2018MNRAS.480.3842O,2022arXiv220515515H}, where the authors concluded that the black hole mass function would only be impacted at the high-mass end.
Diego explored the probability of extreme magnifications of transients (not limited to the transients in this article) in a cosmological context and included the effect of microlenses near critical curves~\citep{2019A&A...625A..84D}.

\subsubsection{GW localization}
To localize a GW signal, one usually needs its EM counterpart. For black hole mergers without
EM counterparts, current localization ability is quite poor.
Hannuksela et al~\cite{2020MNRAS.498.3395H} proposed to localize merging black holes with sub-arcsecond precision using GW lensing.
Strongly lensed GWs could allow us to localize the binary by locating its lensed host galaxy.
The hard part in this approach is there may be many other lensed galaxies within the sky area spanned by the GW observation. Nevertheless, one can
constrain the lensing galaxy's physical properties with both GW and EM observations. They showed that these
simultaneous constraints allow one to localize quadruply lensed waves to one or at most a
few galaxies with the LIGO/Virgo/Kagra network in typical scenarios. Once we identify the
host, we can localize the binary to two sub-arcsec regions within the host galaxy.
Similar idea can be found in~\cite{2021PhRvD.104j3011Y}.

\subsection{Cosmography}
The Hubble constant $H_0$ corresponds to the current expansion rate of the Universe anchoring the distance scale of the Universe.
There is an ongoing debate about the value of $H_0$. The value measured from the local distance ladders based on
Cepheids calibration is in significant statistical disagreement with the value inferred from the cosmic microwave background (CMB) and the
large scale structure. Assuming the flat $\Lambda$CDM model, CMB observations from Planck gave $H_0=67.4\pm0.5\rm{\ km\ s^{-1}Mpc^{-1}}$~\citep{2020A&A...641A...6P}.
Combining CMB temperature plus polarization anisotropy measurements from Atacama Cosmology Telescope and large-scale information from WMAP, Aiola et al~\citep{2020JCAP...12..047A}
gave $H_0=67.6\pm1.1\rm{\ km\ s^{-1}Mpc^{-1}}$ in good agreement with \textit{Planck}.
The Supernovae, $H_0$, for the Equation of State of dark energy (SH0ES) team used SNe Ia calibrated by Cepheids and parallax distances to get
$H_0=74.03\pm1.42\rm{\ km\ s^{-1}Mpc^{-1}}$~\citep{2019ApJ...876...85R} which is in 4.4$\sigma$ tension with \textit{Planck}. The latest analysis from the team gave $H_0=73.04\pm1.04\rm{\ km\ s^{-1}Mpc^{-1}}$~\citep{2021arXiv211204510R} with distance ladders, which showed the tension with \textit{Planck} as large as $5\sigma$.
This discrepancy would either imply
unknown systematic errors in one or both measurements, or new physics beyond the standard model. Multiple independent and precise
measurement of $H_0$ are essential to allow us better understand the current tension.

Strong lensing provides a one-step distance anchor of the Universe. Time delays between multiple lensed images can be used to measure $H_0$:
\begin{equation}
\Delta t_{i,j} = \frac{D_{\mathrm{\Delta t}}}{c}\Delta \phi_{i,j},
\end{equation}
where $i,j$ correspond to two images. $D_{\mathrm{\Delta t}}=D_dD_s(1+z_d)/D_{ds}$ as a combination of three angular diameter distances is called the ``time-delay distance'' and is inversely proportional to $H_0$.  $\Delta \phi_{i,j}$ is the Fermat potential difference, which can be determined by high resolution imaging of the lesning system combined with spectroscopic data on stellar kinematics of the lens galaxy.
In addition, the angular diameter distance to the lens $D_d$ can be obtained with the stellar kinematic information on the lens galaxy available~\citep{2009A&A...507L..49P}, independently of the line-of-sight density fluctuation~\cite{2015JCAP...11..033J}. Although $D_d$ alone gives a relatively weak constraint on $H_0$, it can break degeneracies among cosmological parameters, particularly for models beyond flat $\Lambda$CDM. In summary, the combination of lensing, time delays and lens stellar kinematic data provides  posteriors of $(D_{\Delta t}, D_d)$ which are used in cosmological parameter inference, particularly $H_0$.

Traditional targets are quasars lensed by elliptical galaxies~\citep{2016A&ARv..24...11T}. The time delays are measured with year-long campaigns focused on collecting lightcurves.
To infer $H_0$, one needs at least three ingredients regarding strong lensing: 1) time delays, 2) Fermat potential differences between lensed images, and 3) line-of-sight mass fluctuations.
For currently available lensed quasars, each measurement has a precision at the level of several percents. However, challenges exist for each aspect. For example, the internal mass-sheet degeneracy~\cite{2020ApJ...892L..27B} and the microlensing time delays~\cite{2018MNRAS.473...80T} may bias the results.

Time-delay cosmography with lensed quasars has achieved much progress, especially by programs: H0LiCOW~\cite{2017MNRAS.468.2590S}, COSMOGRAIL~\cite{2005A&A...436...25E}, STRIDES~\cite{2018MNRAS.481.1041T} and SHARP which have now
upgraded/combined in the TDCOSMO collaboration~\cite{2020A&A...639A.101M}. H0LiCOW~\cite{2020MNRAS.498.1420W} using 6 lenses reached a $\sim2.4\%$ precision. Under the assumption that radial mass density
profile of the lens can be described by a power-law the Hubble constant turned out to be $H_0=74.2\pm1.6\rm{\ km\ s^{-1}Mpc^{-1}}$. For the density model being a composite of NFW dark matter halo and stars described by the surface brightness scaled by a constant stellar mass to light ratio the H0LiCOW team yielded $H_0=74.0\pm1.7\rm{\ km\ s^{-1}Mpc^{-1}}$. Afterwards, considering the internal mass-sheet degeneracy by dark matter in lens galaxy,
TDCOSMO~\citep{2020A&A...643A.165B} obtained  $H_0=74.5_{-6.1}^{+5.6}\rm{\ km\ s^{-1}Mpc^{-1}}$ using 7 lenses. Under the assumption that TDCOSMO lenses are consistent with
SLACS, they find $H_0=67.4_{-3.2}^{+4.1}\rm{\ km\ s^{-1}Mpc^{-1}}$. While efforts are being devoted to lensed quasars, more and more attention is being paid to lensed transients.

Lensed transients have the following advantages over lensed quasars from the perspective of cosmography: 1) Time delays are supposed to be measured much better than quasars~\cite{2017NatCo...8.1148L}. For quasars, time delays
are measured with light curve pairs. However, one needs high cadence, long campaign in a survey or monitoring project to guarantee a precise measurement. The lightcuves are heterogeneous and probably impacted by
microlensing effect~\cite{2018MNRAS.473...80T}. For lensed SNe Ia, the well-known light curve shape makes the time-delay measurements possible on shorter monitoring time. For the other transients, the transient nature makes the time delay quite accurate~\cite{2017NatCo...8.1148L}. 2) Lens modelling for the lensed quasars is relatively difficult because AGNs typically outshine their host galaxies by several magnitudes. For lensed transients, they fade away allowing a simpler reconstruction of the lensed hosts~\cite{2017NatCo...8.1148L}. For lensed SNe Ia, their standardizable nature gives them the potential to directly determine the lensing magnification factor $\mu$, which breaks the degeneracy between the lens potential and $H_0$~\cite{2001ApJ...556L..71H,2003MNRAS.338L..25O}.

Works on using lensed GWs to constrain cosmological parameters include:  ~\cite{2017NatCo...8.1148L} with third-generation ground-based detectors, ~\cite{2011MNRAS.415.2773S} with LISA, ~\cite{2021MNRAS.507..761H} with DECIGO. For lensed FRBs, see the works in~\cite{2018NatCo...9.3833L,2022MNRAS.516.1977G}.

\subsubsection{SN Ia time delays}
In principle, for lensed SNe, the method of measuring time delays is  the same as for the lensed quasars with lightcurves.
The sharp rise and decline of SN Ia light curves make them much more promising for meauring time delays.
However, one should take care of microlensing effect. For SNe Ia, the exploding material has a finite size comparable to the Einstein radii of stars in the lens galaxy and its volume is changing with time. Microlensing may thus bias
the measurements like in quasars. An alternative approach is to use color curves rather than light curves.
Goldstein et al.~\cite{2018ApJ...855...22G} studied how to get precise time delays from the chromatically microlensed images of
SNe Ia. They took LSST simulated data as an example and studied the impact of microlensing on time delay measurements of lensed SNe Ia.
They reached an interesting conclusion that microlensing for lensed SNe Ia is achromatic until 3 rest-frame weeks after the
explosion, suggesting the early-time color curves can be a  microlensing-free time delay tool.
While the uncertainty for the light curve fitting is comparable to the  lensed quasars ($4\%$), the uncertainty is much smaller for color curve fitting (at $1\%$ level)
using an underlying unlensed spectral templates in the achromatic phase. For LSST observation strategy, $70\%$ of lensed SNe Ia should be
discovered during the achromatic phase. If prompt multicolor follow-ups can be obtained, time delay measurements can be very accurate.

One of the aims of the program HOLISMOKES~\cite{2020A&A...644A.162S} is studying lensed SNe. The authors studied the achromatic phase of lensed SNe Ia~\cite{2021A&A...646A.110H}.
They investigated in detail the achromatic phase assuming 4 different photosphere evolution models with various microlensing magnification
configurations determined by local environments of the image. Similarly, they found that on average the achromatic phase lasts about 3 rest-frame weeks or longer for
most cases, but the spread is quite large due to different microlensing maps and filter combinations. In some cases, the phase could be as short as just a few days. The phase is longer for smoother microlensing maps, lower macro-magnifications and larger mean Einstein radius of stars. Further, they found 3 independent LSST color curves exhibiting features such as extreme points or turning points within the achromatic phase, which make them promising for time-delay measurements. These curves contain combinations of rest-frame bands u, g, r, and i and to observe them for typical redshifts of lensed SN Ia, it would be ideal to cover (observer-frame) filters r, i, z, y, J, and H.
They developed algorithms of time delay measurements using machine learning~\cite{2022A&A...658A.157H}. Two machine learning approaches were tested: a fully connected neural network and a random forest algorithm. They found that only the random forest algorithm provides a low enough bias to achieve the demands of precision cosmology. More band data and lower observational noise could significantly improve the measurements. In addition to SNe Ia, the team studied microlensing of type II supernovae and time-delay inference through spectroscopic retrieval~\cite{2021A&A...653A..29B}, suggesting that in certain conditions such type of SNe could be cosmological probe as well.

\subsubsection{Lens modelling}

Lensed transients have advantage over lensed quasars in determining the lens model.
It has been well-known that positions of multiple images enable to determine the lens model.
Furthermore, according to the state-of-art techniques, lens modelling depends on the host arcs
which provide hundreds of constraints in pixels.
For lensed quasar systems, the lensed and magnified AGNs are too bright such that the underlying arcs are outshined. In the practical process, one needs to extract the AGN images by fitting a point spread function from a nearby star. However, to balance the field-of-view and the resolution, space telescopes like HST usually give dithered images. One has to drizzle them to a smaller-scale image. AGN images are very sensitive to this process and may be biased from the point spread function leaving a mismatch in the center pixels.
In the case of transients, one can obtain a clean image of the lensed host galaxy, before or after the transient's appearance. Thus, this source of contamination is eliminated.

To support this viewpoint, Liao et al~\cite{2017NatCo...8.1148L} did a preliminary simulation and showed that the lens model inference can be improved considerably without the impact of bright point sources.
Furthermore, Ding et al~\cite{2021MNRAS.504.5621D} gave a quantitative estimation of
the improvement of lens modelling with realistic simulations. They quantified this advantage by comparing the precision
of a lens model obtained from the same lenses with and without point sources. For the same set of lensing parameters with
HST Wide Field Camera 3 observation condition, they simulated realistic mock data sets of 48 quasar lensing systems adding
AGNs in the galaxy center and 48 galaxy-galaxy lensing systems. For the latter, they assumed the transient is not visible but the time
delay and image positions have been or will be measured. Then they modelled the images and compared the inferences of the lens model parameters
and $H_0$. They concluded that the deflector mass slope as the index of lens model determination is improved by a factor of 4.1 for lensed transients.
This results in an improved $H_0$ inference by a factor of 2.9. In addition, they claimed that clean arcs would facilitate the determination of
higher signal-to-noise stellar kinematics of the deflector and thus its mass density profile. This would in turn play a key role in breaking
the internal mass-sheet degeneracy which dominates current time-delay cosmography as a systematic error.

In addition to stellar kinematics information, one of the approaches to break the internal mass-sheet degeneracy is to measure the absolute magnifications for each lensed image.
For SNe Ia, its standard candle nature makes lensing magnification measurements possible when the knowledge about the unlensed apparent brightness
of the explosion is imposed~\cite{2003MNRAS.338L..25O}. In most of the  cases, microlensing would not change the standardizable nature of SNe Ia~\cite{2018MNRAS.478.5081F}.
Birrer et al~\cite{2022ApJ...924....2B} studied measuring $H_0$ with lensed SNe Ia with standardizable magnifications. They presented a hierarchical Bayesian framework
to combine a dataset of SNe that are unlensed and a dataset of lensed SNe with time delays. They showed that future surveys will provide precise constraints on mass profiles and $H_0$. Similar work can be found in~\cite{2022arXiv220201396Q}. The standard candle nature also makes lensed SNe Ia useful to constrain the cosmic curvature~\cite{2019PhRvD.100b3530Q}.

\section{Summaries}
With the increasing data of transient explosive sources by the ongoing and upcoming facilities, strongly lensed transients will be constantly detected in the era of time-domain astronomy.  In this article, we have given a review on the science of strongly lensed transients, especially supernovae in all types, gamma ray bursts with afterglows, fast radio bursts and gravitational waves with electromagnetic counterparts. For lensed transients, wave optics can sometimes be the appropriate description. For example, in the case of a GW signal from a binary star merger lensed by a primordial black hole with a stellar mass scale. Many new works on this topic are constantly emerging while we are writing this article.

Strategies of detecting and verifying lensed transients have been developed in recent years. For SNe Ia, strategies are classified according to resolved or unresolved images. Their standardizable nature is usually used in the lens search. For GRBs and FRBs, the key technique is to identify lensed signals with auto-correlation/cross-correlation algorithms in their multiband lightcurves or dynamic spectra. For GWs, in addition to use the nature of standard sirens like SNe Ia, one could compare the consistency of parameter spaces between multiple signals or identify lensed GWs with corresponding templates. Currently, there are several reported strongly lensed SNe and GRBs while no lensed FRBs and GWs have been reported.

As we summarized in the article, lensed transients as new astrophysical systems are powerful
in studying 1)Fundamental physics:
for example, lensed GWs exhibiting their wave nature have the potential to test GR; 2) Astrophysics: for example, probing dark matter with various sub-galactic scales. 3) Cosmology: time delays and lens model can be better measured with the transients.

This area is new and there are many open questions like: how to deal with microlensing effects in lensed SNe? how to identify lensed FRBs in real datasets? how to consider lensing effects in real GW searches? Dedicated programs/teams on lensed transients are emerging. For example, LIGO-Virgo Collaboration has assembled a GW lensing team and HOLISMOKES program aims at studying lensed SNe. One could note that related papers on arXiv are rapidly increasing in the past several years. With more data, the benefits from these systems will become evident and they will attract more attention.
We believe that lensed transients will deepen our understanding of the Universe and the transients themselves in the coming years.

\acknowledgments
The authors thank the referee for many useful comments and suggestions. This work was supported by the National Natural Science Foundation of China under Grants Nos. 12222302, 11603015, 12021003, 11920101003 and 11633001, and the Strategic Priority Research Program of the Chinese Academy of Sciences, Grant No. XDB23000000.

\bibliography{refs}

\begin{thebibliography}{281}%
\makeatletter
\providecommand \@ifxundefined [1]{%
 \@ifx{#1\undefined}
}%
\providecommand \@ifnum [1]{%
 \ifnum #1\expandafter \@firstoftwo
 \else \expandafter \@secondoftwo
 \fi
}%
\providecommand \@ifx [1]{%
 \ifx #1\expandafter \@firstoftwo
 \else \expandafter \@secondoftwo
 \fi
}%
\providecommand \natexlab [1]{#1}%
\providecommand \enquote  [1]{``#1''}%
\providecommand \bibnamefont  [1]{#1}%
\providecommand \bibfnamefont [1]{#1}%
\providecommand \citenamefont [1]{#1}%
\providecommand \href@noop [0]{\@secondoftwo}%
\providecommand \href [0]{\begingroup \@sanitize@url \@href}%
\providecommand \@href[1]{\@@startlink{#1}\@@href}%
\providecommand \@@href[1]{\endgroup#1\@@endlink}%
\providecommand \@sanitize@url [0]{\catcode `\\12\catcode `\$12\catcode
  `\&12\catcode `\#12\catcode `\^12\catcode `\_12\catcode `\%12\relax}%
\providecommand \@@startlink[1]{}%
\providecommand \@@endlink[0]{}%
\providecommand \url  [0]{\begingroup\@sanitize@url \@url }%
\providecommand \@url [1]{\endgroup\@href {#1}{\urlprefix }}%
\providecommand \urlprefix  [0]{URL }%
\providecommand \Eprint [0]{\href }%
\providecommand \doibase [0]{http://dx.doi.org/}%
\providecommand \selectlanguage [0]{\@gobble}%
\providecommand \bibinfo  [0]{\@secondoftwo}%
\providecommand \bibfield  [0]{\@secondoftwo}%
\providecommand \translation [1]{[#1]}%
\providecommand \BibitemOpen [0]{}%
\providecommand \bibitemStop [0]{}%
\providecommand \bibitemNoStop [0]{.\EOS\space}%
\providecommand \EOS [0]{\spacefactor3000\relax}%
\providecommand \BibitemShut  [1]{\csname bibitem#1\endcsname}%
\let\auto@bib@innerbib\@empty
\bibitem [{\citenamefont {{Treu}}(2010)}]{2010ARA&A..48...87T}%
  \BibitemOpen
  \bibfield  {author} {\bibinfo {author} {\bibfnamefont {T.}~\bibnamefont
  {{Treu}}},\ }\href {\doibase 10.1146/annurev-astro-081309-130924} {\bibfield
  {journal} {\bibinfo  {journal} {\araa}\ }\textbf {\bibinfo {volume} {48}},\
  \bibinfo {pages} {87} (\bibinfo {year} {2010})}\BibitemShut {NoStop}%
\bibitem [{\citenamefont {{Lemon}}\ \emph {et~al.}(2022)\citenamefont
  {{Lemon}}, \citenamefont {{Anguita}}, \citenamefont {{Auger}}, \citenamefont
  {{Courbin}}, \citenamefont {{Galan}}, \citenamefont {{McMahon}},
  \citenamefont {{Neira}}, \citenamefont {{Oguri}}, \citenamefont
  {{Schechter}}, \citenamefont {{Shajib}},\ and\ \citenamefont
  {{Treu}}}]{2022arXiv220607714L}%
  \BibitemOpen
  \bibfield  {author} {\bibinfo {author} {\bibfnamefont {C.}~\bibnamefont
  {{Lemon}}}, \bibinfo {author} {\bibfnamefont {T.}~\bibnamefont {{Anguita}}},
  \bibinfo {author} {\bibfnamefont {M.}~\bibnamefont {{Auger}}}, \bibinfo
  {author} {\bibfnamefont {F.}~\bibnamefont {{Courbin}}}, \bibinfo {author}
  {\bibfnamefont {A.}~\bibnamefont {{Galan}}}, \bibinfo {author} {\bibfnamefont
  {R.}~\bibnamefont {{McMahon}}}, \bibinfo {author} {\bibfnamefont
  {F.}~\bibnamefont {{Neira}}}, \bibinfo {author} {\bibfnamefont
  {M.}~\bibnamefont {{Oguri}}}, \bibinfo {author} {\bibfnamefont
  {P.}~\bibnamefont {{Schechter}}}, \bibinfo {author} {\bibfnamefont
  {A.}~\bibnamefont {{Shajib}}}, \ and\ \bibinfo {author} {\bibfnamefont
  {T.}~\bibnamefont {{Treu}}},\ }\href@noop {} {\bibfield  {journal} {\bibinfo
  {journal} {arXiv e-prints}\ } (\bibinfo {year} {2022})},\ \Eprint
  {http://arxiv.org/abs/2206.07714} {2206.07714} \BibitemShut {NoStop}%
\bibitem [{\citenamefont {{Oguri}}\ and\ \citenamefont
  {{Marshall}}(2010)}]{2010MNRAS.405.2579O}%
  \BibitemOpen
  \bibfield  {author} {\bibinfo {author} {\bibfnamefont {M.}~\bibnamefont
  {{Oguri}}}\ and\ \bibinfo {author} {\bibfnamefont {P.~J.}\ \bibnamefont
  {{Marshall}}},\ }\href {\doibase 10.1111/j.1365-2966.2010.16639.x} {\bibfield
   {journal} {\bibinfo  {journal} {\mnras}\ }\textbf {\bibinfo {volume}
  {405}},\ \bibinfo {pages} {2579} (\bibinfo {year} {2010})}\BibitemShut
  {NoStop}%
\bibitem [{\citenamefont {{Treu}}\ and\ \citenamefont
  {{Marshall}}(2016)}]{2016A&ARv..24...11T}%
  \BibitemOpen
  \bibfield  {author} {\bibinfo {author} {\bibfnamefont {T.}~\bibnamefont
  {{Treu}}}\ and\ \bibinfo {author} {\bibfnamefont {P.~J.}\ \bibnamefont
  {{Marshall}}},\ }\href {\doibase 10.1007/s00159-016-0096-8} {\bibfield
  {journal} {\bibinfo  {journal} {\aapr}\ }\textbf {\bibinfo {volume} {24}},\
  \bibinfo {eid} {11} (\bibinfo {year} {2016})}\BibitemShut {NoStop}%
\bibitem [{\citenamefont {{Refsdal}}(1964)}]{1964MNRAS.128..307R}%
  \BibitemOpen
  \bibfield  {author} {\bibinfo {author} {\bibfnamefont {S.}~\bibnamefont
  {{Refsdal}}},\ }\href {\doibase 10.1093/mnras/128.4.307} {\bibfield
  {journal} {\bibinfo  {journal} {\mnras}\ }\textbf {\bibinfo {volume} {128}},\
  \bibinfo {pages} {307} (\bibinfo {year} {1964})}\BibitemShut {NoStop}%
\bibitem [{\citenamefont {{Tisserand}}\ \emph {et~al.}(2007)\citenamefont
  {{Tisserand}}, \citenamefont {{Le Guillou}}, \citenamefont {{Afonso}},\ and\
  \citenamefont {{et al.}}}]{2007A&A...469..387T}%
  \BibitemOpen
  \bibfield  {author} {\bibinfo {author} {\bibfnamefont {P.}~\bibnamefont
  {{Tisserand}}}, \bibinfo {author} {\bibfnamefont {L.}~\bibnamefont {{Le
  Guillou}}}, \bibinfo {author} {\bibfnamefont {C.}~\bibnamefont {{Afonso}}}, \
  and\ \bibinfo {author} {\bibnamefont {{et al.}}},\ }\href {\doibase
  10.1051/0004-6361:20066017} {\bibfield  {journal} {\bibinfo  {journal}
  {\aap}\ }\textbf {\bibinfo {volume} {469}},\ \bibinfo {pages} {387} (\bibinfo
  {year} {2007})}\BibitemShut {NoStop}%
\bibitem [{\citenamefont {{Niikura}}\ \emph {et~al.}(2019)\citenamefont
  {{Niikura}}, \citenamefont {{Takada}}, \citenamefont {{Yasuda}},\ and\
  \citenamefont {{et al.}}}]{2019NatAs...3..524N}%
  \BibitemOpen
  \bibfield  {author} {\bibinfo {author} {\bibfnamefont {H.}~\bibnamefont
  {{Niikura}}}, \bibinfo {author} {\bibfnamefont {M.}~\bibnamefont {{Takada}}},
  \bibinfo {author} {\bibfnamefont {N.}~\bibnamefont {{Yasuda}}}, \ and\
  \bibinfo {author} {\bibnamefont {{et al.}}},\ }\href {\doibase
  10.1038/s41550-019-0723-1} {\bibfield  {journal} {\bibinfo  {journal} {Nature
  Astronomy}\ }\textbf {\bibinfo {volume} {3}},\ \bibinfo {pages} {524}
  (\bibinfo {year} {2019})}\BibitemShut {NoStop}%
\bibitem [{\citenamefont {{Mao}}\ and\ \citenamefont
  {{Paczynski}}(1991)}]{1991ApJ...374L..37M}%
  \BibitemOpen
  \bibfield  {author} {\bibinfo {author} {\bibfnamefont {S.}~\bibnamefont
  {{Mao}}}\ and\ \bibinfo {author} {\bibfnamefont {B.}~\bibnamefont
  {{Paczynski}}},\ }\href {\doibase 10.1086/186066} {\bibfield  {journal}
  {\bibinfo  {journal} {\apjl}\ }\textbf {\bibinfo {volume} {374}},\ \bibinfo
  {pages} {L37} (\bibinfo {year} {1991})}\BibitemShut {NoStop}%
\bibitem [{\citenamefont {{Kelly}}\ \emph {et~al.}(2015)\citenamefont
  {{Kelly}}, \citenamefont {{Rodney}}, \citenamefont {{Treu}},\ and\
  \citenamefont {{et al.}}}]{2015Sci...347.1123K}%
  \BibitemOpen
  \bibfield  {author} {\bibinfo {author} {\bibfnamefont {P.~L.}\ \bibnamefont
  {{Kelly}}}, \bibinfo {author} {\bibfnamefont {S.~A.}\ \bibnamefont
  {{Rodney}}}, \bibinfo {author} {\bibfnamefont {T.}~\bibnamefont {{Treu}}}, \
  and\ \bibinfo {author} {\bibnamefont {{et al.}}},\ }\href {\doibase
  10.1126/science.aaa3350} {\bibfield  {journal} {\bibinfo  {journal}
  {Science}\ }\textbf {\bibinfo {volume} {347}},\ \bibinfo {pages} {1123}
  (\bibinfo {year} {2015})}\BibitemShut {NoStop}%
\bibitem [{\citenamefont {{Gaia Collaboration}}(2016)}]{2016A&A...595A...2G}%
  \BibitemOpen
  \bibfield  {author} {\bibinfo {author} {\bibnamefont {{Gaia
  Collaboration}}},\ }\href {\doibase 10.1051/0004-6361/201629512} {\bibfield
  {journal} {\bibinfo  {journal} {\aap}\ }\textbf {\bibinfo {volume} {595}},\
  \bibinfo {eid} {A2} (\bibinfo {year} {2016})}\BibitemShut {NoStop}%
\bibitem [{\citenamefont {{Diehl}}\ \emph {et~al.}(2017)\citenamefont
  {{Diehl}}, \citenamefont {{Buckley-Geer}}, \citenamefont {{Lindgren}},\ and\
  \citenamefont {{DES Collaboration}}}]{2017ApJS..232...15D}%
  \BibitemOpen
  \bibfield  {author} {\bibinfo {author} {\bibfnamefont {H.~T.}\ \bibnamefont
  {{Diehl}}}, \bibinfo {author} {\bibfnamefont {E.~J.}\ \bibnamefont
  {{Buckley-Geer}}}, \bibinfo {author} {\bibfnamefont {K.~A.}\ \bibnamefont
  {{Lindgren}}}, \ and\ \bibinfo {author} {\bibnamefont {{DES
  Collaboration}}},\ }\href {\doibase 10.3847/1538-4365/aa8667} {\bibfield
  {journal} {\bibinfo  {journal} {\apjs}\ }\textbf {\bibinfo {volume} {232}},\
  \bibinfo {eid} {15} (\bibinfo {year} {2017})}\BibitemShut {NoStop}%
\bibitem [{\citenamefont {{Sonnenfeld}}\ \emph {et~al.}(2020)\citenamefont
  {{Sonnenfeld}}, \citenamefont {{Verma}}, \citenamefont {{More}},\ and\
  \citenamefont {{et al.}}}]{2020A&A...642A.148S}%
  \BibitemOpen
  \bibfield  {author} {\bibinfo {author} {\bibfnamefont {A.}~\bibnamefont
  {{Sonnenfeld}}}, \bibinfo {author} {\bibfnamefont {A.}~\bibnamefont
  {{Verma}}}, \bibinfo {author} {\bibfnamefont {A.}~\bibnamefont {{More}}}, \
  and\ \bibinfo {author} {\bibnamefont {{et al.}}},\ }\href {\doibase
  10.1051/0004-6361/202038067} {\bibfield  {journal} {\bibinfo  {journal}
  {\aap}\ }\textbf {\bibinfo {volume} {642}},\ \bibinfo {eid} {A148} (\bibinfo
  {year} {2020})}\BibitemShut {NoStop}%
\bibitem [{\citenamefont {{CHIME/FRB
  Collaboration}}(2021)}]{2021ApJS..257...59C}%
  \BibitemOpen
  \bibfield  {author} {\bibinfo {author} {\bibnamefont {{CHIME/FRB
  Collaboration}}},\ }\href {\doibase 10.3847/1538-4365/ac33ab} {\bibfield
  {journal} {\bibinfo  {journal} {\apjs}\ }\textbf {\bibinfo {volume} {257}},\
  \bibinfo {eid} {59} (\bibinfo {year} {2021})}\BibitemShut {NoStop}%
\bibitem [{\citenamefont {{Niu}}\ \emph {et~al.}(2021)\citenamefont {{Niu}},
  \citenamefont {{Aggarwal}}, \citenamefont {{Li}},\ and\ \citenamefont {{et
  al.}}}]{2021arXiv211007418N}%
  \BibitemOpen
  \bibfield  {author} {\bibinfo {author} {\bibfnamefont {C.~H.}\ \bibnamefont
  {{Niu}}}, \bibinfo {author} {\bibfnamefont {K.}~\bibnamefont {{Aggarwal}}},
  \bibinfo {author} {\bibfnamefont {D.}~\bibnamefont {{Li}}}, \ and\ \bibinfo
  {author} {\bibnamefont {{et al.}}},\ }\href@noop {} {\bibfield  {journal}
  {\bibinfo  {journal} {arXiv e-prints}\ } (\bibinfo {year} {2021})},\ \Eprint
  {http://arxiv.org/abs/2110.07418} {2110.07418} \BibitemShut {NoStop}%
\bibitem [{\citenamefont {{The LIGO Scientific Collaboration}}\ \emph
  {et~al.}(2021{\natexlab{a}})\citenamefont {{The LIGO Scientific
  Collaboration}}, \citenamefont {{the Virgo Collaboration}},\ and\
  \citenamefont {{the KAGRA Collaboration}}}]{2021arXiv211103606T}%
  \BibitemOpen
  \bibfield  {author} {\bibinfo {author} {\bibnamefont {{The LIGO Scientific
  Collaboration}}}, \bibinfo {author} {\bibnamefont {{the Virgo
  Collaboration}}}, \ and\ \bibinfo {author} {\bibnamefont {{the KAGRA
  Collaboration}}},\ }\href@noop {} {\bibfield  {journal} {\bibinfo  {journal}
  {arXiv e-prints}\ } (\bibinfo {year} {2021}{\natexlab{a}})},\ \Eprint
  {http://arxiv.org/abs/2111.03606} {2111.03606} \BibitemShut {NoStop}%
\bibitem [{\citenamefont {{Dewdney}}\ \emph {et~al.}(2009)\citenamefont
  {{Dewdney}}, \citenamefont {{Hall}}, \citenamefont {{Schilizzi}},\ and\
  \citenamefont {{Lazio}}}]{2009IEEEP..97.1482D}%
  \BibitemOpen
  \bibfield  {author} {\bibinfo {author} {\bibfnamefont {P.~E.}\ \bibnamefont
  {{Dewdney}}}, \bibinfo {author} {\bibfnamefont {P.~J.}\ \bibnamefont
  {{Hall}}}, \bibinfo {author} {\bibfnamefont {R.~T.}\ \bibnamefont
  {{Schilizzi}}}, \ and\ \bibinfo {author} {\bibfnamefont {T.~J.~L.~W.}\
  \bibnamefont {{Lazio}}},\ }\href {\doibase 10.1109/JPROC.2009.2021005}
  {\bibfield  {journal} {\bibinfo  {journal} {IEEE Proceedings}\ }\textbf
  {\bibinfo {volume} {97}},\ \bibinfo {pages} {1482} (\bibinfo {year}
  {2009})}\BibitemShut {NoStop}%
\bibitem [{\citenamefont {{Regimbau}}\ \emph {et~al.}(2012)\citenamefont
  {{Regimbau}}, \citenamefont {{Dent}}, \citenamefont {{Del Pozzo}},\ and\
  \citenamefont {{et al.}}}]{2012PhRvD..86l2001R}%
  \BibitemOpen
  \bibfield  {author} {\bibinfo {author} {\bibfnamefont {T.}~\bibnamefont
  {{Regimbau}}}, \bibinfo {author} {\bibfnamefont {T.}~\bibnamefont {{Dent}}},
  \bibinfo {author} {\bibfnamefont {W.}~\bibnamefont {{Del Pozzo}}}, \ and\
  \bibinfo {author} {\bibnamefont {{et al.}}},\ }\href {\doibase
  10.1103/PhysRevD.86.122001} {\bibfield  {journal} {\bibinfo  {journal}
  {\prd}\ }\textbf {\bibinfo {volume} {86}},\ \bibinfo {eid} {122001} (\bibinfo
  {year} {2012})}\BibitemShut {NoStop}%
\bibitem [{\citenamefont {{Oguri}}(2019)}]{2019RPPh...82l6901O}%
  \BibitemOpen
  \bibfield  {author} {\bibinfo {author} {\bibfnamefont {M.}~\bibnamefont
  {{Oguri}}},\ }\href {\doibase 10.1088/1361-6633/ab4fc5} {\bibfield  {journal}
  {\bibinfo  {journal} {Reports on Progress in Physics}\ }\textbf {\bibinfo
  {volume} {82}},\ \bibinfo {eid} {126901} (\bibinfo {year}
  {2019})}\BibitemShut {NoStop}%
\bibitem [{\citenamefont {{Schneider}}\ \emph {et~al.}(1992)\citenamefont
  {{Schneider}}, \citenamefont {{Ehlers}},\ and\ \citenamefont
  {{Falco}}}]{Schneider1992LensesBook}%
  \BibitemOpen
  \bibfield  {author} {\bibinfo {author} {\bibfnamefont {P.}~\bibnamefont
  {{Schneider}}}, \bibinfo {author} {\bibfnamefont {J.}~\bibnamefont
  {{Ehlers}}}, \ and\ \bibinfo {author} {\bibfnamefont {E.~E.}\ \bibnamefont
  {{Falco}}},\ }\href {\doibase 10.1007/978-3-662-03758-4} {\emph {\bibinfo
  {title} {{Gravitational Lenses}}}}\ (\bibinfo  {publisher} {Springer},\
  \bibinfo {year} {1992})\BibitemShut {NoStop}%
\bibitem [{\citenamefont {{Matsunaga}}\ and\ \citenamefont
  {{Yamamoto}}(2006)}]{2006JCAP...01..023M}%
  \BibitemOpen
  \bibfield  {author} {\bibinfo {author} {\bibfnamefont {N.}~\bibnamefont
  {{Matsunaga}}}\ and\ \bibinfo {author} {\bibfnamefont {K.}~\bibnamefont
  {{Yamamoto}}},\ }\href {\doibase 10.1088/1475-7516/2006/01/023} {\bibfield
  {journal} {\bibinfo  {journal} {\jcap}\ }\textbf {\bibinfo {volume} {2006}},\
  \bibinfo {eid} {023} (\bibinfo {year} {2006})}\BibitemShut {NoStop}%
\bibitem [{\citenamefont {{Weiler}}\ and\ \citenamefont
  {{Sramek}}(1988)}]{1988ARA&A..26..295W}%
  \BibitemOpen
  \bibfield  {author} {\bibinfo {author} {\bibfnamefont {K.~W.}\ \bibnamefont
  {{Weiler}}}\ and\ \bibinfo {author} {\bibfnamefont {R.~A.}\ \bibnamefont
  {{Sramek}}},\ }\href {\doibase 10.1146/annurev.aa.26.090188.001455}
  {\bibfield  {journal} {\bibinfo  {journal} {\araa}\ }\textbf {\bibinfo
  {volume} {26}},\ \bibinfo {pages} {295} (\bibinfo {year} {1988})}\BibitemShut
  {NoStop}%
\bibitem [{\citenamefont {{Filippenko}}(1997)}]{1997ARA&A..35..309F}%
  \BibitemOpen
  \bibfield  {author} {\bibinfo {author} {\bibfnamefont {A.~V.}\ \bibnamefont
  {{Filippenko}}},\ }\href {\doibase 10.1146/annurev.astro.35.1.309} {\bibfield
   {journal} {\bibinfo  {journal} {\araa}\ }\textbf {\bibinfo {volume} {35}},\
  \bibinfo {pages} {309} (\bibinfo {year} {1997})}\BibitemShut {NoStop}%
\bibitem [{\citenamefont {{Branch}}\ and\ \citenamefont
  {{Tammann}}(1992)}]{1992ARA&A..30..359B}%
  \BibitemOpen
  \bibfield  {author} {\bibinfo {author} {\bibfnamefont {D.}~\bibnamefont
  {{Branch}}}\ and\ \bibinfo {author} {\bibfnamefont {G.~A.}\ \bibnamefont
  {{Tammann}}},\ }\href {\doibase 10.1146/annurev.aa.30.090192.002043}
  {\bibfield  {journal} {\bibinfo  {journal} {\araa}\ }\textbf {\bibinfo
  {volume} {30}},\ \bibinfo {pages} {359} (\bibinfo {year} {1992})}\BibitemShut
  {NoStop}%
\bibitem [{\citenamefont {{Riess}}\ \emph {et~al.}(1998)\citenamefont
  {{Riess}}, \citenamefont {{Filippenko}}, \citenamefont {{Challis}},\ and\
  \citenamefont {{et al.}}}]{1998AJ....116.1009R}%
  \BibitemOpen
  \bibfield  {author} {\bibinfo {author} {\bibfnamefont {A.~G.}\ \bibnamefont
  {{Riess}}}, \bibinfo {author} {\bibfnamefont {A.~V.}\ \bibnamefont
  {{Filippenko}}}, \bibinfo {author} {\bibfnamefont {P.}~\bibnamefont
  {{Challis}}}, \ and\ \bibinfo {author} {\bibnamefont {{et al.}}},\ }\href
  {\doibase 10.1086/300499} {\bibfield  {journal} {\bibinfo  {journal} {\aj}\
  }\textbf {\bibinfo {volume} {116}},\ \bibinfo {pages} {1009} (\bibinfo {year}
  {1998})}\BibitemShut {NoStop}%
\bibitem [{\citenamefont {{Perlmutter}}\ \emph {et~al.}(1999)\citenamefont
  {{Perlmutter}}, \citenamefont {{Aldering}}, \citenamefont {{Goldhaber}},\
  and\ \citenamefont {{et al.}}}]{1999ApJ...517..565P}%
  \BibitemOpen
  \bibfield  {author} {\bibinfo {author} {\bibfnamefont {S.}~\bibnamefont
  {{Perlmutter}}}, \bibinfo {author} {\bibfnamefont {G.}~\bibnamefont
  {{Aldering}}}, \bibinfo {author} {\bibfnamefont {G.}~\bibnamefont
  {{Goldhaber}}}, \ and\ \bibinfo {author} {\bibnamefont {{et al.}}},\ }\href
  {\doibase 10.1086/307221} {\bibfield  {journal} {\bibinfo  {journal} {\apj}\
  }\textbf {\bibinfo {volume} {517}},\ \bibinfo {pages} {565} (\bibinfo {year}
  {1999})}\BibitemShut {NoStop}%
\bibitem [{\citenamefont {{Pakmor}}\ \emph {et~al.}(2011)\citenamefont
  {{Pakmor}}, \citenamefont {{Hachinger}}, \citenamefont {{R{\"o}pke}},\ and\
  \citenamefont {{Hillebrandt}}}]{2011A&A...528A.117P}%
  \BibitemOpen
  \bibfield  {author} {\bibinfo {author} {\bibfnamefont {R.}~\bibnamefont
  {{Pakmor}}}, \bibinfo {author} {\bibfnamefont {S.}~\bibnamefont
  {{Hachinger}}}, \bibinfo {author} {\bibfnamefont {F.~K.}\ \bibnamefont
  {{R{\"o}pke}}}, \ and\ \bibinfo {author} {\bibfnamefont {W.}~\bibnamefont
  {{Hillebrandt}}},\ }\href {\doibase 10.1051/0004-6361/201015653} {\bibfield
  {journal} {\bibinfo  {journal} {\aap}\ }\textbf {\bibinfo {volume} {528}},\
  \bibinfo {eid} {A117} (\bibinfo {year} {2011})}\BibitemShut {NoStop}%
\bibitem [{\citenamefont {{Maoz}}\ \emph {et~al.}(2014)\citenamefont {{Maoz}},
  \citenamefont {{Mannucci}},\ and\ \citenamefont
  {{Nelemans}}}]{2014ARA&A..52..107M}%
  \BibitemOpen
  \bibfield  {author} {\bibinfo {author} {\bibfnamefont {D.}~\bibnamefont
  {{Maoz}}}, \bibinfo {author} {\bibfnamefont {F.}~\bibnamefont {{Mannucci}}},
  \ and\ \bibinfo {author} {\bibfnamefont {G.}~\bibnamefont {{Nelemans}}},\
  }\href {\doibase 10.1146/annurev-astro-082812-141031} {\bibfield  {journal}
  {\bibinfo  {journal} {\araa}\ }\textbf {\bibinfo {volume} {52}},\ \bibinfo
  {pages} {107} (\bibinfo {year} {2014})}\BibitemShut {NoStop}%
\bibitem [{\citenamefont {{Scolnic}}\ \emph {et~al.}(2018)\citenamefont
  {{Scolnic}}, \citenamefont {{Jones}}, \citenamefont {{Rest}},\ and\
  \citenamefont {{et al.}}}]{2018ApJ...859..101S}%
  \BibitemOpen
  \bibfield  {author} {\bibinfo {author} {\bibfnamefont {D.~M.}\ \bibnamefont
  {{Scolnic}}}, \bibinfo {author} {\bibfnamefont {D.~O.}\ \bibnamefont
  {{Jones}}}, \bibinfo {author} {\bibfnamefont {A.}~\bibnamefont {{Rest}}}, \
  and\ \bibinfo {author} {\bibnamefont {{et al.}}},\ }\href {\doibase
  10.3847/1538-4357/aab9bb} {\bibfield  {journal} {\bibinfo  {journal} {\apj}\
  }\textbf {\bibinfo {volume} {859}},\ \bibinfo {eid} {101} (\bibinfo {year}
  {2018})}\BibitemShut {NoStop}%
\bibitem [{\citenamefont {{Porciani}}\ and\ \citenamefont
  {{Madau}}(2000)}]{2000ApJ...532..679P}%
  \BibitemOpen
  \bibfield  {author} {\bibinfo {author} {\bibfnamefont {C.}~\bibnamefont
  {{Porciani}}}\ and\ \bibinfo {author} {\bibfnamefont {P.}~\bibnamefont
  {{Madau}}},\ }\href {\doibase 10.1086/308587} {\bibfield  {journal} {\bibinfo
   {journal} {\apj}\ }\textbf {\bibinfo {volume} {532}},\ \bibinfo {pages}
  {679} (\bibinfo {year} {2000})}\BibitemShut {NoStop}%
\bibitem [{\citenamefont {{Holz}}(2001)}]{2001ApJ...556L..71H}%
  \BibitemOpen
  \bibfield  {author} {\bibinfo {author} {\bibfnamefont {D.~E.}\ \bibnamefont
  {{Holz}}},\ }\href {\doibase 10.1086/322947} {\bibfield  {journal} {\bibinfo
  {journal} {\apjl}\ }\textbf {\bibinfo {volume} {556}},\ \bibinfo {pages}
  {L71} (\bibinfo {year} {2001})}\BibitemShut {NoStop}%
\bibitem [{\citenamefont {{Wang}}(2000)}]{2000ApJ...531..676W}%
  \BibitemOpen
  \bibfield  {author} {\bibinfo {author} {\bibfnamefont {Y.}~\bibnamefont
  {{Wang}}},\ }\href {\doibase 10.1086/308500} {\bibfield  {journal} {\bibinfo
  {journal} {\apj}\ }\textbf {\bibinfo {volume} {531}},\ \bibinfo {pages} {676}
  (\bibinfo {year} {2000})}\BibitemShut {NoStop}%
\bibitem [{\citenamefont {{Goobar}}\ \emph
  {et~al.}(2002{\natexlab{a}})\citenamefont {{Goobar}}, \citenamefont
  {{M{\"o}rtsell}}, \citenamefont {{Amanullah}},\ and\ \citenamefont
  {{Nugent}}}]{2002A&A...393...25G}%
  \BibitemOpen
  \bibfield  {author} {\bibinfo {author} {\bibfnamefont {A.}~\bibnamefont
  {{Goobar}}}, \bibinfo {author} {\bibfnamefont {E.}~\bibnamefont
  {{M{\"o}rtsell}}}, \bibinfo {author} {\bibfnamefont {R.}~\bibnamefont
  {{Amanullah}}}, \ and\ \bibinfo {author} {\bibfnamefont {P.}~\bibnamefont
  {{Nugent}}},\ }\href {\doibase 10.1051/0004-6361:20020987} {\bibfield
  {journal} {\bibinfo  {journal} {\aap}\ }\textbf {\bibinfo {volume} {393}},\
  \bibinfo {pages} {25} (\bibinfo {year} {2002}{\natexlab{a}})}\BibitemShut
  {NoStop}%
\bibitem [{\citenamefont {{Goobar}}\ \emph
  {et~al.}(2002{\natexlab{b}})\citenamefont {{Goobar}}, \citenamefont
  {{M{\"o}rtsell}}, \citenamefont {{Amanullah}},\ and\ \citenamefont {{et
  al.}}}]{2002A&A...392..757G}%
  \BibitemOpen
  \bibfield  {author} {\bibinfo {author} {\bibfnamefont {A.}~\bibnamefont
  {{Goobar}}}, \bibinfo {author} {\bibfnamefont {E.}~\bibnamefont
  {{M{\"o}rtsell}}}, \bibinfo {author} {\bibfnamefont {R.}~\bibnamefont
  {{Amanullah}}}, \ and\ \bibinfo {author} {\bibnamefont {{et al.}}},\ }\href
  {\doibase 10.1051/0004-6361:20020930} {\bibfield  {journal} {\bibinfo
  {journal} {\aap}\ }\textbf {\bibinfo {volume} {392}},\ \bibinfo {pages} {757}
  (\bibinfo {year} {2002}{\natexlab{b}})}\BibitemShut {NoStop}%
\bibitem [{\citenamefont {{Quimby}}\ \emph {et~al.}(2014)\citenamefont
  {{Quimby}}, \citenamefont {{Oguri}}, \citenamefont {{More}},\ and\
  \citenamefont {{et al.}}}]{2014Sci...344..396Q}%
  \BibitemOpen
  \bibfield  {author} {\bibinfo {author} {\bibfnamefont {R.~M.}\ \bibnamefont
  {{Quimby}}}, \bibinfo {author} {\bibfnamefont {M.}~\bibnamefont {{Oguri}}},
  \bibinfo {author} {\bibfnamefont {A.}~\bibnamefont {{More}}}, \ and\ \bibinfo
  {author} {\bibnamefont {{et al.}}},\ }\href {\doibase
  10.1126/science.1250903} {\bibfield  {journal} {\bibinfo  {journal}
  {Science}\ }\textbf {\bibinfo {volume} {344}},\ \bibinfo {pages} {396}
  (\bibinfo {year} {2014})}\BibitemShut {NoStop}%
\bibitem [{\citenamefont {{Goldstein}}\ and\ \citenamefont
  {{Nugent}}(2017)}]{2017ApJ...834L...5G}%
  \BibitemOpen
  \bibfield  {author} {\bibinfo {author} {\bibfnamefont {D.~A.}\ \bibnamefont
  {{Goldstein}}}\ and\ \bibinfo {author} {\bibfnamefont {P.~E.}\ \bibnamefont
  {{Nugent}}},\ }\href {\doibase 10.3847/2041-8213/834/1/L5} {\bibfield
  {journal} {\bibinfo  {journal} {\apjl}\ }\textbf {\bibinfo {volume} {834}},\
  \bibinfo {eid} {L5} (\bibinfo {year} {2017})}\BibitemShut {NoStop}%
\bibitem [{\citenamefont {{Lee}}(2018)}]{2018RNAAS...2..186L}%
  \BibitemOpen
  \bibfield  {author} {\bibinfo {author} {\bibfnamefont {C.-H.}\ \bibnamefont
  {{Lee}}},\ }\href {\doibase 10.3847/2515-5172/aae588} {\bibfield  {journal}
  {\bibinfo  {journal} {Research Notes of the American Astronomical Society}\
  }\textbf {\bibinfo {volume} {2}},\ \bibinfo {eid} {186} (\bibinfo {year}
  {2018})}\BibitemShut {NoStop}%
\bibitem [{\citenamefont {{Bag}}\ \emph {et~al.}(2021)\citenamefont {{Bag}},
  \citenamefont {{Kim}}, \citenamefont {{Linder}},\ and\ \citenamefont
  {{Shafieloo}}}]{2021ApJ...910...65B}%
  \BibitemOpen
  \bibfield  {author} {\bibinfo {author} {\bibfnamefont {S.}~\bibnamefont
  {{Bag}}}, \bibinfo {author} {\bibfnamefont {A.~G.}\ \bibnamefont {{Kim}}},
  \bibinfo {author} {\bibfnamefont {E.~V.}\ \bibnamefont {{Linder}}}, \ and\
  \bibinfo {author} {\bibfnamefont {A.}~\bibnamefont {{Shafieloo}}},\ }\href
  {\doibase 10.3847/1538-4357/abe238} {\bibfield  {journal} {\bibinfo
  {journal} {\apj}\ }\textbf {\bibinfo {volume} {910}},\ \bibinfo {eid} {65}
  (\bibinfo {year} {2021})}\BibitemShut {NoStop}%
\bibitem [{\citenamefont {{Denissenya}}\ \emph {et~al.}(2022)\citenamefont
  {{Denissenya}}, \citenamefont {{Bag}}, \citenamefont {{Kim}}, \citenamefont
  {{Linder}},\ and\ \citenamefont {{Shafieloo}}}]{2022MNRAS.511.1210D}%
  \BibitemOpen
  \bibfield  {author} {\bibinfo {author} {\bibfnamefont {M.}~\bibnamefont
  {{Denissenya}}}, \bibinfo {author} {\bibfnamefont {S.}~\bibnamefont {{Bag}}},
  \bibinfo {author} {\bibfnamefont {A.~G.}\ \bibnamefont {{Kim}}}, \bibinfo
  {author} {\bibfnamefont {E.~V.}\ \bibnamefont {{Linder}}}, \ and\ \bibinfo
  {author} {\bibfnamefont {A.}~\bibnamefont {{Shafieloo}}},\ }\href {\doibase
  10.1093/mnras/stac143} {\bibfield  {journal} {\bibinfo  {journal} {\mnras}\
  }\textbf {\bibinfo {volume} {511}},\ \bibinfo {pages} {1210} (\bibinfo {year}
  {2022})}\BibitemShut {NoStop}%
\bibitem [{\citenamefont {{Wojtak}}\ \emph {et~al.}(2019)\citenamefont
  {{Wojtak}}, \citenamefont {{Hjorth}},\ and\ \citenamefont
  {{Gall}}}]{2019MNRAS.487.3342W}%
  \BibitemOpen
  \bibfield  {author} {\bibinfo {author} {\bibfnamefont {R.}~\bibnamefont
  {{Wojtak}}}, \bibinfo {author} {\bibfnamefont {J.}~\bibnamefont {{Hjorth}}},
  \ and\ \bibinfo {author} {\bibfnamefont {C.}~\bibnamefont {{Gall}}},\ }\href
  {\doibase 10.1093/mnras/stz1516} {\bibfield  {journal} {\bibinfo  {journal}
  {\mnras}\ }\textbf {\bibinfo {volume} {487}},\ \bibinfo {pages} {3342}
  (\bibinfo {year} {2019})}\BibitemShut {NoStop}%
\bibitem [{\citenamefont {{Holwerda}}\ \emph {et~al.}(2021)\citenamefont
  {{Holwerda}}, \citenamefont {{Knabel}}, \citenamefont {{Steele}},\ and\
  \citenamefont {{et al.}}}]{2021MNRAS.505.1316H}%
  \BibitemOpen
  \bibfield  {author} {\bibinfo {author} {\bibfnamefont {B.~W.}\ \bibnamefont
  {{Holwerda}}}, \bibinfo {author} {\bibfnamefont {S.}~\bibnamefont
  {{Knabel}}}, \bibinfo {author} {\bibfnamefont {R.~C.}\ \bibnamefont
  {{Steele}}}, \ and\ \bibinfo {author} {\bibnamefont {{et al.}}},\ }\href
  {\doibase 10.1093/mnras/stab1370} {\bibfield  {journal} {\bibinfo  {journal}
  {\mnras}\ }\textbf {\bibinfo {volume} {505}},\ \bibinfo {pages} {1316}
  (\bibinfo {year} {2021})}\BibitemShut {NoStop}%
\bibitem [{\citenamefont {{Sullivan}}\ \emph {et~al.}(2000)\citenamefont
  {{Sullivan}}, \citenamefont {{Ellis}}, \citenamefont {{Nugent}},
  \citenamefont {{Smail}},\ and\ \citenamefont
  {{Madau}}}]{2000MNRAS.319..549S}%
  \BibitemOpen
  \bibfield  {author} {\bibinfo {author} {\bibfnamefont {M.}~\bibnamefont
  {{Sullivan}}}, \bibinfo {author} {\bibfnamefont {R.}~\bibnamefont {{Ellis}}},
  \bibinfo {author} {\bibfnamefont {P.}~\bibnamefont {{Nugent}}}, \bibinfo
  {author} {\bibfnamefont {I.}~\bibnamefont {{Smail}}}, \ and\ \bibinfo
  {author} {\bibfnamefont {P.}~\bibnamefont {{Madau}}},\ }\href {\doibase
  10.1046/j.1365-8711.2000.03875.x} {\bibfield  {journal} {\bibinfo  {journal}
  {\mnras}\ }\textbf {\bibinfo {volume} {319}},\ \bibinfo {pages} {549}
  (\bibinfo {year} {2000})}\BibitemShut {NoStop}%
\bibitem [{\citenamefont {{Saini}}\ \emph {et~al.}(2000)\citenamefont
  {{Saini}}, \citenamefont {{Raychaudhury}},\ and\ \citenamefont
  {{Shchekinov}}}]{2000A&A...363..349S}%
  \BibitemOpen
  \bibfield  {author} {\bibinfo {author} {\bibfnamefont {T.~D.}\ \bibnamefont
  {{Saini}}}, \bibinfo {author} {\bibfnamefont {S.}~\bibnamefont
  {{Raychaudhury}}}, \ and\ \bibinfo {author} {\bibfnamefont {Y.~A.}\
  \bibnamefont {{Shchekinov}}},\ }\href@noop {} {\bibfield  {journal} {\bibinfo
   {journal} {\aap}\ }\textbf {\bibinfo {volume} {363}},\ \bibinfo {pages}
  {349} (\bibinfo {year} {2000})}\BibitemShut {NoStop}%
\bibitem [{\citenamefont {{Stanishev}}\ \emph {et~al.}(2009)\citenamefont
  {{Stanishev}}, \citenamefont {{Goobar}}, \citenamefont {{Paech}},\ and\
  \citenamefont {{et al}}}]{2009A&A...507...61S}%
  \BibitemOpen
  \bibfield  {author} {\bibinfo {author} {\bibfnamefont {V.}~\bibnamefont
  {{Stanishev}}}, \bibinfo {author} {\bibfnamefont {A.}~\bibnamefont
  {{Goobar}}}, \bibinfo {author} {\bibfnamefont {K.}~\bibnamefont {{Paech}}}, \
  and\ \bibinfo {author} {\bibnamefont {{et al}}},\ }\href {\doibase
  10.1051/0004-6361/200911982} {\bibfield  {journal} {\bibinfo  {journal}
  {\aap}\ }\textbf {\bibinfo {volume} {507}},\ \bibinfo {pages} {61} (\bibinfo
  {year} {2009})}\BibitemShut {NoStop}%
\bibitem [{\citenamefont {{Petrushevska}}\ \emph
  {et~al.}(2018{\natexlab{a}})\citenamefont {{Petrushevska}}, \citenamefont
  {{Goobar}}, \citenamefont {{Lagattuta}},\ and\ \citenamefont {{et
  al.}}}]{2018A&A...614A.103P}%
  \BibitemOpen
  \bibfield  {author} {\bibinfo {author} {\bibfnamefont {T.}~\bibnamefont
  {{Petrushevska}}}, \bibinfo {author} {\bibfnamefont {A.}~\bibnamefont
  {{Goobar}}}, \bibinfo {author} {\bibfnamefont {D.~J.}\ \bibnamefont
  {{Lagattuta}}}, \ and\ \bibinfo {author} {\bibnamefont {{et al.}}},\ }\href
  {\doibase 10.1051/0004-6361/201731552} {\bibfield  {journal} {\bibinfo
  {journal} {\aap}\ }\textbf {\bibinfo {volume} {614}},\ \bibinfo {eid} {A103}
  (\bibinfo {year} {2018}{\natexlab{a}})}\BibitemShut {NoStop}%
\bibitem [{\citenamefont {{Shu}}\ \emph {et~al.}(2018)\citenamefont {{Shu}},
  \citenamefont {{Bolton}}, \citenamefont {{Mao}},\ and\ \citenamefont {{et
  al.}}}]{2018ApJ...864...91S}%
  \BibitemOpen
  \bibfield  {author} {\bibinfo {author} {\bibfnamefont {Y.}~\bibnamefont
  {{Shu}}}, \bibinfo {author} {\bibfnamefont {A.~S.}\ \bibnamefont {{Bolton}}},
  \bibinfo {author} {\bibfnamefont {S.}~\bibnamefont {{Mao}}}, \ and\ \bibinfo
  {author} {\bibnamefont {{et al.}}},\ }\href {\doibase
  10.3847/1538-4357/aad5ea} {\bibfield  {journal} {\bibinfo  {journal} {\apj}\
  }\textbf {\bibinfo {volume} {864}},\ \bibinfo {eid} {91} (\bibinfo {year}
  {2018})}\BibitemShut {NoStop}%
\bibitem [{\citenamefont {{Nordin}}\ \emph {et~al.}(2014)\citenamefont
  {{Nordin}}, \citenamefont {{Rubin}}, \citenamefont {{Richard}}, \citenamefont
  {{et al.}},\ and\ \citenamefont {{Supernova Cosmology
  Project}}}]{2014MNRAS.440.2742N}%
  \BibitemOpen
  \bibfield  {author} {\bibinfo {author} {\bibfnamefont {J.}~\bibnamefont
  {{Nordin}}}, \bibinfo {author} {\bibfnamefont {D.}~\bibnamefont {{Rubin}}},
  \bibinfo {author} {\bibfnamefont {J.}~\bibnamefont {{Richard}}}, \bibinfo
  {author} {\bibnamefont {{et al.}}}, \ and\ \bibinfo {author} {\bibnamefont
  {{Supernova Cosmology Project}}},\ }\href {\doibase 10.1093/mnras/stu376}
  {\bibfield  {journal} {\bibinfo  {journal} {\mnras}\ }\textbf {\bibinfo
  {volume} {440}},\ \bibinfo {pages} {2742} (\bibinfo {year}
  {2014})}\BibitemShut {NoStop}%
\bibitem [{\citenamefont {{Patel}}\ \emph {et~al.}(2014)\citenamefont
  {{Patel}}, \citenamefont {{McCully}}, \citenamefont {{Jha}},\ and\
  \citenamefont {{et al.}}}]{2014ApJ...786....9P}%
  \BibitemOpen
  \bibfield  {author} {\bibinfo {author} {\bibfnamefont {B.}~\bibnamefont
  {{Patel}}}, \bibinfo {author} {\bibfnamefont {C.}~\bibnamefont {{McCully}}},
  \bibinfo {author} {\bibfnamefont {S.~W.}\ \bibnamefont {{Jha}}}, \ and\
  \bibinfo {author} {\bibnamefont {{et al.}}},\ }\href {\doibase
  10.1088/0004-637X/786/1/9} {\bibfield  {journal} {\bibinfo  {journal} {\apj}\
  }\textbf {\bibinfo {volume} {786}},\ \bibinfo {eid} {9} (\bibinfo {year}
  {2014})}\BibitemShut {NoStop}%
\bibitem [{\citenamefont {{Rodney}}\ \emph {et~al.}(2015)\citenamefont
  {{Rodney}}, \citenamefont {{Patel}}, \citenamefont {{Scolnic}},\ and\
  \citenamefont {{et al.}}}]{2015ApJ...811...70R}%
  \BibitemOpen
  \bibfield  {author} {\bibinfo {author} {\bibfnamefont {S.~A.}\ \bibnamefont
  {{Rodney}}}, \bibinfo {author} {\bibfnamefont {B.}~\bibnamefont {{Patel}}},
  \bibinfo {author} {\bibfnamefont {D.}~\bibnamefont {{Scolnic}}}, \ and\
  \bibinfo {author} {\bibnamefont {{et al.}}},\ }\href {\doibase
  10.1088/0004-637X/811/1/70} {\bibfield  {journal} {\bibinfo  {journal}
  {\apj}\ }\textbf {\bibinfo {volume} {811}},\ \bibinfo {eid} {70} (\bibinfo
  {year} {2015})}\BibitemShut {NoStop}%
\bibitem [{\citenamefont {{Goobar}}\ \emph {et~al.}(2009)\citenamefont
  {{Goobar}}, \citenamefont {{Paech}}, \citenamefont {{Stanishev}},\ and\
  \citenamefont {{et al.}}}]{2009A&A...507...71G}%
  \BibitemOpen
  \bibfield  {author} {\bibinfo {author} {\bibfnamefont {A.}~\bibnamefont
  {{Goobar}}}, \bibinfo {author} {\bibfnamefont {K.}~\bibnamefont {{Paech}}},
  \bibinfo {author} {\bibfnamefont {V.}~\bibnamefont {{Stanishev}}}, \ and\
  \bibinfo {author} {\bibnamefont {{et al.}}},\ }\href {\doibase
  10.1051/0004-6361/200811254} {\bibfield  {journal} {\bibinfo  {journal}
  {\aap}\ }\textbf {\bibinfo {volume} {507}},\ \bibinfo {pages} {71} (\bibinfo
  {year} {2009})}\BibitemShut {NoStop}%
\bibitem [{\citenamefont {{Amanullah}}\ \emph {et~al.}(2011)\citenamefont
  {{Amanullah}}, \citenamefont {{Goobar}}, \citenamefont {{Cl{\'e}ment}},\ and\
  \citenamefont {{et al.}}}]{2011ApJ...742L...7A}%
  \BibitemOpen
  \bibfield  {author} {\bibinfo {author} {\bibfnamefont {R.}~\bibnamefont
  {{Amanullah}}}, \bibinfo {author} {\bibfnamefont {A.}~\bibnamefont
  {{Goobar}}}, \bibinfo {author} {\bibfnamefont {B.}~\bibnamefont
  {{Cl{\'e}ment}}}, \ and\ \bibinfo {author} {\bibnamefont {{et al.}}},\ }\href
  {\doibase 10.1088/2041-8205/742/1/L7} {\bibfield  {journal} {\bibinfo
  {journal} {\apjl}\ }\textbf {\bibinfo {volume} {742}},\ \bibinfo {eid} {L7}
  (\bibinfo {year} {2011})}\BibitemShut {NoStop}%
\bibitem [{\citenamefont {{Rubin}}\ \emph {et~al.}(2018)\citenamefont
  {{Rubin}}, \citenamefont {{Hayden}}, \citenamefont {{Huang}},\ and\
  \citenamefont {{et al.}}}]{2018ApJ...866...65R}%
  \BibitemOpen
  \bibfield  {author} {\bibinfo {author} {\bibfnamefont {D.}~\bibnamefont
  {{Rubin}}}, \bibinfo {author} {\bibfnamefont {B.}~\bibnamefont {{Hayden}}},
  \bibinfo {author} {\bibfnamefont {X.}~\bibnamefont {{Huang}}}, \ and\
  \bibinfo {author} {\bibnamefont {{et al.}}},\ }\href {\doibase
  10.3847/1538-4357/aad565} {\bibfield  {journal} {\bibinfo  {journal} {\apj}\
  }\textbf {\bibinfo {volume} {866}},\ \bibinfo {eid} {65} (\bibinfo {year}
  {2018})}\BibitemShut {NoStop}%
\bibitem [{\citenamefont {{Ryczanowski}}\ \emph {et~al.}(2020)\citenamefont
  {{Ryczanowski}}, \citenamefont {{Smith}}, \citenamefont {{Bianconi}},\ and\
  \citenamefont {{et al.}}}]{2020MNRAS.495.1666R}%
  \BibitemOpen
  \bibfield  {author} {\bibinfo {author} {\bibfnamefont {D.}~\bibnamefont
  {{Ryczanowski}}}, \bibinfo {author} {\bibfnamefont {G.~P.}\ \bibnamefont
  {{Smith}}}, \bibinfo {author} {\bibfnamefont {M.}~\bibnamefont {{Bianconi}}},
  \ and\ \bibinfo {author} {\bibnamefont {{et al.}}},\ }\href {\doibase
  10.1093/mnras/staa1274} {\bibfield  {journal} {\bibinfo  {journal} {\mnras}\
  }\textbf {\bibinfo {volume} {495}},\ \bibinfo {pages} {1666} (\bibinfo {year}
  {2020})}\BibitemShut {NoStop}%
\bibitem [{\citenamefont {{Chornock}}\ \emph {et~al.}(2013)\citenamefont
  {{Chornock}}, \citenamefont {{Berger}}, \citenamefont {{Rest}},\ and\
  \citenamefont {{et al.}}}]{2013ApJ...767..162C}%
  \BibitemOpen
  \bibfield  {author} {\bibinfo {author} {\bibfnamefont {R.}~\bibnamefont
  {{Chornock}}}, \bibinfo {author} {\bibfnamefont {E.}~\bibnamefont
  {{Berger}}}, \bibinfo {author} {\bibfnamefont {A.}~\bibnamefont {{Rest}}}, \
  and\ \bibinfo {author} {\bibnamefont {{et al.}}},\ }\href {\doibase
  10.1088/0004-637X/767/2/162} {\bibfield  {journal} {\bibinfo  {journal}
  {\apj}\ }\textbf {\bibinfo {volume} {767}},\ \bibinfo {eid} {162} (\bibinfo
  {year} {2013})}\BibitemShut {NoStop}%
\bibitem [{\citenamefont {{Quimby}}\ \emph {et~al.}(2013)\citenamefont
  {{Quimby}}, \citenamefont {{Werner}}, \citenamefont {{Oguri}},\ and\
  \citenamefont {{et al.}}}]{2013ApJ...768L..20Q}%
  \BibitemOpen
  \bibfield  {author} {\bibinfo {author} {\bibfnamefont {R.~M.}\ \bibnamefont
  {{Quimby}}}, \bibinfo {author} {\bibfnamefont {M.~C.}\ \bibnamefont
  {{Werner}}}, \bibinfo {author} {\bibfnamefont {M.}~\bibnamefont {{Oguri}}}, \
  and\ \bibinfo {author} {\bibnamefont {{et al.}}},\ }\href {\doibase
  10.1088/2041-8205/768/1/L20} {\bibfield  {journal} {\bibinfo  {journal}
  {\apjl}\ }\textbf {\bibinfo {volume} {768}},\ \bibinfo {eid} {L20} (\bibinfo
  {year} {2013})}\BibitemShut {NoStop}%
\bibitem [{\citenamefont {{Treu}}\ \emph {et~al.}(2015)\citenamefont {{Treu}},
  \citenamefont {{Schmidt}}, \citenamefont {{Brammer}},\ and\ \citenamefont
  {{et al.}}}]{2015ApJ...812..114T}%
  \BibitemOpen
  \bibfield  {author} {\bibinfo {author} {\bibfnamefont {T.}~\bibnamefont
  {{Treu}}}, \bibinfo {author} {\bibfnamefont {K.~B.}\ \bibnamefont
  {{Schmidt}}}, \bibinfo {author} {\bibfnamefont {G.~B.}\ \bibnamefont
  {{Brammer}}}, \ and\ \bibinfo {author} {\bibnamefont {{et al.}}},\ }\href
  {\doibase 10.1088/0004-637X/812/2/114} {\bibfield  {journal} {\bibinfo
  {journal} {\apj}\ }\textbf {\bibinfo {volume} {812}},\ \bibinfo {eid} {114}
  (\bibinfo {year} {2015})}\BibitemShut {NoStop}%
\bibitem [{\citenamefont {{Lotz}}\ \emph {et~al.}(2017)\citenamefont {{Lotz}},
  \citenamefont {{Koekemoer}}, \citenamefont {{Coe}},\ and\ \citenamefont {{et
  al.}}}]{2017ApJ...837...97L}%
  \BibitemOpen
  \bibfield  {author} {\bibinfo {author} {\bibfnamefont {J.~M.}\ \bibnamefont
  {{Lotz}}}, \bibinfo {author} {\bibfnamefont {A.}~\bibnamefont {{Koekemoer}}},
  \bibinfo {author} {\bibfnamefont {D.}~\bibnamefont {{Coe}}}, \ and\ \bibinfo
  {author} {\bibnamefont {{et al.}}},\ }\href {\doibase
  10.3847/1538-4357/837/1/97} {\bibfield  {journal} {\bibinfo  {journal}
  {\apj}\ }\textbf {\bibinfo {volume} {837}},\ \bibinfo {eid} {97} (\bibinfo
  {year} {2017})}\BibitemShut {NoStop}%
\bibitem [{\citenamefont {{Kelly}}\ \emph
  {et~al.}(2016{\natexlab{a}})\citenamefont {{Kelly}}, \citenamefont
  {{Brammer}}, \citenamefont {{Selsing}},\ and\ \citenamefont {{et
  al.}}}]{2016ApJ...831..205K}%
  \BibitemOpen
  \bibfield  {author} {\bibinfo {author} {\bibfnamefont {P.~L.}\ \bibnamefont
  {{Kelly}}}, \bibinfo {author} {\bibfnamefont {G.}~\bibnamefont {{Brammer}}},
  \bibinfo {author} {\bibfnamefont {J.}~\bibnamefont {{Selsing}}}, \ and\
  \bibinfo {author} {\bibnamefont {{et al.}}},\ }\href {\doibase
  10.3847/0004-637X/831/2/205} {\bibfield  {journal} {\bibinfo  {journal}
  {\apj}\ }\textbf {\bibinfo {volume} {831}},\ \bibinfo {eid} {205} (\bibinfo
  {year} {2016}{\natexlab{a}})}\BibitemShut {NoStop}%
\bibitem [{\citenamefont {{Rodney}}\ \emph {et~al.}(2016)\citenamefont
  {{Rodney}}, \citenamefont {{Strolger}}, \citenamefont {{Kelly}},\ and\
  \citenamefont {{et al.}}}]{2016ApJ...820...50R}%
  \BibitemOpen
  \bibfield  {author} {\bibinfo {author} {\bibfnamefont {S.~A.}\ \bibnamefont
  {{Rodney}}}, \bibinfo {author} {\bibfnamefont {L.~G.}\ \bibnamefont
  {{Strolger}}}, \bibinfo {author} {\bibfnamefont {P.~L.}\ \bibnamefont
  {{Kelly}}}, \ and\ \bibinfo {author} {\bibnamefont {{et al.}}},\ }\href
  {\doibase 10.3847/0004-637X/820/1/50} {\bibfield  {journal} {\bibinfo
  {journal} {\apj}\ }\textbf {\bibinfo {volume} {820}},\ \bibinfo {eid} {50}
  (\bibinfo {year} {2016})}\BibitemShut {NoStop}%
\bibitem [{\citenamefont {{Oguri}}(2015)}]{2015MNRAS.449L..86O}%
  \BibitemOpen
  \bibfield  {author} {\bibinfo {author} {\bibfnamefont {M.}~\bibnamefont
  {{Oguri}}},\ }\href {\doibase 10.1093/mnrasl/slv025} {\bibfield  {journal}
  {\bibinfo  {journal} {\mnras}\ }\textbf {\bibinfo {volume} {449}},\ \bibinfo
  {pages} {L86} (\bibinfo {year} {2015})}\BibitemShut {NoStop}%
\bibitem [{\citenamefont {{Sharon}}\ and\ \citenamefont
  {{Johnson}}(2015)}]{2015ApJ...800L..26S}%
  \BibitemOpen
  \bibfield  {author} {\bibinfo {author} {\bibfnamefont {K.}~\bibnamefont
  {{Sharon}}}\ and\ \bibinfo {author} {\bibfnamefont {T.~L.}\ \bibnamefont
  {{Johnson}}},\ }\href {\doibase 10.1088/2041-8205/800/2/L26} {\bibfield
  {journal} {\bibinfo  {journal} {\apjl}\ }\textbf {\bibinfo {volume} {800}},\
  \bibinfo {eid} {L26} (\bibinfo {year} {2015})}\BibitemShut {NoStop}%
\bibitem [{\citenamefont {{Treu}}\ \emph {et~al.}(2016)\citenamefont {{Treu}},
  \citenamefont {{Brammer}}, \citenamefont {{Diego}},\ and\ \citenamefont {{et
  al.}}}]{2016ApJ...817...60T}%
  \BibitemOpen
  \bibfield  {author} {\bibinfo {author} {\bibfnamefont {T.}~\bibnamefont
  {{Treu}}}, \bibinfo {author} {\bibfnamefont {G.}~\bibnamefont {{Brammer}}},
  \bibinfo {author} {\bibfnamefont {J.~M.}\ \bibnamefont {{Diego}}}, \ and\
  \bibinfo {author} {\bibnamefont {{et al.}}},\ }\href {\doibase
  10.3847/0004-637X/817/1/60} {\bibfield  {journal} {\bibinfo  {journal}
  {\apj}\ }\textbf {\bibinfo {volume} {817}},\ \bibinfo {eid} {60} (\bibinfo
  {year} {2016})}\BibitemShut {NoStop}%
\bibitem [{\citenamefont {{Diego}}\ \emph {et~al.}(2016)\citenamefont
  {{Diego}}, \citenamefont {{Broadhurst}}, \citenamefont {{Chen}},\ and\
  \citenamefont {{et al.}}}]{2016MNRAS.456..356D}%
  \BibitemOpen
  \bibfield  {author} {\bibinfo {author} {\bibfnamefont {J.~M.}\ \bibnamefont
  {{Diego}}}, \bibinfo {author} {\bibfnamefont {T.}~\bibnamefont
  {{Broadhurst}}}, \bibinfo {author} {\bibfnamefont {C.}~\bibnamefont
  {{Chen}}}, \ and\ \bibinfo {author} {\bibnamefont {{et al.}}},\ }\href
  {\doibase 10.1093/mnras/stv2638} {\bibfield  {journal} {\bibinfo  {journal}
  {\mnras}\ }\textbf {\bibinfo {volume} {456}},\ \bibinfo {pages} {356}
  (\bibinfo {year} {2016})}\BibitemShut {NoStop}%
\bibitem [{\citenamefont {{Jauzac}}\ \emph {et~al.}(2016)\citenamefont
  {{Jauzac}}, \citenamefont {{Richard}}, \citenamefont {{Limousin}},\ and\
  \citenamefont {{et al.}}}]{2016MNRAS.457.2029J}%
  \BibitemOpen
  \bibfield  {author} {\bibinfo {author} {\bibfnamefont {M.}~\bibnamefont
  {{Jauzac}}}, \bibinfo {author} {\bibfnamefont {J.}~\bibnamefont {{Richard}}},
  \bibinfo {author} {\bibfnamefont {M.}~\bibnamefont {{Limousin}}}, \ and\
  \bibinfo {author} {\bibnamefont {{et al.}}},\ }\href {\doibase
  10.1093/mnras/stw069} {\bibfield  {journal} {\bibinfo  {journal} {\mnras}\
  }\textbf {\bibinfo {volume} {457}},\ \bibinfo {pages} {2029} (\bibinfo {year}
  {2016})}\BibitemShut {NoStop}%
\bibitem [{\citenamefont {{Grillo}}\ \emph {et~al.}(2016)\citenamefont
  {{Grillo}}, \citenamefont {{Karman}}, \citenamefont {{Suyu}},\ and\
  \citenamefont {{et al.}}}]{2016ApJ...822...78G}%
  \BibitemOpen
  \bibfield  {author} {\bibinfo {author} {\bibfnamefont {C.}~\bibnamefont
  {{Grillo}}}, \bibinfo {author} {\bibfnamefont {W.}~\bibnamefont {{Karman}}},
  \bibinfo {author} {\bibfnamefont {S.~H.}\ \bibnamefont {{Suyu}}}, \ and\
  \bibinfo {author} {\bibnamefont {{et al.}}},\ }\href {\doibase
  10.3847/0004-637X/822/2/78} {\bibfield  {journal} {\bibinfo  {journal}
  {\apj}\ }\textbf {\bibinfo {volume} {822}},\ \bibinfo {eid} {78} (\bibinfo
  {year} {2016})}\BibitemShut {NoStop}%
\bibitem [{\citenamefont {{Kelly}}\ \emph
  {et~al.}(2016{\natexlab{b}})\citenamefont {{Kelly}}, \citenamefont
  {{Rodney}}, \citenamefont {{Treu}},\ and\ \citenamefont {{et
  al.}}}]{2016ApJ...819L...8K}%
  \BibitemOpen
  \bibfield  {author} {\bibinfo {author} {\bibfnamefont {P.~L.}\ \bibnamefont
  {{Kelly}}}, \bibinfo {author} {\bibfnamefont {S.~A.}\ \bibnamefont
  {{Rodney}}}, \bibinfo {author} {\bibfnamefont {T.}~\bibnamefont {{Treu}}}, \
  and\ \bibinfo {author} {\bibnamefont {{et al.}}},\ }\href {\doibase
  10.3847/2041-8205/819/1/L8} {\bibfield  {journal} {\bibinfo  {journal}
  {\apjl}\ }\textbf {\bibinfo {volume} {819}},\ \bibinfo {eid} {L8} (\bibinfo
  {year} {2016}{\natexlab{b}})}\BibitemShut {NoStop}%
\bibitem [{\citenamefont {{Kawamata}}\ \emph {et~al.}(2016)\citenamefont
  {{Kawamata}}, \citenamefont {{Oguri}}, \citenamefont {{Ishigaki}},
  \citenamefont {{Shimasaku}},\ and\ \citenamefont
  {{Ouchi}}}]{2016ApJ...819..114K}%
  \BibitemOpen
  \bibfield  {author} {\bibinfo {author} {\bibfnamefont {R.}~\bibnamefont
  {{Kawamata}}}, \bibinfo {author} {\bibfnamefont {M.}~\bibnamefont {{Oguri}}},
  \bibinfo {author} {\bibfnamefont {M.}~\bibnamefont {{Ishigaki}}}, \bibinfo
  {author} {\bibfnamefont {K.}~\bibnamefont {{Shimasaku}}}, \ and\ \bibinfo
  {author} {\bibfnamefont {M.}~\bibnamefont {{Ouchi}}},\ }\href {\doibase
  10.3847/0004-637X/819/2/114} {\bibfield  {journal} {\bibinfo  {journal}
  {\apj}\ }\textbf {\bibinfo {volume} {819}},\ \bibinfo {eid} {114} (\bibinfo
  {year} {2016})}\BibitemShut {NoStop}%
\bibitem [{\citenamefont {{Vega-Ferrero}}\ \emph {et~al.}(2018)\citenamefont
  {{Vega-Ferrero}}, \citenamefont {{Diego}}, \citenamefont {{Miranda}},\ and\
  \citenamefont {{Bernstein}}}]{2018ApJ...853L..31V}%
  \BibitemOpen
  \bibfield  {author} {\bibinfo {author} {\bibfnamefont {J.}~\bibnamefont
  {{Vega-Ferrero}}}, \bibinfo {author} {\bibfnamefont {J.~M.}\ \bibnamefont
  {{Diego}}}, \bibinfo {author} {\bibfnamefont {V.}~\bibnamefont {{Miranda}}},
  \ and\ \bibinfo {author} {\bibfnamefont {G.~M.}\ \bibnamefont
  {{Bernstein}}},\ }\href {\doibase 10.3847/2041-8213/aaa95f} {\bibfield
  {journal} {\bibinfo  {journal} {\apjl}\ }\textbf {\bibinfo {volume} {853}},\
  \bibinfo {eid} {L31} (\bibinfo {year} {2018})}\BibitemShut {NoStop}%
\bibitem [{\citenamefont {{Grillo}}\ \emph {et~al.}(2018)\citenamefont
  {{Grillo}}, \citenamefont {{Rosati}}, \citenamefont {{Suyu}},\ and\
  \citenamefont {{et al.}}}]{2018ApJ...860...94G}%
  \BibitemOpen
  \bibfield  {author} {\bibinfo {author} {\bibfnamefont {C.}~\bibnamefont
  {{Grillo}}}, \bibinfo {author} {\bibfnamefont {P.}~\bibnamefont {{Rosati}}},
  \bibinfo {author} {\bibfnamefont {S.~H.}\ \bibnamefont {{Suyu}}}, \ and\
  \bibinfo {author} {\bibnamefont {{et al.}}},\ }\href {\doibase
  10.3847/1538-4357/aac2c9} {\bibfield  {journal} {\bibinfo  {journal} {\apj}\
  }\textbf {\bibinfo {volume} {860}},\ \bibinfo {eid} {94} (\bibinfo {year}
  {2018})}\BibitemShut {NoStop}%
\bibitem [{\citenamefont {{Williams}}\ and\ \citenamefont
  {{Liesenborgs}}(2019)}]{2019MNRAS.482.5666W}%
  \BibitemOpen
  \bibfield  {author} {\bibinfo {author} {\bibfnamefont {L.~L.~R.}\
  \bibnamefont {{Williams}}}\ and\ \bibinfo {author} {\bibfnamefont
  {J.}~\bibnamefont {{Liesenborgs}}},\ }\href {\doibase 10.1093/mnras/sty3113}
  {\bibfield  {journal} {\bibinfo  {journal} {\mnras}\ }\textbf {\bibinfo
  {volume} {482}},\ \bibinfo {pages} {5666} (\bibinfo {year}
  {2019})}\BibitemShut {NoStop}%
\bibitem [{\citenamefont {{Grillo}}\ \emph {et~al.}(2020)\citenamefont
  {{Grillo}}, \citenamefont {{Rosati}}, \citenamefont {{Suyu}},\ and\
  \citenamefont {{et al.}}}]{2020ApJ...898...87G}%
  \BibitemOpen
  \bibfield  {author} {\bibinfo {author} {\bibfnamefont {C.}~\bibnamefont
  {{Grillo}}}, \bibinfo {author} {\bibfnamefont {P.}~\bibnamefont {{Rosati}}},
  \bibinfo {author} {\bibfnamefont {S.~H.}\ \bibnamefont {{Suyu}}}, \ and\
  \bibinfo {author} {\bibnamefont {{et al.}}},\ }\href {\doibase
  10.3847/1538-4357/ab9a4c} {\bibfield  {journal} {\bibinfo  {journal} {\apj}\
  }\textbf {\bibinfo {volume} {898}},\ \bibinfo {eid} {87} (\bibinfo {year}
  {2020})}\BibitemShut {NoStop}%
\bibitem [{\citenamefont {{Goobar}}\ \emph {et~al.}(2017)\citenamefont
  {{Goobar}}, \citenamefont {{Amanullah}}, \citenamefont {{Kulkarni}},\ and\
  \citenamefont {{et al.}}}]{2017Sci...356..291G}%
  \BibitemOpen
  \bibfield  {author} {\bibinfo {author} {\bibfnamefont {A.}~\bibnamefont
  {{Goobar}}}, \bibinfo {author} {\bibfnamefont {R.}~\bibnamefont
  {{Amanullah}}}, \bibinfo {author} {\bibfnamefont {S.~R.}\ \bibnamefont
  {{Kulkarni}}}, \ and\ \bibinfo {author} {\bibnamefont {{et al.}}},\ }\href
  {\doibase 10.1126/science.aal2729} {\bibfield  {journal} {\bibinfo  {journal}
  {Science}\ }\textbf {\bibinfo {volume} {356}},\ \bibinfo {pages} {291}
  (\bibinfo {year} {2017})}\BibitemShut {NoStop}%
\bibitem [{\citenamefont {{Kulkarni}}(2013)}]{2013ATel.4807....1K}%
  \BibitemOpen
  \bibfield  {author} {\bibinfo {author} {\bibfnamefont {S.~R.}\ \bibnamefont
  {{Kulkarni}}},\ }\href@noop {} {\bibfield  {journal} {\bibinfo  {journal}
  {The Astronomer's Telegram}\ }\textbf {\bibinfo {volume} {4807}},\ \bibinfo
  {pages} {1} (\bibinfo {year} {2013})}\BibitemShut {NoStop}%
\bibitem [{\citenamefont {{Dhawan}}\ \emph {et~al.}(2020)\citenamefont
  {{Dhawan}}, \citenamefont {{Johansson}}, \citenamefont {{Goobar}},\ and\
  \citenamefont {{et al.}}}]{2020MNRAS.491.2639D}%
  \BibitemOpen
  \bibfield  {author} {\bibinfo {author} {\bibfnamefont {S.}~\bibnamefont
  {{Dhawan}}}, \bibinfo {author} {\bibfnamefont {J.}~\bibnamefont
  {{Johansson}}}, \bibinfo {author} {\bibfnamefont {A.}~\bibnamefont
  {{Goobar}}}, \ and\ \bibinfo {author} {\bibnamefont {{et al.}}},\ }\href
  {\doibase 10.1093/mnras/stz2965} {\bibfield  {journal} {\bibinfo  {journal}
  {\mnras}\ }\textbf {\bibinfo {volume} {491}},\ \bibinfo {pages} {2639}
  (\bibinfo {year} {2020})}\BibitemShut {NoStop}%
\bibitem [{\citenamefont {{More}}\ \emph {et~al.}(2017)\citenamefont {{More}},
  \citenamefont {{Suyu}}, \citenamefont {{Oguri}}, \citenamefont {{More}},\
  and\ \citenamefont {{Lee}}}]{2017ApJ...835L..25M}%
  \BibitemOpen
  \bibfield  {author} {\bibinfo {author} {\bibfnamefont {A.}~\bibnamefont
  {{More}}}, \bibinfo {author} {\bibfnamefont {S.~H.}\ \bibnamefont {{Suyu}}},
  \bibinfo {author} {\bibfnamefont {M.}~\bibnamefont {{Oguri}}}, \bibinfo
  {author} {\bibfnamefont {S.}~\bibnamefont {{More}}}, \ and\ \bibinfo {author}
  {\bibfnamefont {C.-H.}\ \bibnamefont {{Lee}}},\ }\href {\doibase
  10.3847/2041-8213/835/2/L25} {\bibfield  {journal} {\bibinfo  {journal}
  {\apjl}\ }\textbf {\bibinfo {volume} {835}},\ \bibinfo {eid} {L25} (\bibinfo
  {year} {2017})}\BibitemShut {NoStop}%
\bibitem [{\citenamefont {{M{\"o}rtsell}}\ \emph {et~al.}(2020)\citenamefont
  {{M{\"o}rtsell}}, \citenamefont {{Johansson}}, \citenamefont {{Dhawan}},\
  and\ \citenamefont {{et al.}}}]{2020MNRAS.496.3270M}%
  \BibitemOpen
  \bibfield  {author} {\bibinfo {author} {\bibfnamefont {E.}~\bibnamefont
  {{M{\"o}rtsell}}}, \bibinfo {author} {\bibfnamefont {J.}~\bibnamefont
  {{Johansson}}}, \bibinfo {author} {\bibfnamefont {S.}~\bibnamefont
  {{Dhawan}}}, \ and\ \bibinfo {author} {\bibnamefont {{et al.}}},\ }\href
  {\doibase 10.1093/mnras/staa1600} {\bibfield  {journal} {\bibinfo  {journal}
  {\mnras}\ }\textbf {\bibinfo {volume} {496}},\ \bibinfo {pages} {3270}
  (\bibinfo {year} {2020})}\BibitemShut {NoStop}%
\bibitem [{\citenamefont {{Diego}}\ \emph {et~al.}(2022)\citenamefont
  {{Diego}}, \citenamefont {{Bernstein}}, \citenamefont {{Chen}}, \citenamefont
  {{Goobar}}, \citenamefont {{Johansson}}, \citenamefont {{Kelly}},
  \citenamefont {{M{\"o}rtsell}},\ and\ \citenamefont
  {{Nightingale}}}]{2022A&A...662A..34D}%
  \BibitemOpen
  \bibfield  {author} {\bibinfo {author} {\bibfnamefont {J.~M.}\ \bibnamefont
  {{Diego}}}, \bibinfo {author} {\bibfnamefont {G.}~\bibnamefont
  {{Bernstein}}}, \bibinfo {author} {\bibfnamefont {W.}~\bibnamefont {{Chen}}},
  \bibinfo {author} {\bibfnamefont {A.}~\bibnamefont {{Goobar}}}, \bibinfo
  {author} {\bibfnamefont {J.~P.}\ \bibnamefont {{Johansson}}}, \bibinfo
  {author} {\bibfnamefont {P.~L.}\ \bibnamefont {{Kelly}}}, \bibinfo {author}
  {\bibfnamefont {E.}~\bibnamefont {{M{\"o}rtsell}}}, \ and\ \bibinfo {author}
  {\bibfnamefont {J.~W.}\ \bibnamefont {{Nightingale}}},\ }\href {\doibase
  10.1051/0004-6361/202143009} {\bibfield  {journal} {\bibinfo  {journal}
  {\aap}\ }\textbf {\bibinfo {volume} {662}},\ \bibinfo {eid} {A34} (\bibinfo
  {year} {2022})}\BibitemShut {NoStop}%
\bibitem [{\citenamefont {{Yahalomi}}\ \emph {et~al.}(2017)\citenamefont
  {{Yahalomi}}, \citenamefont {{Schechter}},\ and\ \citenamefont
  {{Wambsganss}}}]{2017arXiv171107919Y}%
  \BibitemOpen
  \bibfield  {author} {\bibinfo {author} {\bibfnamefont {D.~A.}\ \bibnamefont
  {{Yahalomi}}}, \bibinfo {author} {\bibfnamefont {P.~L.}\ \bibnamefont
  {{Schechter}}}, \ and\ \bibinfo {author} {\bibfnamefont {J.}~\bibnamefont
  {{Wambsganss}}},\ }\href@noop {} {\bibfield  {journal} {\bibinfo  {journal}
  {arXiv e-prints}\ } (\bibinfo {year} {2017})},\ \Eprint
  {http://arxiv.org/abs/1711.07919} {1711.07919} \BibitemShut {NoStop}%
\bibitem [{\citenamefont {{Foxley-Marrable}}\ \emph {et~al.}(2018)\citenamefont
  {{Foxley-Marrable}}, \citenamefont {{Collett}}, \citenamefont {{Vernardos}},
  \citenamefont {{Goldstein}},\ and\ \citenamefont
  {{Bacon}}}]{2018MNRAS.478.5081F}%
  \BibitemOpen
  \bibfield  {author} {\bibinfo {author} {\bibfnamefont {M.}~\bibnamefont
  {{Foxley-Marrable}}}, \bibinfo {author} {\bibfnamefont {T.~E.}\ \bibnamefont
  {{Collett}}}, \bibinfo {author} {\bibfnamefont {G.}~\bibnamefont
  {{Vernardos}}}, \bibinfo {author} {\bibfnamefont {D.~A.}\ \bibnamefont
  {{Goldstein}}}, \ and\ \bibinfo {author} {\bibfnamefont {D.}~\bibnamefont
  {{Bacon}}},\ }\href {\doibase 10.1093/mnras/sty1346} {\bibfield  {journal}
  {\bibinfo  {journal} {\mnras}\ }\textbf {\bibinfo {volume} {478}},\ \bibinfo
  {pages} {5081} (\bibinfo {year} {2018})}\BibitemShut {NoStop}%
\bibitem [{\citenamefont {{Rodney}}\ \emph {et~al.}(2021)\citenamefont
  {{Rodney}}, \citenamefont {{Brammer}}, \citenamefont {{Pierel}},\ and\
  \citenamefont {{et al.}}}]{2021NatAs...5.1118R}%
  \BibitemOpen
  \bibfield  {author} {\bibinfo {author} {\bibfnamefont {S.~A.}\ \bibnamefont
  {{Rodney}}}, \bibinfo {author} {\bibfnamefont {G.~B.}\ \bibnamefont
  {{Brammer}}}, \bibinfo {author} {\bibfnamefont {J.~D.~R.}\ \bibnamefont
  {{Pierel}}}, \ and\ \bibinfo {author} {\bibnamefont {{et al.}}},\ }\href
  {\doibase 10.1038/s41550-021-01450-9} {\bibfield  {journal} {\bibinfo
  {journal} {Nature Astronomy}\ }\textbf {\bibinfo {volume} {5}},\ \bibinfo
  {pages} {1118} (\bibinfo {year} {2021})}\BibitemShut {NoStop}%
\bibitem [{\citenamefont {{Akhshik}}\ \emph {et~al.}(2021)\citenamefont
  {{Akhshik}}, \citenamefont {{Whitaker}}, \citenamefont {{Leja}},\ and\
  \citenamefont {{et al.}}}]{2021ApJ...907L...8A}%
  \BibitemOpen
  \bibfield  {author} {\bibinfo {author} {\bibfnamefont {M.}~\bibnamefont
  {{Akhshik}}}, \bibinfo {author} {\bibfnamefont {K.~E.}\ \bibnamefont
  {{Whitaker}}}, \bibinfo {author} {\bibfnamefont {J.}~\bibnamefont {{Leja}}},
  \ and\ \bibinfo {author} {\bibnamefont {{et al.}}},\ }\href {\doibase
  10.3847/2041-8213/abd416} {\bibfield  {journal} {\bibinfo  {journal} {\apjl}\
  }\textbf {\bibinfo {volume} {907}},\ \bibinfo {eid} {L8} (\bibinfo {year}
  {2021})}\BibitemShut {NoStop}%
\bibitem [{\citenamefont {{Fishman}}\ and\ \citenamefont
  {{Meegan}}(1995)}]{1995ARA&A..33..415F}%
  \BibitemOpen
  \bibfield  {author} {\bibinfo {author} {\bibfnamefont {G.~J.}\ \bibnamefont
  {{Fishman}}}\ and\ \bibinfo {author} {\bibfnamefont {C.~A.}\ \bibnamefont
  {{Meegan}}},\ }\href {\doibase 10.1146/annurev.aa.33.090195.002215}
  {\bibfield  {journal} {\bibinfo  {journal} {\araa}\ }\textbf {\bibinfo
  {volume} {33}},\ \bibinfo {pages} {415} (\bibinfo {year} {1995})}\BibitemShut
  {NoStop}%
\bibitem [{\citenamefont {{M{\'e}sz{\'a}ros}}(2002)}]{2002ARA&A..40..137M}%
  \BibitemOpen
  \bibfield  {author} {\bibinfo {author} {\bibfnamefont {P.}~\bibnamefont
  {{M{\'e}sz{\'a}ros}}},\ }\href {\doibase
  10.1146/annurev.astro.40.060401.093821} {\bibfield  {journal} {\bibinfo
  {journal} {\araa}\ }\textbf {\bibinfo {volume} {40}},\ \bibinfo {pages} {137}
  (\bibinfo {year} {2002})}\BibitemShut {NoStop}%
\bibitem [{\citenamefont {{van Paradijs}}\ \emph {et~al.}(2000)\citenamefont
  {{van Paradijs}}, \citenamefont {{Kouveliotou}},\ and\ \citenamefont
  {{Wijers}}}]{2000ARA&A..38..379V}%
  \BibitemOpen
  \bibfield  {author} {\bibinfo {author} {\bibfnamefont {J.}~\bibnamefont {{van
  Paradijs}}}, \bibinfo {author} {\bibfnamefont {C.}~\bibnamefont
  {{Kouveliotou}}}, \ and\ \bibinfo {author} {\bibfnamefont {R.~A.~M.~J.}\
  \bibnamefont {{Wijers}}},\ }\href {\doibase 10.1146/annurev.astro.38.1.379}
  {\bibfield  {journal} {\bibinfo  {journal} {\araa}\ }\textbf {\bibinfo
  {volume} {38}},\ \bibinfo {pages} {379} (\bibinfo {year} {2000})}\BibitemShut
  {NoStop}%
\bibitem [{\citenamefont {{Klebesadel}}\ \emph {et~al.}(1973)\citenamefont
  {{Klebesadel}}, \citenamefont {{Strong}},\ and\ \citenamefont
  {{Olson}}}]{1973ApJ...182L..85K}%
  \BibitemOpen
  \bibfield  {author} {\bibinfo {author} {\bibfnamefont {R.~W.}\ \bibnamefont
  {{Klebesadel}}}, \bibinfo {author} {\bibfnamefont {I.~B.}\ \bibnamefont
  {{Strong}}}, \ and\ \bibinfo {author} {\bibfnamefont {R.~A.}\ \bibnamefont
  {{Olson}}},\ }\href {\doibase 10.1086/181225} {\bibfield  {journal} {\bibinfo
   {journal} {\apjl}\ }\textbf {\bibinfo {volume} {182}},\ \bibinfo {pages}
  {L85} (\bibinfo {year} {1973})}\BibitemShut {NoStop}%
\bibitem [{\citenamefont {{Meegan}}\ \emph {et~al.}(1992)\citenamefont
  {{Meegan}}, \citenamefont {{Fishman}}, \citenamefont {{Wilson}},\ and\
  \citenamefont {{et al.}}}]{1992Natur.355..143M}%
  \BibitemOpen
  \bibfield  {author} {\bibinfo {author} {\bibfnamefont {C.~A.}\ \bibnamefont
  {{Meegan}}}, \bibinfo {author} {\bibfnamefont {G.~J.}\ \bibnamefont
  {{Fishman}}}, \bibinfo {author} {\bibfnamefont {R.~B.}\ \bibnamefont
  {{Wilson}}}, \ and\ \bibinfo {author} {\bibnamefont {{et al.}}},\ }\href
  {\doibase 10.1038/355143a0} {\bibfield  {journal} {\bibinfo  {journal}
  {\nat}\ }\textbf {\bibinfo {volume} {355}},\ \bibinfo {pages} {143} (\bibinfo
  {year} {1992})}\BibitemShut {NoStop}%
\bibitem [{\citenamefont {{Gehrels}}\ \emph {et~al.}(2004)\citenamefont
  {{Gehrels}}, \citenamefont {{Chincarini}}, \citenamefont {{Giommi}},\ and\
  \citenamefont {{et al.}}}]{2004ApJ...611.1005G}%
  \BibitemOpen
  \bibfield  {author} {\bibinfo {author} {\bibfnamefont {N.}~\bibnamefont
  {{Gehrels}}}, \bibinfo {author} {\bibfnamefont {G.}~\bibnamefont
  {{Chincarini}}}, \bibinfo {author} {\bibfnamefont {P.}~\bibnamefont
  {{Giommi}}}, \ and\ \bibinfo {author} {\bibnamefont {{et al.}}},\ }\href
  {\doibase 10.1086/422091} {\bibfield  {journal} {\bibinfo  {journal} {\apj}\
  }\textbf {\bibinfo {volume} {611}},\ \bibinfo {pages} {1005} (\bibinfo {year}
  {2004})}\BibitemShut {NoStop}%
\bibitem [{\citenamefont {{Atwood}}\ \emph {et~al.}(2009)\citenamefont
  {{Atwood}}, \citenamefont {{Abdo}}, \citenamefont {{Ackermann}},\ and\
  \citenamefont {{et al.}}}]{2009ApJ...697.1071A}%
  \BibitemOpen
  \bibfield  {author} {\bibinfo {author} {\bibfnamefont {W.~B.}\ \bibnamefont
  {{Atwood}}}, \bibinfo {author} {\bibfnamefont {A.~A.}\ \bibnamefont
  {{Abdo}}}, \bibinfo {author} {\bibfnamefont {M.}~\bibnamefont {{Ackermann}}},
  \ and\ \bibinfo {author} {\bibnamefont {{et al.}}},\ }\href {\doibase
  10.1088/0004-637X/697/2/1071} {\bibfield  {journal} {\bibinfo  {journal}
  {\apj}\ }\textbf {\bibinfo {volume} {697}},\ \bibinfo {pages} {1071}
  (\bibinfo {year} {2009})}\BibitemShut {NoStop}%
\bibitem [{\citenamefont {{Meegan}}\ \emph {et~al.}(2009)\citenamefont
  {{Meegan}}, \citenamefont {{Lichti}}, \citenamefont {{Bhat}},\ and\
  \citenamefont {{et al.}}}]{2009ApJ...702..791M}%
  \BibitemOpen
  \bibfield  {author} {\bibinfo {author} {\bibfnamefont {C.}~\bibnamefont
  {{Meegan}}}, \bibinfo {author} {\bibfnamefont {G.}~\bibnamefont {{Lichti}}},
  \bibinfo {author} {\bibfnamefont {P.~N.}\ \bibnamefont {{Bhat}}}, \ and\
  \bibinfo {author} {\bibnamefont {{et al.}}},\ }\href {\doibase
  10.1088/0004-637X/702/1/791} {\bibfield  {journal} {\bibinfo  {journal}
  {\apj}\ }\textbf {\bibinfo {volume} {702}},\ \bibinfo {pages} {791} (\bibinfo
  {year} {2009})}\BibitemShut {NoStop}%
\bibitem [{\citenamefont {{Woosley}}\ and\ \citenamefont
  {{Bloom}}(2006)}]{2006ARA&A..44..507W}%
  \BibitemOpen
  \bibfield  {author} {\bibinfo {author} {\bibfnamefont {S.~E.}\ \bibnamefont
  {{Woosley}}}\ and\ \bibinfo {author} {\bibfnamefont {J.~S.}\ \bibnamefont
  {{Bloom}}},\ }\href {\doibase 10.1146/annurev.astro.43.072103.150558}
  {\bibfield  {journal} {\bibinfo  {journal} {\araa}\ }\textbf {\bibinfo
  {volume} {44}},\ \bibinfo {pages} {507} (\bibinfo {year} {2006})}\BibitemShut
  {NoStop}%
\bibitem [{\citenamefont {{Berger}}(2014)}]{2014ARA&A..52...43B}%
  \BibitemOpen
  \bibfield  {author} {\bibinfo {author} {\bibfnamefont {E.}~\bibnamefont
  {{Berger}}},\ }\href {\doibase 10.1146/annurev-astro-081913-035926}
  {\bibfield  {journal} {\bibinfo  {journal} {\araa}\ }\textbf {\bibinfo
  {volume} {52}},\ \bibinfo {pages} {43} (\bibinfo {year} {2014})}\BibitemShut
  {NoStop}%
\bibitem [{\citenamefont {{Cucchiara}}\ \emph {et~al.}(2011)\citenamefont
  {{Cucchiara}}, \citenamefont {{Levan}}, \citenamefont {{Fox}},\ and\
  \citenamefont {{et al.}}}]{2011ApJ...736....7C}%
  \BibitemOpen
  \bibfield  {author} {\bibinfo {author} {\bibfnamefont {A.}~\bibnamefont
  {{Cucchiara}}}, \bibinfo {author} {\bibfnamefont {A.~J.}\ \bibnamefont
  {{Levan}}}, \bibinfo {author} {\bibfnamefont {D.~B.}\ \bibnamefont {{Fox}}},
  \ and\ \bibinfo {author} {\bibnamefont {{et al.}}},\ }\href {\doibase
  10.1088/0004-637X/736/1/7} {\bibfield  {journal} {\bibinfo  {journal} {\apj}\
  }\textbf {\bibinfo {volume} {736}},\ \bibinfo {eid} {7} (\bibinfo {year}
  {2011})}\BibitemShut {NoStop}%
\bibitem [{\citenamefont {{Paczynski}}(1986)}]{1986ApJ...308L..43P}%
  \BibitemOpen
  \bibfield  {author} {\bibinfo {author} {\bibfnamefont {B.}~\bibnamefont
  {{Paczynski}}},\ }\href {\doibase 10.1086/184740} {\bibfield  {journal}
  {\bibinfo  {journal} {\apjl}\ }\textbf {\bibinfo {volume} {308}},\ \bibinfo
  {pages} {L43} (\bibinfo {year} {1986})}\BibitemShut {NoStop}%
\bibitem [{\citenamefont {{Paczynski}}(1987)}]{1987ApJ...317L..51P}%
  \BibitemOpen
  \bibfield  {author} {\bibinfo {author} {\bibfnamefont {B.}~\bibnamefont
  {{Paczynski}}},\ }\href {\doibase 10.1086/184911} {\bibfield  {journal}
  {\bibinfo  {journal} {\apjl}\ }\textbf {\bibinfo {volume} {317}},\ \bibinfo
  {pages} {L51} (\bibinfo {year} {1987})}\BibitemShut {NoStop}%
\bibitem [{\citenamefont {{Mao}}(1993)}]{1993ApJ...402..382M}%
  \BibitemOpen
  \bibfield  {author} {\bibinfo {author} {\bibfnamefont {S.}~\bibnamefont
  {{Mao}}},\ }\href {\doibase 10.1086/172142} {\bibfield  {journal} {\bibinfo
  {journal} {\apj}\ }\textbf {\bibinfo {volume} {402}},\ \bibinfo {pages} {382}
  (\bibinfo {year} {1993})}\BibitemShut {NoStop}%
\bibitem [{\citenamefont {{Wambsganss}}(1993)}]{1993ApJ...406...29W}%
  \BibitemOpen
  \bibfield  {author} {\bibinfo {author} {\bibfnamefont {J.}~\bibnamefont
  {{Wambsganss}}},\ }\href {\doibase 10.1086/172416} {\bibfield  {journal}
  {\bibinfo  {journal} {\apj}\ }\textbf {\bibinfo {volume} {406}},\ \bibinfo
  {pages} {29} (\bibinfo {year} {1993})}\BibitemShut {NoStop}%
\bibitem [{\citenamefont {{Nowak}}\ and\ \citenamefont
  {{Grossman}}(1994)}]{1994ApJ...435..557N}%
  \BibitemOpen
  \bibfield  {author} {\bibinfo {author} {\bibfnamefont {M.~A.}\ \bibnamefont
  {{Nowak}}}\ and\ \bibinfo {author} {\bibfnamefont {S.~A.}\ \bibnamefont
  {{Grossman}}},\ }\href {\doibase 10.1086/174837} {\bibfield  {journal}
  {\bibinfo  {journal} {\apj}\ }\textbf {\bibinfo {volume} {435}},\ \bibinfo
  {pages} {557} (\bibinfo {year} {1994})}\BibitemShut {NoStop}%
\bibitem [{\citenamefont {{Grossman}}\ and\ \citenamefont
  {{Nowak}}(1994)}]{1994ApJ...435..548G}%
  \BibitemOpen
  \bibfield  {author} {\bibinfo {author} {\bibfnamefont {S.~A.}\ \bibnamefont
  {{Grossman}}}\ and\ \bibinfo {author} {\bibfnamefont {M.~A.}\ \bibnamefont
  {{Nowak}}},\ }\href {\doibase 10.1086/174836} {\bibfield  {journal} {\bibinfo
   {journal} {\apj}\ }\textbf {\bibinfo {volume} {435}},\ \bibinfo {pages}
  {548} (\bibinfo {year} {1994})}\BibitemShut {NoStop}%
\bibitem [{\citenamefont {{Quashnock}}\ and\ \citenamefont
  {{Lamb}}(1993)}]{1993MNRAS.265L..59Q}%
  \BibitemOpen
  \bibfield  {author} {\bibinfo {author} {\bibfnamefont {J.~M.}\ \bibnamefont
  {{Quashnock}}}\ and\ \bibinfo {author} {\bibfnamefont {D.~Q.}\ \bibnamefont
  {{Lamb}}},\ }\href {\doibase 10.1093/mnras/265.1.L59} {\bibfield  {journal}
  {\bibinfo  {journal} {\mnras}\ }\textbf {\bibinfo {volume} {265}},\ \bibinfo
  {pages} {L59} (\bibinfo {year} {1993})}\BibitemShut {NoStop}%
\bibitem [{\citenamefont {{Tegmark}}\ \emph {et~al.}(1996)\citenamefont
  {{Tegmark}}, \citenamefont {{Hartmann}}, \citenamefont {{Briggs}},
  \citenamefont {{Hakkila}},\ and\ \citenamefont
  {{Meegan}}}]{1996ApJ...466..757T}%
  \BibitemOpen
  \bibfield  {author} {\bibinfo {author} {\bibfnamefont {M.}~\bibnamefont
  {{Tegmark}}}, \bibinfo {author} {\bibfnamefont {D.~H.}\ \bibnamefont
  {{Hartmann}}}, \bibinfo {author} {\bibfnamefont {M.~S.}\ \bibnamefont
  {{Briggs}}}, \bibinfo {author} {\bibfnamefont {J.}~\bibnamefont {{Hakkila}}},
  \ and\ \bibinfo {author} {\bibfnamefont {C.~A.}\ \bibnamefont {{Meegan}}},\
  }\href {\doibase 10.1086/177549} {\bibfield  {journal} {\bibinfo  {journal}
  {\apj}\ }\textbf {\bibinfo {volume} {466}},\ \bibinfo {pages} {757} (\bibinfo
  {year} {1996})}\BibitemShut {NoStop}%
\bibitem [{\citenamefont {{Gorosabel}}\ \emph {et~al.}(1998)\citenamefont
  {{Gorosabel}}, \citenamefont {{Castro-Tirado}}, \citenamefont {{Brandt}},\
  and\ \citenamefont {{Lund}}}]{1998A&A...336...57G}%
  \BibitemOpen
  \bibfield  {author} {\bibinfo {author} {\bibfnamefont {J.}~\bibnamefont
  {{Gorosabel}}}, \bibinfo {author} {\bibfnamefont {A.~J.}\ \bibnamefont
  {{Castro-Tirado}}}, \bibinfo {author} {\bibfnamefont {S.}~\bibnamefont
  {{Brandt}}}, \ and\ \bibinfo {author} {\bibfnamefont {N.}~\bibnamefont
  {{Lund}}},\ }\href@noop {} {\bibfield  {journal} {\bibinfo  {journal} {\aap}\
  }\textbf {\bibinfo {volume} {336}},\ \bibinfo {pages} {57} (\bibinfo {year}
  {1998})}\BibitemShut {NoStop}%
\bibitem [{\citenamefont {{Nemiroff}}\ \emph {et~al.}(1994)\citenamefont
  {{Nemiroff}}, \citenamefont {{Wickramasinghe}}, \citenamefont {{Norris}},\
  and\ \citenamefont {{et al.}}}]{1994ApJ...432..478N}%
  \BibitemOpen
  \bibfield  {author} {\bibinfo {author} {\bibfnamefont {R.~J.}\ \bibnamefont
  {{Nemiroff}}}, \bibinfo {author} {\bibfnamefont {W.~A.~D.~T.}\ \bibnamefont
  {{Wickramasinghe}}}, \bibinfo {author} {\bibfnamefont {J.~P.}\ \bibnamefont
  {{Norris}}}, \ and\ \bibinfo {author} {\bibnamefont {{et al.}}},\ }\href
  {\doibase 10.1086/174587} {\bibfield  {journal} {\bibinfo  {journal} {\apj}\
  }\textbf {\bibinfo {volume} {432}},\ \bibinfo {pages} {478} (\bibinfo {year}
  {1994})}\BibitemShut {NoStop}%
\bibitem [{\citenamefont {{Nemiroff}}\ \emph {et~al.}(2001)\citenamefont
  {{Nemiroff}}, \citenamefont {{Marani}}, \citenamefont {{Norris}},\ and\
  \citenamefont {{Bonnell}}}]{2001PhRvL..86..580N}%
  \BibitemOpen
  \bibfield  {author} {\bibinfo {author} {\bibfnamefont {R.~J.}\ \bibnamefont
  {{Nemiroff}}}, \bibinfo {author} {\bibfnamefont {G.~F.}\ \bibnamefont
  {{Marani}}}, \bibinfo {author} {\bibfnamefont {J.~P.}\ \bibnamefont
  {{Norris}}}, \ and\ \bibinfo {author} {\bibfnamefont {J.~T.}\ \bibnamefont
  {{Bonnell}}},\ }\href {\doibase 10.1103/PhysRevLett.86.580} {\bibfield
  {journal} {\bibinfo  {journal} {\prl}\ }\textbf {\bibinfo {volume} {86}},\
  \bibinfo {pages} {580} (\bibinfo {year} {2001})}\BibitemShut {NoStop}%
\bibitem [{\citenamefont {{Veres}}\ \emph {et~al.}(2009)\citenamefont
  {{Veres}}, \citenamefont {{Bagoly}}, \citenamefont {{Horvath}}, \citenamefont
  {{Meszaros}},\ and\ \citenamefont {{Balazs}}}]{2009arXiv0912.3928V}%
  \BibitemOpen
  \bibfield  {author} {\bibinfo {author} {\bibfnamefont {P.}~\bibnamefont
  {{Veres}}}, \bibinfo {author} {\bibfnamefont {Z.}~\bibnamefont {{Bagoly}}},
  \bibinfo {author} {\bibfnamefont {I.}~\bibnamefont {{Horvath}}}, \bibinfo
  {author} {\bibfnamefont {A.}~\bibnamefont {{Meszaros}}}, \ and\ \bibinfo
  {author} {\bibfnamefont {L.~G.}\ \bibnamefont {{Balazs}}},\ }\href@noop {}
  {\bibfield  {journal} {\bibinfo  {journal} {arXiv e-prints}\ } (\bibinfo
  {year} {2009})},\ \Eprint {http://arxiv.org/abs/0912.3928} {0912.3928}
  \BibitemShut {NoStop}%
\bibitem [{\citenamefont {{Hurley}}\ \emph {et~al.}(2019)\citenamefont
  {{Hurley}}, \citenamefont {{Tsvetkova}}, \citenamefont {{Svinkin}},\ and\
  \citenamefont {{et al.}}}]{2019ApJ...871..121H}%
  \BibitemOpen
  \bibfield  {author} {\bibinfo {author} {\bibfnamefont {K.}~\bibnamefont
  {{Hurley}}}, \bibinfo {author} {\bibfnamefont {A.~E.}\ \bibnamefont
  {{Tsvetkova}}}, \bibinfo {author} {\bibfnamefont {D.~S.}\ \bibnamefont
  {{Svinkin}}}, \ and\ \bibinfo {author} {\bibnamefont {{et al.}}},\ }\href
  {\doibase 10.3847/1538-4357/aaf645} {\bibfield  {journal} {\bibinfo
  {journal} {\apj}\ }\textbf {\bibinfo {volume} {871}},\ \bibinfo {eid} {121}
  (\bibinfo {year} {2019})}\BibitemShut {NoStop}%
\bibitem [{\citenamefont {{Ahlgren}}\ and\ \citenamefont
  {{Larsson}}(2020)}]{2020ApJ...897..178A}%
  \BibitemOpen
  \bibfield  {author} {\bibinfo {author} {\bibfnamefont {B.}~\bibnamefont
  {{Ahlgren}}}\ and\ \bibinfo {author} {\bibfnamefont {J.}~\bibnamefont
  {{Larsson}}},\ }\href {\doibase 10.3847/1538-4357/ab9b8a} {\bibfield
  {journal} {\bibinfo  {journal} {\apj}\ }\textbf {\bibinfo {volume} {897}},\
  \bibinfo {eid} {178} (\bibinfo {year} {2020})}\BibitemShut {NoStop}%
\bibitem [{\citenamefont {{Hirose}}\ \emph {et~al.}(2006)\citenamefont
  {{Hirose}}, \citenamefont {{Umemura}}, \citenamefont {{Yonehara}},\ and\
  \citenamefont {{Sato}}}]{2006ApJ...650..252H}%
  \BibitemOpen
  \bibfield  {author} {\bibinfo {author} {\bibfnamefont {Y.}~\bibnamefont
  {{Hirose}}}, \bibinfo {author} {\bibfnamefont {M.}~\bibnamefont {{Umemura}}},
  \bibinfo {author} {\bibfnamefont {A.}~\bibnamefont {{Yonehara}}}, \ and\
  \bibinfo {author} {\bibfnamefont {J.}~\bibnamefont {{Sato}}},\ }\href
  {\doibase 10.1086/506169} {\bibfield  {journal} {\bibinfo  {journal} {\apj}\
  }\textbf {\bibinfo {volume} {650}},\ \bibinfo {pages} {252} (\bibinfo {year}
  {2006})}\BibitemShut {NoStop}%
\bibitem [{\citenamefont {{Ji}}\ \emph {et~al.}(2018)\citenamefont {{Ji}},
  \citenamefont {{Kovetz}},\ and\ \citenamefont
  {{Kamionkowski}}}]{2018PhRvD..98l3523J}%
  \BibitemOpen
  \bibfield  {author} {\bibinfo {author} {\bibfnamefont {L.}~\bibnamefont
  {{Ji}}}, \bibinfo {author} {\bibfnamefont {E.~D.}\ \bibnamefont {{Kovetz}}},
  \ and\ \bibinfo {author} {\bibfnamefont {M.}~\bibnamefont {{Kamionkowski}}},\
  }\href {\doibase 10.1103/PhysRevD.98.123523} {\bibfield  {journal} {\bibinfo
  {journal} {\prd}\ }\textbf {\bibinfo {volume} {98}},\ \bibinfo {eid} {123523}
  (\bibinfo {year} {2018})}\BibitemShut {NoStop}%
\bibitem [{\citenamefont {{Loeb}}\ and\ \citenamefont
  {{Perna}}(1998)}]{1998ApJ...495..597L}%
  \BibitemOpen
  \bibfield  {author} {\bibinfo {author} {\bibfnamefont {A.}~\bibnamefont
  {{Loeb}}}\ and\ \bibinfo {author} {\bibfnamefont {R.}~\bibnamefont
  {{Perna}}},\ }\href {\doibase 10.1086/305337} {\bibfield  {journal} {\bibinfo
   {journal} {\apj}\ }\textbf {\bibinfo {volume} {495}},\ \bibinfo {pages}
  {597} (\bibinfo {year} {1998})}\BibitemShut {NoStop}%
\bibitem [{\citenamefont {{Koopmans}}\ and\ \citenamefont
  {{Wambsganss}}(2001)}]{2001MNRAS.325.1317K}%
  \BibitemOpen
  \bibfield  {author} {\bibinfo {author} {\bibfnamefont {L.~V.~E.}\
  \bibnamefont {{Koopmans}}}\ and\ \bibinfo {author} {\bibfnamefont
  {J.}~\bibnamefont {{Wambsganss}}},\ }\href {\doibase
  10.1046/j.1365-8711.2001.04279.x} {\bibfield  {journal} {\bibinfo  {journal}
  {\mnras}\ }\textbf {\bibinfo {volume} {325}},\ \bibinfo {pages} {1317}
  (\bibinfo {year} {2001})}\BibitemShut {NoStop}%
\bibitem [{\citenamefont {{Mao}}\ and\ \citenamefont
  {{Loeb}}(2001)}]{2001ApJ...547L..97M}%
  \BibitemOpen
  \bibfield  {author} {\bibinfo {author} {\bibfnamefont {S.}~\bibnamefont
  {{Mao}}}\ and\ \bibinfo {author} {\bibfnamefont {A.}~\bibnamefont {{Loeb}}},\
  }\href {\doibase 10.1086/318912} {\bibfield  {journal} {\bibinfo  {journal}
  {\apjl}\ }\textbf {\bibinfo {volume} {547}},\ \bibinfo {pages} {L97}
  (\bibinfo {year} {2001})}\BibitemShut {NoStop}%
\bibitem [{\citenamefont {{Granot}}\ and\ \citenamefont
  {{Loeb}}(2001)}]{2001ApJ...551L..63G}%
  \BibitemOpen
  \bibfield  {author} {\bibinfo {author} {\bibfnamefont {J.}~\bibnamefont
  {{Granot}}}\ and\ \bibinfo {author} {\bibfnamefont {A.}~\bibnamefont
  {{Loeb}}},\ }\href {\doibase 10.1086/319843} {\bibfield  {journal} {\bibinfo
  {journal} {\apjl}\ }\textbf {\bibinfo {volume} {551}},\ \bibinfo {pages}
  {L63} (\bibinfo {year} {2001})}\BibitemShut {NoStop}%
\bibitem [{\citenamefont {{Gaudi}}\ and\ \citenamefont
  {{Loeb}}(2001)}]{2001ApJ...558..643G}%
  \BibitemOpen
  \bibfield  {author} {\bibinfo {author} {\bibfnamefont {B.~S.}\ \bibnamefont
  {{Gaudi}}}\ and\ \bibinfo {author} {\bibfnamefont {A.}~\bibnamefont
  {{Loeb}}},\ }\href {\doibase 10.1086/322289} {\bibfield  {journal} {\bibinfo
  {journal} {\apj}\ }\textbf {\bibinfo {volume} {558}},\ \bibinfo {pages} {643}
  (\bibinfo {year} {2001})}\BibitemShut {NoStop}%
\bibitem [{\citenamefont {{Ioka}}\ and\ \citenamefont
  {{Nakamura}}(2001)}]{2001ApJ...561..703I}%
  \BibitemOpen
  \bibfield  {author} {\bibinfo {author} {\bibfnamefont {K.}~\bibnamefont
  {{Ioka}}}\ and\ \bibinfo {author} {\bibfnamefont {T.}~\bibnamefont
  {{Nakamura}}},\ }\href {\doibase 10.1086/322427} {\bibfield  {journal}
  {\bibinfo  {journal} {\apj}\ }\textbf {\bibinfo {volume} {561}},\ \bibinfo
  {pages} {703} (\bibinfo {year} {2001})}\BibitemShut {NoStop}%
\bibitem [{\citenamefont {{Baltz}}\ and\ \citenamefont
  {{Hui}}(2005)}]{2005ApJ...618..403B}%
  \BibitemOpen
  \bibfield  {author} {\bibinfo {author} {\bibfnamefont {E.~A.}\ \bibnamefont
  {{Baltz}}}\ and\ \bibinfo {author} {\bibfnamefont {L.}~\bibnamefont
  {{Hui}}},\ }\href {\doibase 10.1086/425954} {\bibfield  {journal} {\bibinfo
  {journal} {\apj}\ }\textbf {\bibinfo {volume} {618}},\ \bibinfo {pages} {403}
  (\bibinfo {year} {2005})}\BibitemShut {NoStop}%
\bibitem [{\citenamefont {{Chen}}\ \emph {et~al.}(2022)\citenamefont {{Chen}},
  \citenamefont {{Wen}}, \citenamefont {{Gao}},\ and\ \citenamefont {{et
  al.}}}]{2022ApJ...924...49C}%
  \BibitemOpen
  \bibfield  {author} {\bibinfo {author} {\bibfnamefont {S.}~\bibnamefont
  {{Chen}}}, \bibinfo {author} {\bibfnamefont {X.}~\bibnamefont {{Wen}}},
  \bibinfo {author} {\bibfnamefont {H.}~\bibnamefont {{Gao}}}, \ and\ \bibinfo
  {author} {\bibnamefont {{et al.}}},\ }\href {\doibase
  10.3847/1538-4357/ac31ad} {\bibfield  {journal} {\bibinfo  {journal} {\apj}\
  }\textbf {\bibinfo {volume} {924}},\ \bibinfo {eid} {49} (\bibinfo {year}
  {2022})}\BibitemShut {NoStop}%
\bibitem [{\citenamefont {{Paynter}}\ \emph {et~al.}(2021)\citenamefont
  {{Paynter}}, \citenamefont {{Webster}},\ and\ \citenamefont
  {{Thrane}}}]{2021NatAs...5..560P}%
  \BibitemOpen
  \bibfield  {author} {\bibinfo {author} {\bibfnamefont {J.}~\bibnamefont
  {{Paynter}}}, \bibinfo {author} {\bibfnamefont {R.}~\bibnamefont
  {{Webster}}}, \ and\ \bibinfo {author} {\bibfnamefont {E.}~\bibnamefont
  {{Thrane}}},\ }\href {\doibase 10.1038/s41550-021-01307-1} {\bibfield
  {journal} {\bibinfo  {journal} {Nature Astronomy}\ }\textbf {\bibinfo
  {volume} {5}},\ \bibinfo {pages} {560} (\bibinfo {year} {2021})}\BibitemShut
  {NoStop}%
\bibitem [{\citenamefont {{Veres}}\ \emph {et~al.}(2020)\citenamefont
  {{Veres}}, \citenamefont {{Meegan}},\ and\ \citenamefont {{Fermi GBM
  Team}}}]{2020GCN.28135....1V}%
  \BibitemOpen
  \bibfield  {author} {\bibinfo {author} {\bibfnamefont {P.}~\bibnamefont
  {{Veres}}}, \bibinfo {author} {\bibfnamefont {C.}~\bibnamefont {{Meegan}}}, \
  and\ \bibinfo {author} {\bibnamefont {{Fermi GBM Team}}},\ }\href@noop {}
  {\bibfield  {journal} {\bibinfo  {journal} {GRB Coordinates Network}\
  }\textbf {\bibinfo {volume} {28135}},\ \bibinfo {pages} {1} (\bibinfo {year}
  {2020})}\BibitemShut {NoStop}%
\bibitem [{\citenamefont {{Barthelmy}}\ \emph {et~al.}(2020)\citenamefont
  {{Barthelmy}}, \citenamefont {{Cummings}}, \citenamefont {{Krimm}},\ and\
  \citenamefont {{et al.}}}]{2020GCN.28136....1B}%
  \BibitemOpen
  \bibfield  {author} {\bibinfo {author} {\bibfnamefont {S.~D.}\ \bibnamefont
  {{Barthelmy}}}, \bibinfo {author} {\bibfnamefont {J.~R.}\ \bibnamefont
  {{Cummings}}}, \bibinfo {author} {\bibfnamefont {H.~A.}\ \bibnamefont
  {{Krimm}}}, \ and\ \bibinfo {author} {\bibnamefont {{et al.}}},\ }\href@noop
  {} {\bibfield  {journal} {\bibinfo  {journal} {GRB Coordinates Network}\
  }\textbf {\bibinfo {volume} {28136}},\ \bibinfo {pages} {1} (\bibinfo {year}
  {2020})}\BibitemShut {NoStop}%
\bibitem [{\citenamefont {{Xue}}\ \emph {et~al.}(2020)\citenamefont {{Xue}},
  \citenamefont {{Xiao}}, \citenamefont {{Yi}},\ and\ \citenamefont
  {{Insight-HXMT Team}}}]{2020GCN.28145....1X}%
  \BibitemOpen
  \bibfield  {author} {\bibinfo {author} {\bibfnamefont {W.~C.}\ \bibnamefont
  {{Xue}}}, \bibinfo {author} {\bibfnamefont {S.}~\bibnamefont {{Xiao}}},
  \bibinfo {author} {\bibfnamefont {Q.~B.}\ \bibnamefont {{Yi}}}, \ and\
  \bibinfo {author} {\bibnamefont {{Insight-HXMT Team}}},\ }\href@noop {}
  {\bibfield  {journal} {\bibinfo  {journal} {GRB Coordinates Network}\
  }\textbf {\bibinfo {volume} {28145}},\ \bibinfo {pages} {1} (\bibinfo {year}
  {2020})}\BibitemShut {NoStop}%
\bibitem [{\citenamefont {{Wang}}\ \emph
  {et~al.}(2021{\natexlab{a}})\citenamefont {{Wang}}, \citenamefont {{Jiang}},
  \citenamefont {{Li}},\ and\ \citenamefont {{et al.}}}]{2021ApJ...918L..34W}%
  \BibitemOpen
  \bibfield  {author} {\bibinfo {author} {\bibfnamefont {Y.}~\bibnamefont
  {{Wang}}}, \bibinfo {author} {\bibfnamefont {L.-Y.}\ \bibnamefont {{Jiang}}},
  \bibinfo {author} {\bibfnamefont {C.-K.}\ \bibnamefont {{Li}}}, \ and\
  \bibinfo {author} {\bibnamefont {{et al.}}},\ }\href {\doibase
  10.3847/2041-8213/ac1ff9} {\bibfield  {journal} {\bibinfo  {journal} {\apjl}\
  }\textbf {\bibinfo {volume} {918}},\ \bibinfo {eid} {L34} (\bibinfo {year}
  {2021}{\natexlab{a}})}\BibitemShut {NoStop}%
\bibitem [{\citenamefont {{Yang}}\ \emph {et~al.}(2021)\citenamefont {{Yang}},
  \citenamefont {{L{\"u}}}, \citenamefont {{Yuan}},\ and\ \citenamefont {{et
  al.}}}]{2021ApJ...921L..29Y}%
  \BibitemOpen
  \bibfield  {author} {\bibinfo {author} {\bibfnamefont {X.}~\bibnamefont
  {{Yang}}}, \bibinfo {author} {\bibfnamefont {H.-J.}\ \bibnamefont
  {{L{\"u}}}}, \bibinfo {author} {\bibfnamefont {H.-Y.}\ \bibnamefont
  {{Yuan}}}, \ and\ \bibinfo {author} {\bibnamefont {{et al.}}},\ }\href
  {\doibase 10.3847/2041-8213/ac2f39} {\bibfield  {journal} {\bibinfo
  {journal} {\apjl}\ }\textbf {\bibinfo {volume} {921}},\ \bibinfo {eid} {L29}
  (\bibinfo {year} {2021})}\BibitemShut {NoStop}%
\bibitem [{\citenamefont {{Lin}}\ \emph {et~al.}(2022)\citenamefont {{Lin}},
  \citenamefont {{Li}}, \citenamefont {{Gao}},\ and\ \citenamefont {{et
  al.}}}]{2022ApJ...931....4L}%
  \BibitemOpen
  \bibfield  {author} {\bibinfo {author} {\bibfnamefont {S.-J.}\ \bibnamefont
  {{Lin}}}, \bibinfo {author} {\bibfnamefont {A.}~\bibnamefont {{Li}}},
  \bibinfo {author} {\bibfnamefont {H.}~\bibnamefont {{Gao}}}, \ and\ \bibinfo
  {author} {\bibnamefont {{et al.}}},\ }\href {\doibase
  10.3847/1538-4357/ac6505} {\bibfield  {journal} {\bibinfo  {journal} {\apj}\
  }\textbf {\bibinfo {volume} {931}},\ \bibinfo {eid} {4} (\bibinfo {year}
  {2022})}\BibitemShut {NoStop}%
\bibitem [{\citenamefont {{Veres}}\ \emph {et~al.}(2021)\citenamefont
  {{Veres}}, \citenamefont {{Bhat}}, \citenamefont {{Fraija}},\ and\
  \citenamefont {{Lesage}}}]{2021ApJ...921L..30V}%
  \BibitemOpen
  \bibfield  {author} {\bibinfo {author} {\bibfnamefont {P.}~\bibnamefont
  {{Veres}}}, \bibinfo {author} {\bibfnamefont {N.}~\bibnamefont {{Bhat}}},
  \bibinfo {author} {\bibfnamefont {N.}~\bibnamefont {{Fraija}}}, \ and\
  \bibinfo {author} {\bibfnamefont {S.}~\bibnamefont {{Lesage}}},\ }\href
  {\doibase 10.3847/2041-8213/ac2ee6} {\bibfield  {journal} {\bibinfo
  {journal} {\apjl}\ }\textbf {\bibinfo {volume} {921}},\ \bibinfo {eid} {L30}
  (\bibinfo {year} {2021})}\BibitemShut {NoStop}%
\bibitem [{\citenamefont {{Cordes}}\ and\ \citenamefont
  {{Chatterjee}}(2019)}]{2019ARA&A..57..417C}%
  \BibitemOpen
  \bibfield  {author} {\bibinfo {author} {\bibfnamefont {J.~M.}\ \bibnamefont
  {{Cordes}}}\ and\ \bibinfo {author} {\bibfnamefont {S.}~\bibnamefont
  {{Chatterjee}}},\ }\href {\doibase 10.1146/annurev-astro-091918-104501}
  {\bibfield  {journal} {\bibinfo  {journal} {\araa}\ }\textbf {\bibinfo
  {volume} {57}},\ \bibinfo {pages} {417} (\bibinfo {year} {2019})}\BibitemShut
  {NoStop}%
\bibitem [{\citenamefont {{Petroff}}\ \emph {et~al.}(2019)\citenamefont
  {{Petroff}}, \citenamefont {{Hessels}},\ and\ \citenamefont
  {{Lorimer}}}]{2019A&ARv..27....4P}%
  \BibitemOpen
  \bibfield  {author} {\bibinfo {author} {\bibfnamefont {E.}~\bibnamefont
  {{Petroff}}}, \bibinfo {author} {\bibfnamefont {J.~W.~T.}\ \bibnamefont
  {{Hessels}}}, \ and\ \bibinfo {author} {\bibfnamefont {D.~R.}\ \bibnamefont
  {{Lorimer}}},\ }\href {\doibase 10.1007/s00159-019-0116-6} {\bibfield
  {journal} {\bibinfo  {journal} {\aapr}\ }\textbf {\bibinfo {volume} {27}},\
  \bibinfo {eid} {4} (\bibinfo {year} {2019})}\BibitemShut {NoStop}%
\bibitem [{\citenamefont {{Lorimer}}\ \emph {et~al.}(2007)\citenamefont
  {{Lorimer}}, \citenamefont {{Bailes}}, \citenamefont {{McLaughlin}},
  \citenamefont {{Narkevic}},\ and\ \citenamefont
  {{Crawford}}}]{2007Sci...318..777L}%
  \BibitemOpen
  \bibfield  {author} {\bibinfo {author} {\bibfnamefont {D.~R.}\ \bibnamefont
  {{Lorimer}}}, \bibinfo {author} {\bibfnamefont {M.}~\bibnamefont {{Bailes}}},
  \bibinfo {author} {\bibfnamefont {M.~A.}\ \bibnamefont {{McLaughlin}}},
  \bibinfo {author} {\bibfnamefont {D.~J.}\ \bibnamefont {{Narkevic}}}, \ and\
  \bibinfo {author} {\bibfnamefont {F.}~\bibnamefont {{Crawford}}},\ }\href
  {\doibase 10.1126/science.1147532} {\bibfield  {journal} {\bibinfo  {journal}
  {Science}\ }\textbf {\bibinfo {volume} {318}},\ \bibinfo {pages} {777}
  (\bibinfo {year} {2007})}\BibitemShut {NoStop}%
\bibitem [{\citenamefont {{Thornton}}\ \emph {et~al.}(2013)\citenamefont
  {{Thornton}}, \citenamefont {{Stappers}}, \citenamefont {{Bailes}},\ and\
  \citenamefont {{et al.}}}]{2013Sci...341...53T}%
  \BibitemOpen
  \bibfield  {author} {\bibinfo {author} {\bibfnamefont {D.}~\bibnamefont
  {{Thornton}}}, \bibinfo {author} {\bibfnamefont {B.}~\bibnamefont
  {{Stappers}}}, \bibinfo {author} {\bibfnamefont {M.}~\bibnamefont
  {{Bailes}}}, \ and\ \bibinfo {author} {\bibnamefont {{et al.}}},\ }\href
  {\doibase 10.1126/science.1236789} {\bibfield  {journal} {\bibinfo  {journal}
  {Science}\ }\textbf {\bibinfo {volume} {341}},\ \bibinfo {pages} {53}
  (\bibinfo {year} {2013})}\BibitemShut {NoStop}%
\bibitem [{\citenamefont {{Champion}}\ \emph {et~al.}(2016)\citenamefont
  {{Champion}}, \citenamefont {{Petroff}}, \citenamefont {{Kramer}},\ and\
  \citenamefont {{et al.}}}]{2016MNRAS.460L..30C}%
  \BibitemOpen
  \bibfield  {author} {\bibinfo {author} {\bibfnamefont {D.~J.}\ \bibnamefont
  {{Champion}}}, \bibinfo {author} {\bibfnamefont {E.}~\bibnamefont
  {{Petroff}}}, \bibinfo {author} {\bibfnamefont {M.}~\bibnamefont {{Kramer}}},
  \ and\ \bibinfo {author} {\bibnamefont {{et al.}}},\ }\href {\doibase
  10.1093/mnrasl/slw069} {\bibfield  {journal} {\bibinfo  {journal} {\mnras}\
  }\textbf {\bibinfo {volume} {460}},\ \bibinfo {pages} {L30} (\bibinfo {year}
  {2016})}\BibitemShut {NoStop}%
\bibitem [{\citenamefont {{Caleb}}\ \emph {et~al.}(2016)\citenamefont
  {{Caleb}}, \citenamefont {{Flynn}}, \citenamefont {{Bailes}},\ and\
  \citenamefont {{et al.}}}]{2016MNRAS.458..718C}%
  \BibitemOpen
  \bibfield  {author} {\bibinfo {author} {\bibfnamefont {M.}~\bibnamefont
  {{Caleb}}}, \bibinfo {author} {\bibfnamefont {C.}~\bibnamefont {{Flynn}}},
  \bibinfo {author} {\bibfnamefont {M.}~\bibnamefont {{Bailes}}}, \ and\
  \bibinfo {author} {\bibnamefont {{et al.}}},\ }\href {\doibase
  10.1093/mnras/stw109} {\bibfield  {journal} {\bibinfo  {journal} {\mnras}\
  }\textbf {\bibinfo {volume} {458}},\ \bibinfo {pages} {718} (\bibinfo {year}
  {2016})}\BibitemShut {NoStop}%
\bibitem [{\citenamefont {{Shannon}}\ \emph {et~al.}(2018)\citenamefont
  {{Shannon}}, \citenamefont {{Macquart}}, \citenamefont {{Bannister}},\ and\
  \citenamefont {{et al.}}}]{2018Natur.562..386S}%
  \BibitemOpen
  \bibfield  {author} {\bibinfo {author} {\bibfnamefont {R.~M.}\ \bibnamefont
  {{Shannon}}}, \bibinfo {author} {\bibfnamefont {J.~P.}\ \bibnamefont
  {{Macquart}}}, \bibinfo {author} {\bibfnamefont {K.~W.}\ \bibnamefont
  {{Bannister}}}, \ and\ \bibinfo {author} {\bibnamefont {{et al.}}},\ }\href
  {\doibase 10.1038/s41586-018-0588-y} {\bibfield  {journal} {\bibinfo
  {journal} {\nat}\ }\textbf {\bibinfo {volume} {562}},\ \bibinfo {pages} {386}
  (\bibinfo {year} {2018})}\BibitemShut {NoStop}%
\bibitem [{\citenamefont {{Tendulkar}}\ \emph {et~al.}(2017)\citenamefont
  {{Tendulkar}}, \citenamefont {{Bassa}}, \citenamefont {{Cordes}},\ and\
  \citenamefont {{et al.}}}]{2017ApJ...834L...7T}%
  \BibitemOpen
  \bibfield  {author} {\bibinfo {author} {\bibfnamefont {S.~P.}\ \bibnamefont
  {{Tendulkar}}}, \bibinfo {author} {\bibfnamefont {C.~G.}\ \bibnamefont
  {{Bassa}}}, \bibinfo {author} {\bibfnamefont {J.~M.}\ \bibnamefont
  {{Cordes}}}, \ and\ \bibinfo {author} {\bibnamefont {{et al.}}},\ }\href
  {\doibase 10.3847/2041-8213/834/2/L7} {\bibfield  {journal} {\bibinfo
  {journal} {\apjl}\ }\textbf {\bibinfo {volume} {834}},\ \bibinfo {eid} {L7}
  (\bibinfo {year} {2017})}\BibitemShut {NoStop}%
\bibitem [{\citenamefont {{Spitler}}\ \emph {et~al.}(2016)\citenamefont
  {{Spitler}}, \citenamefont {{Scholz}}, \citenamefont {{Hessels}},\ and\
  \citenamefont {{et al.}}}]{2016Natur.531..202S}%
  \BibitemOpen
  \bibfield  {author} {\bibinfo {author} {\bibfnamefont {L.~G.}\ \bibnamefont
  {{Spitler}}}, \bibinfo {author} {\bibfnamefont {P.}~\bibnamefont {{Scholz}}},
  \bibinfo {author} {\bibfnamefont {J.~W.~T.}\ \bibnamefont {{Hessels}}}, \
  and\ \bibinfo {author} {\bibnamefont {{et al.}}},\ }\href {\doibase
  10.1038/nature17168} {\bibfield  {journal} {\bibinfo  {journal} {\nat}\
  }\textbf {\bibinfo {volume} {531}},\ \bibinfo {pages} {202} (\bibinfo {year}
  {2016})}\BibitemShut {NoStop}%
\bibitem [{\citenamefont {{Zhang}}(2018)}]{2018ApJ...867L..21Z}%
  \BibitemOpen
  \bibfield  {author} {\bibinfo {author} {\bibfnamefont {B.}~\bibnamefont
  {{Zhang}}},\ }\href {\doibase 10.3847/2041-8213/aae8e3} {\bibfield  {journal}
  {\bibinfo  {journal} {\apjl}\ }\textbf {\bibinfo {volume} {867}},\ \bibinfo
  {eid} {L21} (\bibinfo {year} {2018})}\BibitemShut {NoStop}%
\bibitem [{\citenamefont {{Mu{\~n}oz}}\ \emph {et~al.}(2016)\citenamefont
  {{Mu{\~n}oz}}, \citenamefont {{Kovetz}}, \citenamefont {{Dai}},\ and\
  \citenamefont {{Kamionkowski}}}]{2016PhRvL.117i1301M}%
  \BibitemOpen
  \bibfield  {author} {\bibinfo {author} {\bibfnamefont {J.~B.}\ \bibnamefont
  {{Mu{\~n}oz}}}, \bibinfo {author} {\bibfnamefont {E.~D.}\ \bibnamefont
  {{Kovetz}}}, \bibinfo {author} {\bibfnamefont {L.}~\bibnamefont {{Dai}}}, \
  and\ \bibinfo {author} {\bibfnamefont {M.}~\bibnamefont {{Kamionkowski}}},\
  }\href {\doibase 10.1103/PhysRevLett.117.091301} {\bibfield  {journal}
  {\bibinfo  {journal} {\prl}\ }\textbf {\bibinfo {volume} {117}},\ \bibinfo
  {eid} {091301} (\bibinfo {year} {2016})}\BibitemShut {NoStop}%
\bibitem [{\citenamefont {{Liao}}\ \emph
  {et~al.}(2020{\natexlab{a}})\citenamefont {{Liao}}, \citenamefont {{Zhang}},
  \citenamefont {{Li}},\ and\ \citenamefont {{Gao}}}]{2020ApJ...896L..11L}%
  \BibitemOpen
  \bibfield  {author} {\bibinfo {author} {\bibfnamefont {K.}~\bibnamefont
  {{Liao}}}, \bibinfo {author} {\bibfnamefont {S.~B.}\ \bibnamefont {{Zhang}}},
  \bibinfo {author} {\bibfnamefont {Z.}~\bibnamefont {{Li}}}, \ and\ \bibinfo
  {author} {\bibfnamefont {H.}~\bibnamefont {{Gao}}},\ }\href {\doibase
  10.3847/2041-8213/ab963e} {\bibfield  {journal} {\bibinfo  {journal} {\apjl}\
  }\textbf {\bibinfo {volume} {896}},\ \bibinfo {eid} {L11} (\bibinfo {year}
  {2020}{\natexlab{a}})}\BibitemShut {NoStop}%
\bibitem [{\citenamefont {{Zhou}}\ \emph
  {et~al.}(2022{\natexlab{a}})\citenamefont {{Zhou}}, \citenamefont {{Li}},
  \citenamefont {{Huang}}, \citenamefont {{Gao}},\ and\ \citenamefont
  {{Huang}}}]{2022MNRAS.511.1141Z}%
  \BibitemOpen
  \bibfield  {author} {\bibinfo {author} {\bibfnamefont {H.}~\bibnamefont
  {{Zhou}}}, \bibinfo {author} {\bibfnamefont {Z.}~\bibnamefont {{Li}}},
  \bibinfo {author} {\bibfnamefont {Z.}~\bibnamefont {{Huang}}}, \bibinfo
  {author} {\bibfnamefont {H.}~\bibnamefont {{Gao}}}, \ and\ \bibinfo {author}
  {\bibfnamefont {L.}~\bibnamefont {{Huang}}},\ }\href {\doibase
  10.1093/mnras/stac139} {\bibfield  {journal} {\bibinfo  {journal} {\mnras}\
  }\textbf {\bibinfo {volume} {511}},\ \bibinfo {pages} {1141} (\bibinfo {year}
  {2022}{\natexlab{a}})}\BibitemShut {NoStop}%
\bibitem [{\citenamefont {{Zhou}}\ \emph
  {et~al.}(2022{\natexlab{b}})\citenamefont {{Zhou}}, \citenamefont {{Li}},
  \citenamefont {{Liao}},\ and\ \citenamefont {{et
  al.}}}]{2022ApJ...928..124Z}%
  \BibitemOpen
  \bibfield  {author} {\bibinfo {author} {\bibfnamefont {H.}~\bibnamefont
  {{Zhou}}}, \bibinfo {author} {\bibfnamefont {Z.}~\bibnamefont {{Li}}},
  \bibinfo {author} {\bibfnamefont {K.}~\bibnamefont {{Liao}}}, \ and\ \bibinfo
  {author} {\bibnamefont {{et al.}}},\ }\href {\doibase
  10.3847/1538-4357/ac510d} {\bibfield  {journal} {\bibinfo  {journal} {\apj}\
  }\textbf {\bibinfo {volume} {928}},\ \bibinfo {eid} {124} (\bibinfo {year}
  {2022}{\natexlab{b}})}\BibitemShut {NoStop}%
\bibitem [{\citenamefont {{Sammons}}\ \emph {et~al.}(2020)\citenamefont
  {{Sammons}}, \citenamefont {{Macquart}}, \citenamefont {{Ekers}},\ and\
  \citenamefont {{et al.}}}]{2020ApJ...900..122S}%
  \BibitemOpen
  \bibfield  {author} {\bibinfo {author} {\bibfnamefont {M.~W.}\ \bibnamefont
  {{Sammons}}}, \bibinfo {author} {\bibfnamefont {J.-P.}\ \bibnamefont
  {{Macquart}}}, \bibinfo {author} {\bibfnamefont {R.~D.}\ \bibnamefont
  {{Ekers}}}, \ and\ \bibinfo {author} {\bibnamefont {{et al.}}},\ }\href
  {\doibase 10.3847/1538-4357/aba7bb} {\bibfield  {journal} {\bibinfo
  {journal} {\apj}\ }\textbf {\bibinfo {volume} {900}},\ \bibinfo {eid} {122}
  (\bibinfo {year} {2020})}\BibitemShut {NoStop}%
\bibitem [{\citenamefont {{Leung}}\ \emph {et~al.}(2022)\citenamefont
  {{Leung}}, \citenamefont {{Kader}}, \citenamefont {{Masui}},\ and\
  \citenamefont {{et al.}}}]{2022arXiv220406001L}%
  \BibitemOpen
  \bibfield  {author} {\bibinfo {author} {\bibfnamefont {C.}~\bibnamefont
  {{Leung}}}, \bibinfo {author} {\bibfnamefont {Z.}~\bibnamefont {{Kader}}},
  \bibinfo {author} {\bibfnamefont {K.~W.}\ \bibnamefont {{Masui}}}, \ and\
  \bibinfo {author} {\bibnamefont {{et al.}}},\ }\href@noop {} {\bibfield
  {journal} {\bibinfo  {journal} {arXiv e-prints}\ } (\bibinfo {year}
  {2022})},\ \Eprint {http://arxiv.org/abs/2204.06001} {2204.06001}
  \BibitemShut {NoStop}%
\bibitem [{\citenamefont {{Er}}\ and\ \citenamefont
  {{Mao}}(2022)}]{2022MNRAS.516.2218E}%
  \BibitemOpen
  \bibfield  {author} {\bibinfo {author} {\bibfnamefont {X.}~\bibnamefont
  {{Er}}}\ and\ \bibinfo {author} {\bibfnamefont {S.}~\bibnamefont {{Mao}}},\
  }\href {\doibase 10.1093/mnras/stac2323} {\bibfield  {journal} {\bibinfo
  {journal} {\mnras}\ }\textbf {\bibinfo {volume} {516}},\ \bibinfo {pages}
  {2218} (\bibinfo {year} {2022})}\BibitemShut {NoStop}%
\bibitem [{\citenamefont {{Er}}\ and\ \citenamefont
  {{Mao}}(2014)}]{2014MNRAS.437.2180E}%
  \BibitemOpen
  \bibfield  {author} {\bibinfo {author} {\bibfnamefont {X.}~\bibnamefont
  {{Er}}}\ and\ \bibinfo {author} {\bibfnamefont {S.}~\bibnamefont {{Mao}}},\
  }\href {\doibase 10.1093/mnras/stt2043} {\bibfield  {journal} {\bibinfo
  {journal} {\mnras}\ }\textbf {\bibinfo {volume} {437}},\ \bibinfo {pages}
  {2180} (\bibinfo {year} {2014})}\BibitemShut {NoStop}%
\bibitem [{\citenamefont {{Connor}}\ and\ \citenamefont
  {{Ravi}}(2022)}]{2022arXiv220614310C}%
  \BibitemOpen
  \bibfield  {author} {\bibinfo {author} {\bibfnamefont {L.}~\bibnamefont
  {{Connor}}}\ and\ \bibinfo {author} {\bibfnamefont {V.}~\bibnamefont
  {{Ravi}}},\ }\href@noop {} {\bibfield  {journal} {\bibinfo  {journal} {arXiv
  e-prints}\ } (\bibinfo {year} {2022})},\ \Eprint
  {http://arxiv.org/abs/2206.14310} {2206.14310} \BibitemShut {NoStop}%
\bibitem [{\citenamefont {{Xiao}}\ \emph {et~al.}(2022)\citenamefont {{Xiao}},
  \citenamefont {{Dai}},\ and\ \citenamefont
  {{McQuinn}}}]{2022arXiv220613534X}%
  \BibitemOpen
  \bibfield  {author} {\bibinfo {author} {\bibfnamefont {H.}~\bibnamefont
  {{Xiao}}}, \bibinfo {author} {\bibfnamefont {L.}~\bibnamefont {{Dai}}}, \
  and\ \bibinfo {author} {\bibfnamefont {M.}~\bibnamefont {{McQuinn}}},\
  }\href@noop {} {\bibfield  {journal} {\bibinfo  {journal} {arXiv e-prints}\ }
  (\bibinfo {year} {2022})},\ \Eprint {http://arxiv.org/abs/2206.13534}
  {2206.13534} \BibitemShut {NoStop}%
\bibitem [{\citenamefont {{Sathyaprakash}}\ and\ \citenamefont
  {{Schutz}}(2009)}]{2009LRR....12....2S}%
  \BibitemOpen
  \bibfield  {author} {\bibinfo {author} {\bibfnamefont {B.~S.}\ \bibnamefont
  {{Sathyaprakash}}}\ and\ \bibinfo {author} {\bibfnamefont {B.~F.}\
  \bibnamefont {{Schutz}}},\ }\href {\doibase 10.12942/lrr-2009-2} {\bibfield
  {journal} {\bibinfo  {journal} {Living Reviews in Relativity}\ }\textbf
  {\bibinfo {volume} {12}},\ \bibinfo {eid} {2} (\bibinfo {year}
  {2009})}\BibitemShut {NoStop}%
\bibitem [{\citenamefont {{Abbott}}\ \emph {et~al.}(2016)\citenamefont
  {{Abbott}}, \citenamefont {{Abbott}}, \citenamefont {{Abbott}},\ and\
  \citenamefont {{et al.}}}]{2016PhRvL.116f1102A}%
  \BibitemOpen
  \bibfield  {author} {\bibinfo {author} {\bibfnamefont {B.~P.}\ \bibnamefont
  {{Abbott}}}, \bibinfo {author} {\bibfnamefont {R.}~\bibnamefont {{Abbott}}},
  \bibinfo {author} {\bibfnamefont {T.~D.~M.}\ \bibnamefont {{Abbott}}}, \ and\
  \bibinfo {author} {\bibnamefont {{et al.}}},\ }\href {\doibase
  10.1103/PhysRevLett.116.061102} {\bibfield  {journal} {\bibinfo  {journal}
  {\prl}\ }\textbf {\bibinfo {volume} {116}},\ \bibinfo {eid} {061102}
  (\bibinfo {year} {2016})}\BibitemShut {NoStop}%
\bibitem [{\citenamefont {{Abbott}}\ \emph
  {et~al.}(2017{\natexlab{a}})\citenamefont {{Abbott}}, \citenamefont
  {{Abbott}}, \citenamefont {{Abbott}},\ and\ \citenamefont {{et
  al.}}}]{2017PhRvL.119p1101A}%
  \BibitemOpen
  \bibfield  {author} {\bibinfo {author} {\bibfnamefont {B.~P.}\ \bibnamefont
  {{Abbott}}}, \bibinfo {author} {\bibfnamefont {R.}~\bibnamefont {{Abbott}}},
  \bibinfo {author} {\bibfnamefont {T.~D.}\ \bibnamefont {{Abbott}}}, \ and\
  \bibinfo {author} {\bibnamefont {{et al.}}},\ }\href {\doibase
  10.1103/PhysRevLett.119.161101} {\bibfield  {journal} {\bibinfo  {journal}
  {\prl}\ }\textbf {\bibinfo {volume} {119}},\ \bibinfo {eid} {161101}
  (\bibinfo {year} {2017}{\natexlab{a}})}\BibitemShut {NoStop}%
\bibitem [{\citenamefont {{The LIGO Scientific Collaboration}}\ \emph
  {et~al.}(2021{\natexlab{b}})\citenamefont {{The LIGO Scientific
  Collaboration}}, \citenamefont {{the Virgo Collaboration}},\ and\
  \citenamefont {{the KAGRA Collaboration}}}]{2021ApJ...915L...5A}%
  \BibitemOpen
  \bibfield  {author} {\bibinfo {author} {\bibnamefont {{The LIGO Scientific
  Collaboration}}}, \bibinfo {author} {\bibnamefont {{the Virgo
  Collaboration}}}, \ and\ \bibinfo {author} {\bibnamefont {{the KAGRA
  Collaboration}}},\ }\href {\doibase 10.3847/2041-8213/ac082e} {\bibfield
  {journal} {\bibinfo  {journal} {\apjl}\ }\textbf {\bibinfo {volume} {915}},\
  \bibinfo {eid} {L5} (\bibinfo {year} {2021}{\natexlab{b}})}\BibitemShut
  {NoStop}%
\bibitem [{\citenamefont {{Abbott}}\ \emph
  {et~al.}(2017{\natexlab{b}})\citenamefont {{Abbott}}, \citenamefont
  {{Abbott}}, \citenamefont {{Abbott}},\ and\ \citenamefont {{et
  al.}}}]{2017CQGra..34d4001A}%
  \BibitemOpen
  \bibfield  {author} {\bibinfo {author} {\bibfnamefont {B.~P.}\ \bibnamefont
  {{Abbott}}}, \bibinfo {author} {\bibfnamefont {R.}~\bibnamefont {{Abbott}}},
  \bibinfo {author} {\bibfnamefont {T.~D.}\ \bibnamefont {{Abbott}}}, \ and\
  \bibinfo {author} {\bibnamefont {{et al.}}},\ }\href {\doibase
  10.1088/1361-6382/aa51f4} {\bibfield  {journal} {\bibinfo  {journal}
  {Classical and Quantum Gravity}\ }\textbf {\bibinfo {volume} {34}},\ \bibinfo
  {eid} {044001} (\bibinfo {year} {2017}{\natexlab{b}})}\BibitemShut {NoStop}%
\bibitem [{\citenamefont {{Wang}}\ \emph {et~al.}(1996)\citenamefont {{Wang}},
  \citenamefont {{Stebbins}},\ and\ \citenamefont
  {{Turner}}}]{1996PhRvL..77.2875W}%
  \BibitemOpen
  \bibfield  {author} {\bibinfo {author} {\bibfnamefont {Y.}~\bibnamefont
  {{Wang}}}, \bibinfo {author} {\bibfnamefont {A.}~\bibnamefont {{Stebbins}}},
  \ and\ \bibinfo {author} {\bibfnamefont {E.~L.}\ \bibnamefont {{Turner}}},\
  }\href {\doibase 10.1103/PhysRevLett.77.2875} {\bibfield  {journal} {\bibinfo
   {journal} {\prl}\ }\textbf {\bibinfo {volume} {77}},\ \bibinfo {pages}
  {2875} (\bibinfo {year} {1996})}\BibitemShut {NoStop}%
\bibitem [{\citenamefont {{Nakamura}}(1998)}]{1998PhRvL..80.1138N}%
  \BibitemOpen
  \bibfield  {author} {\bibinfo {author} {\bibfnamefont {T.~T.}\ \bibnamefont
  {{Nakamura}}},\ }\href {\doibase 10.1103/PhysRevLett.80.1138} {\bibfield
  {journal} {\bibinfo  {journal} {\prl}\ }\textbf {\bibinfo {volume} {80}},\
  \bibinfo {pages} {1138} (\bibinfo {year} {1998})}\BibitemShut {NoStop}%
\bibitem [{\citenamefont {{Takahashi}}\ and\ \citenamefont
  {{Nakamura}}(2003)}]{2003ApJ...595.1039T}%
  \BibitemOpen
  \bibfield  {author} {\bibinfo {author} {\bibfnamefont {R.}~\bibnamefont
  {{Takahashi}}}\ and\ \bibinfo {author} {\bibfnamefont {T.}~\bibnamefont
  {{Nakamura}}},\ }\href {\doibase 10.1086/377430} {\bibfield  {journal}
  {\bibinfo  {journal} {\apj}\ }\textbf {\bibinfo {volume} {595}},\ \bibinfo
  {pages} {1039} (\bibinfo {year} {2003})}\BibitemShut {NoStop}%
\bibitem [{\citenamefont {{Hou}}\ \emph {et~al.}(2020)\citenamefont {{Hou}},
  \citenamefont {{Fan}}, \citenamefont {{Liao}},\ and\ \citenamefont
  {{Zhu}}}]{2020PhRvD.101f4011H}%
  \BibitemOpen
  \bibfield  {author} {\bibinfo {author} {\bibfnamefont {S.}~\bibnamefont
  {{Hou}}}, \bibinfo {author} {\bibfnamefont {X.-L.}\ \bibnamefont {{Fan}}},
  \bibinfo {author} {\bibfnamefont {K.}~\bibnamefont {{Liao}}}, \ and\ \bibinfo
  {author} {\bibfnamefont {Z.-H.}\ \bibnamefont {{Zhu}}},\ }\href {\doibase
  10.1103/PhysRevD.101.064011} {\bibfield  {journal} {\bibinfo  {journal}
  {\prd}\ }\textbf {\bibinfo {volume} {101}},\ \bibinfo {eid} {064011}
  (\bibinfo {year} {2020})}\BibitemShut {NoStop}%
\bibitem [{\citenamefont {{Smith}}\ \emph {et~al.}(2018)\citenamefont
  {{Smith}}, \citenamefont {{Jauzac}}, \citenamefont {{Veitch}},\ and\
  \citenamefont {{et al.}}}]{2018MNRAS.475.3823S}%
  \BibitemOpen
  \bibfield  {author} {\bibinfo {author} {\bibfnamefont {G.~P.}\ \bibnamefont
  {{Smith}}}, \bibinfo {author} {\bibfnamefont {M.}~\bibnamefont {{Jauzac}}},
  \bibinfo {author} {\bibfnamefont {J.}~\bibnamefont {{Veitch}}}, \ and\
  \bibinfo {author} {\bibnamefont {{et al.}}},\ }\href {\doibase
  10.1093/mnras/sty031} {\bibfield  {journal} {\bibinfo  {journal} {\mnras}\
  }\textbf {\bibinfo {volume} {475}},\ \bibinfo {pages} {3823} (\bibinfo {year}
  {2018})}\BibitemShut {NoStop}%
\bibitem [{\citenamefont {{Bianconi}}\ \emph {et~al.}(2022)\citenamefont
  {{Bianconi}}, \citenamefont {{Smith}}, \citenamefont {{Nicholl}},\ and\
  \citenamefont {{et al.}}}]{2022arXiv220412978B}%
  \BibitemOpen
  \bibfield  {author} {\bibinfo {author} {\bibfnamefont {M.}~\bibnamefont
  {{Bianconi}}}, \bibinfo {author} {\bibfnamefont {G.~P.}\ \bibnamefont
  {{Smith}}}, \bibinfo {author} {\bibfnamefont {M.}~\bibnamefont {{Nicholl}}},
  \ and\ \bibinfo {author} {\bibnamefont {{et al.}}},\ }\href@noop {}
  {\bibfield  {journal} {\bibinfo  {journal} {arXiv e-prints}\ } (\bibinfo
  {year} {2022})},\ \Eprint {http://arxiv.org/abs/2204.12978} {2204.12978}
  \BibitemShut {NoStop}%
\bibitem [{\citenamefont {{Wong}}\ \emph {et~al.}(2021)\citenamefont {{Wong}},
  \citenamefont {{Chan}}, \citenamefont {{Wong}}, \citenamefont {{Lo}},\ and\
  \citenamefont {{Li}}}]{2021arXiv211205932W}%
  \BibitemOpen
  \bibfield  {author} {\bibinfo {author} {\bibfnamefont {H.~W.~Y.}\
  \bibnamefont {{Wong}}}, \bibinfo {author} {\bibfnamefont {L.~W.~L.}\
  \bibnamefont {{Chan}}}, \bibinfo {author} {\bibfnamefont {I.~C.~F.}\
  \bibnamefont {{Wong}}}, \bibinfo {author} {\bibfnamefont {R.~K.~L.}\
  \bibnamefont {{Lo}}}, \ and\ \bibinfo {author} {\bibfnamefont {T.~G.~F.}\
  \bibnamefont {{Li}}},\ }\href@noop {} {\bibfield  {journal} {\bibinfo
  {journal} {arXiv e-prints}\ } (\bibinfo {year} {2021})},\ \Eprint
  {http://arxiv.org/abs/2112.05932} {2112.05932} \BibitemShut {NoStop}%
\bibitem [{\citenamefont {{Haris}}\ \emph {et~al.}(2018)\citenamefont
  {{Haris}}, \citenamefont {{Mehta}}, \citenamefont {{Kumar}}, \citenamefont
  {{Venumadhav}},\ and\ \citenamefont {{Ajith}}}]{2018arXiv180707062H}%
  \BibitemOpen
  \bibfield  {author} {\bibinfo {author} {\bibfnamefont {K.}~\bibnamefont
  {{Haris}}}, \bibinfo {author} {\bibfnamefont {A.~K.}\ \bibnamefont
  {{Mehta}}}, \bibinfo {author} {\bibfnamefont {S.}~\bibnamefont {{Kumar}}},
  \bibinfo {author} {\bibfnamefont {T.}~\bibnamefont {{Venumadhav}}}, \ and\
  \bibinfo {author} {\bibfnamefont {P.}~\bibnamefont {{Ajith}}},\ }\href@noop
  {} {\bibfield  {journal} {\bibinfo  {journal} {arXiv e-prints}\ } (\bibinfo
  {year} {2018})},\ \Eprint {http://arxiv.org/abs/1807.07062} {1807.07062}
  \BibitemShut {NoStop}%
\bibitem [{\citenamefont {{Janquart}}\ \emph {et~al.}(2021)\citenamefont
  {{Janquart}}, \citenamefont {{Seo}}, \citenamefont {{Hannuksela}},
  \citenamefont {{Li}},\ and\ \citenamefont {{Van Den
  Broeck}}}]{2021ApJ...923L...1J}%
  \BibitemOpen
  \bibfield  {author} {\bibinfo {author} {\bibfnamefont {J.}~\bibnamefont
  {{Janquart}}}, \bibinfo {author} {\bibfnamefont {E.}~\bibnamefont {{Seo}}},
  \bibinfo {author} {\bibfnamefont {O.~A.}\ \bibnamefont {{Hannuksela}}},
  \bibinfo {author} {\bibfnamefont {T.~G.~F.}\ \bibnamefont {{Li}}}, \ and\
  \bibinfo {author} {\bibfnamefont {C.}~\bibnamefont {{Van Den Broeck}}},\
  }\href {\doibase 10.3847/2041-8213/ac3bcf} {\bibfield  {journal} {\bibinfo
  {journal} {\apjl}\ }\textbf {\bibinfo {volume} {923}},\ \bibinfo {eid} {L1}
  (\bibinfo {year} {2021})}\BibitemShut {NoStop}%
\bibitem [{\citenamefont {{Dai}}\ and\ \citenamefont
  {{Venumadhav}}(2017)}]{2017arXiv170204724D}%
  \BibitemOpen
  \bibfield  {author} {\bibinfo {author} {\bibfnamefont {L.}~\bibnamefont
  {{Dai}}}\ and\ \bibinfo {author} {\bibfnamefont {T.}~\bibnamefont
  {{Venumadhav}}},\ }\href@noop {} {\bibfield  {journal} {\bibinfo  {journal}
  {arXiv e-prints}\ } (\bibinfo {year} {2017})},\ \Eprint
  {http://arxiv.org/abs/1702.04724} {1702.04724} \BibitemShut {NoStop}%
\bibitem [{\citenamefont {{Wang}}\ \emph
  {et~al.}(2021{\natexlab{b}})\citenamefont {{Wang}}, \citenamefont {{Lo}},
  \citenamefont {{Li}},\ and\ \citenamefont {{Chen}}}]{2021PhRvD.103j4055W}%
  \BibitemOpen
  \bibfield  {author} {\bibinfo {author} {\bibfnamefont {Y.}~\bibnamefont
  {{Wang}}}, \bibinfo {author} {\bibfnamefont {R.~K.~L.}\ \bibnamefont {{Lo}}},
  \bibinfo {author} {\bibfnamefont {A.~K.~Y.}\ \bibnamefont {{Li}}}, \ and\
  \bibinfo {author} {\bibfnamefont {Y.}~\bibnamefont {{Chen}}},\ }\href
  {\doibase 10.1103/PhysRevD.103.104055} {\bibfield  {journal} {\bibinfo
  {journal} {\prd}\ }\textbf {\bibinfo {volume} {103}},\ \bibinfo {eid}
  {104055} (\bibinfo {year} {2021}{\natexlab{b}})}\BibitemShut {NoStop}%
\bibitem [{\citenamefont {{Cao}}\ \emph {et~al.}(2014)\citenamefont {{Cao}},
  \citenamefont {{Li}},\ and\ \citenamefont {{Wang}}}]{2014PhRvD..90f2003C}%
  \BibitemOpen
  \bibfield  {author} {\bibinfo {author} {\bibfnamefont {Z.}~\bibnamefont
  {{Cao}}}, \bibinfo {author} {\bibfnamefont {L.-F.}\ \bibnamefont {{Li}}}, \
  and\ \bibinfo {author} {\bibfnamefont {Y.}~\bibnamefont {{Wang}}},\ }\href
  {\doibase 10.1103/PhysRevD.90.062003} {\bibfield  {journal} {\bibinfo
  {journal} {\prd}\ }\textbf {\bibinfo {volume} {90}},\ \bibinfo {eid} {062003}
  (\bibinfo {year} {2014})}\BibitemShut {NoStop}%
\bibitem [{\citenamefont {{Guo}}\ and\ \citenamefont
  {{Lu}}(2020)}]{2020PhRvD.102l4076G}%
  \BibitemOpen
  \bibfield  {author} {\bibinfo {author} {\bibfnamefont {X.}~\bibnamefont
  {{Guo}}}\ and\ \bibinfo {author} {\bibfnamefont {Y.}~\bibnamefont {{Lu}}},\
  }\href {\doibase 10.1103/PhysRevD.102.124076} {\bibfield  {journal} {\bibinfo
   {journal} {\prd}\ }\textbf {\bibinfo {volume} {102}},\ \bibinfo {eid}
  {124076} (\bibinfo {year} {2020})}\BibitemShut {NoStop}%
\bibitem [{\citenamefont {{Li}}\ \emph {et~al.}(2019)\citenamefont {{Li}},
  \citenamefont {{Lo}}, \citenamefont {{Sachdev}},\ and\ \citenamefont {{et
  al.}}}]{2019arXiv190406020L}%
  \BibitemOpen
  \bibfield  {author} {\bibinfo {author} {\bibfnamefont {A.~K.~Y.}\
  \bibnamefont {{Li}}}, \bibinfo {author} {\bibfnamefont {R.~K.~L.}\
  \bibnamefont {{Lo}}}, \bibinfo {author} {\bibfnamefont {S.}~\bibnamefont
  {{Sachdev}}}, \ and\ \bibinfo {author} {\bibnamefont {{et al.}}},\
  }\href@noop {} {\bibfield  {journal} {\bibinfo  {journal} {arXiv e-prints}\ }
  (\bibinfo {year} {2019})},\ \Eprint {http://arxiv.org/abs/1904.06020}
  {1904.06020} \BibitemShut {NoStop}%
\bibitem [{\citenamefont {{Smith}}\ \emph {et~al.}(2022)\citenamefont
  {{Smith}}, \citenamefont {{Robertson}}, \citenamefont {{Mahler}},\ and\
  \citenamefont {{et al.}}}]{2022arXiv220412977S}%
  \BibitemOpen
  \bibfield  {author} {\bibinfo {author} {\bibfnamefont {G.~P.}\ \bibnamefont
  {{Smith}}}, \bibinfo {author} {\bibfnamefont {A.}~\bibnamefont
  {{Robertson}}}, \bibinfo {author} {\bibfnamefont {G.}~\bibnamefont
  {{Mahler}}}, \ and\ \bibinfo {author} {\bibnamefont {{et al.}}},\ }\href@noop
  {} {\bibfield  {journal} {\bibinfo  {journal} {arXiv e-prints}\ } (\bibinfo
  {year} {2022})},\ \Eprint {http://arxiv.org/abs/2204.12977} {2204.12977}
  \BibitemShut {NoStop}%
\bibitem [{\citenamefont {{Wempe}}\ \emph {et~al.}(2022)\citenamefont
  {{Wempe}}, \citenamefont {{Koopmans}}, \citenamefont {{Wierda}},
  \citenamefont {{Akseli Hannuksela}},\ and\ \citenamefont {{van den
  Broeck}}}]{2022arXiv220408732W}%
  \BibitemOpen
  \bibfield  {author} {\bibinfo {author} {\bibfnamefont {E.}~\bibnamefont
  {{Wempe}}}, \bibinfo {author} {\bibfnamefont {L.~V.~E.}\ \bibnamefont
  {{Koopmans}}}, \bibinfo {author} {\bibfnamefont {A.~R. A.~C.}\ \bibnamefont
  {{Wierda}}}, \bibinfo {author} {\bibfnamefont {O.}~\bibnamefont {{Akseli
  Hannuksela}}}, \ and\ \bibinfo {author} {\bibfnamefont {C.}~\bibnamefont
  {{van den Broeck}}},\ }\href@noop {} {\bibfield  {journal} {\bibinfo
  {journal} {arXiv e-prints}\ } (\bibinfo {year} {2022})},\ \Eprint
  {http://arxiv.org/abs/2204.08732} {2204.08732} \BibitemShut {NoStop}%
\bibitem [{\citenamefont {{Janquart}}\ \emph {et~al.}(2022)\citenamefont
  {{Janquart}}, \citenamefont {{More}},\ and\ \citenamefont {{Van Den
  Broeck}}}]{2022arXiv220511499J}%
  \BibitemOpen
  \bibfield  {author} {\bibinfo {author} {\bibfnamefont {J.}~\bibnamefont
  {{Janquart}}}, \bibinfo {author} {\bibfnamefont {A.}~\bibnamefont {{More}}},
  \ and\ \bibinfo {author} {\bibfnamefont {C.}~\bibnamefont {{Van Den
  Broeck}}},\ }\href@noop {} {\bibfield  {journal} {\bibinfo  {journal} {arXiv
  e-prints}\ } (\bibinfo {year} {2022})},\ \Eprint
  {http://arxiv.org/abs/2205.11499} {2205.11499} \BibitemShut {NoStop}%
\bibitem [{\citenamefont {{Ng}}\ \emph {et~al.}(2018)\citenamefont {{Ng}},
  \citenamefont {{Wong}}, \citenamefont {{Broadhurst}},\ and\ \citenamefont
  {{Li}}}]{2018PhRvD..97b3012N}%
  \BibitemOpen
  \bibfield  {author} {\bibinfo {author} {\bibfnamefont {K.~K.~Y.}\
  \bibnamefont {{Ng}}}, \bibinfo {author} {\bibfnamefont {K.~W.~K.}\
  \bibnamefont {{Wong}}}, \bibinfo {author} {\bibfnamefont {T.}~\bibnamefont
  {{Broadhurst}}}, \ and\ \bibinfo {author} {\bibfnamefont {T.~G.~F.}\
  \bibnamefont {{Li}}},\ }\href {\doibase 10.1103/PhysRevD.97.023012}
  {\bibfield  {journal} {\bibinfo  {journal} {\prd}\ }\textbf {\bibinfo
  {volume} {97}},\ \bibinfo {eid} {023012} (\bibinfo {year}
  {2018})}\BibitemShut {NoStop}%
\bibitem [{\citenamefont {{Wierda}}\ \emph {et~al.}(2021)\citenamefont
  {{Wierda}}, \citenamefont {{Wempe}}, \citenamefont {{Hannuksela}},
  \citenamefont {{Koopmans}},\ and\ \citenamefont {{Van Den
  Broeck}}}]{2021ApJ...921..154W}%
  \BibitemOpen
  \bibfield  {author} {\bibinfo {author} {\bibfnamefont {A.~R. A.~C.}\
  \bibnamefont {{Wierda}}}, \bibinfo {author} {\bibfnamefont {E.}~\bibnamefont
  {{Wempe}}}, \bibinfo {author} {\bibfnamefont {O.~A.}\ \bibnamefont
  {{Hannuksela}}}, \bibinfo {author} {\bibfnamefont {L.~V.~E.}\ \bibnamefont
  {{Koopmans}}}, \ and\ \bibinfo {author} {\bibfnamefont {C.}~\bibnamefont
  {{Van Den Broeck}}},\ }\href {\doibase 10.3847/1538-4357/ac1bb4} {\bibfield
  {journal} {\bibinfo  {journal} {\apj}\ }\textbf {\bibinfo {volume} {921}},\
  \bibinfo {eid} {154} (\bibinfo {year} {2021})}\BibitemShut {NoStop}%
\bibitem [{\citenamefont {{Pi{\'o}rkowska}}\ \emph {et~al.}(2013)\citenamefont
  {{Pi{\'o}rkowska}}, \citenamefont {{Biesiada}},\ and\ \citenamefont
  {{Zhu}}}]{2013JCAP...10..022P}%
  \BibitemOpen
  \bibfield  {author} {\bibinfo {author} {\bibfnamefont {A.}~\bibnamefont
  {{Pi{\'o}rkowska}}}, \bibinfo {author} {\bibfnamefont {M.}~\bibnamefont
  {{Biesiada}}}, \ and\ \bibinfo {author} {\bibfnamefont {Z.-H.}\ \bibnamefont
  {{Zhu}}},\ }\href {\doibase 10.1088/1475-7516/2013/10/022} {\bibfield
  {journal} {\bibinfo  {journal} {\jcap}\ }\textbf {\bibinfo {volume} {2013}},\
  \bibinfo {eid} {022} (\bibinfo {year} {2013})}\BibitemShut {NoStop}%
\bibitem [{\citenamefont {{Ding}}\ \emph {et~al.}(2015)\citenamefont {{Ding}},
  \citenamefont {{Biesiada}},\ and\ \citenamefont
  {{Zhu}}}]{2015JCAP...12..006D}%
  \BibitemOpen
  \bibfield  {author} {\bibinfo {author} {\bibfnamefont {X.}~\bibnamefont
  {{Ding}}}, \bibinfo {author} {\bibfnamefont {M.}~\bibnamefont {{Biesiada}}},
  \ and\ \bibinfo {author} {\bibfnamefont {Z.-H.}\ \bibnamefont {{Zhu}}},\
  }\href {\doibase 10.1088/1475-7516/2015/12/006} {\bibfield  {journal}
  {\bibinfo  {journal} {\jcap}\ }\textbf {\bibinfo {volume} {2015}},\ \bibinfo
  {eid} {006} (\bibinfo {year} {2015})}\BibitemShut {NoStop}%
\bibitem [{\citenamefont {{Li}}\ \emph
  {et~al.}(2018{\natexlab{a}})\citenamefont {{Li}}, \citenamefont {{Mao}},
  \citenamefont {{Zhao}},\ and\ \citenamefont {{Lu}}}]{2018MNRAS.476.2220L}%
  \BibitemOpen
  \bibfield  {author} {\bibinfo {author} {\bibfnamefont {S.-S.}\ \bibnamefont
  {{Li}}}, \bibinfo {author} {\bibfnamefont {S.}~\bibnamefont {{Mao}}},
  \bibinfo {author} {\bibfnamefont {Y.}~\bibnamefont {{Zhao}}}, \ and\ \bibinfo
  {author} {\bibfnamefont {Y.}~\bibnamefont {{Lu}}},\ }\href {\doibase
  10.1093/mnras/sty411} {\bibfield  {journal} {\bibinfo  {journal} {\mnras}\
  }\textbf {\bibinfo {volume} {476}},\ \bibinfo {pages} {2220} (\bibinfo {year}
  {2018}{\natexlab{a}})}\BibitemShut {NoStop}%
\bibitem [{\citenamefont {{Yang}}\ \emph {et~al.}(2022)\citenamefont {{Yang}},
  \citenamefont {{Wu}}, \citenamefont {{Liao}},\ and\ \citenamefont {{et
  al.}}}]{2022MNRAS.509.3772Y}%
  \BibitemOpen
  \bibfield  {author} {\bibinfo {author} {\bibfnamefont {L.}~\bibnamefont
  {{Yang}}}, \bibinfo {author} {\bibfnamefont {S.}~\bibnamefont {{Wu}}},
  \bibinfo {author} {\bibfnamefont {K.}~\bibnamefont {{Liao}}}, \ and\ \bibinfo
  {author} {\bibnamefont {{et al.}}},\ }\href {\doibase 10.1093/mnras/stab3298}
  {\bibfield  {journal} {\bibinfo  {journal} {\mnras}\ }\textbf {\bibinfo
  {volume} {509}},\ \bibinfo {pages} {3772} (\bibinfo {year}
  {2022})}\BibitemShut {NoStop}%
\bibitem [{\citenamefont {{Sereno}}\ \emph {et~al.}(2010)\citenamefont
  {{Sereno}}, \citenamefont {{Sesana}}, \citenamefont {{Bleuler}},\ and\
  \citenamefont {{et al.}}}]{2010PhRvL.105y1101S}%
  \BibitemOpen
  \bibfield  {author} {\bibinfo {author} {\bibfnamefont {M.}~\bibnamefont
  {{Sereno}}}, \bibinfo {author} {\bibfnamefont {A.}~\bibnamefont {{Sesana}}},
  \bibinfo {author} {\bibfnamefont {A.}~\bibnamefont {{Bleuler}}}, \ and\
  \bibinfo {author} {\bibnamefont {{et al.}}},\ }\href {\doibase
  10.1103/PhysRevLett.105.251101} {\bibfield  {journal} {\bibinfo  {journal}
  {\prl}\ }\textbf {\bibinfo {volume} {105}},\ \bibinfo {eid} {251101}
  (\bibinfo {year} {2010})}\BibitemShut {NoStop}%
\bibitem [{\citenamefont {{{\c{C}}al{\i}{\c{s}}kan}}\ \emph
  {et~al.}(2022)\citenamefont {{{\c{C}}al{\i}{\c{s}}kan}}, \citenamefont
  {{Ji}}, \citenamefont {{Cotesta}}, \citenamefont {{Berti}}, \citenamefont
  {{Kamionkowski}},\ and\ \citenamefont {{Marsat}}}]{2022arXiv220602803C}%
  \BibitemOpen
  \bibfield  {author} {\bibinfo {author} {\bibfnamefont {M.}~\bibnamefont
  {{{\c{C}}al{\i}{\c{s}}kan}}}, \bibinfo {author} {\bibfnamefont
  {L.}~\bibnamefont {{Ji}}}, \bibinfo {author} {\bibfnamefont {R.}~\bibnamefont
  {{Cotesta}}}, \bibinfo {author} {\bibfnamefont {E.}~\bibnamefont {{Berti}}},
  \bibinfo {author} {\bibfnamefont {M.}~\bibnamefont {{Kamionkowski}}}, \ and\
  \bibinfo {author} {\bibfnamefont {S.}~\bibnamefont {{Marsat}}},\ }\href@noop
  {} {\bibfield  {journal} {\bibinfo  {journal} {arXiv e-prints}\ } (\bibinfo
  {year} {2022})},\ \Eprint {http://arxiv.org/abs/2206.02803} {2206.02803}
  \BibitemShut {NoStop}%
\bibitem [{\citenamefont {{Hou}}\ \emph
  {et~al.}(2021{\natexlab{a}})\citenamefont {{Hou}}, \citenamefont {{Li}},
  \citenamefont {{Yu}},\ and\ \citenamefont {{et al.}}}]{2021PhRvD.103d4005H}%
  \BibitemOpen
  \bibfield  {author} {\bibinfo {author} {\bibfnamefont {S.}~\bibnamefont
  {{Hou}}}, \bibinfo {author} {\bibfnamefont {P.}~\bibnamefont {{Li}}},
  \bibinfo {author} {\bibfnamefont {H.}~\bibnamefont {{Yu}}}, \ and\ \bibinfo
  {author} {\bibnamefont {{et al.}}},\ }\href {\doibase
  10.1103/PhysRevD.103.044005} {\bibfield  {journal} {\bibinfo  {journal}
  {\prd}\ }\textbf {\bibinfo {volume} {103}},\ \bibinfo {eid} {044005}
  (\bibinfo {year} {2021}{\natexlab{a}})}\BibitemShut {NoStop}%
\bibitem [{\citenamefont {{Hannuksela}}\ \emph {et~al.}(2019)\citenamefont
  {{Hannuksela}}, \citenamefont {{Haris}}, \citenamefont {{Ng}},\ and\
  \citenamefont {{et al.}}}]{2019ApJ...874L...2H}%
  \BibitemOpen
  \bibfield  {author} {\bibinfo {author} {\bibfnamefont {O.~A.}\ \bibnamefont
  {{Hannuksela}}}, \bibinfo {author} {\bibfnamefont {K.}~\bibnamefont
  {{Haris}}}, \bibinfo {author} {\bibfnamefont {K.~K.~Y.}\ \bibnamefont
  {{Ng}}}, \ and\ \bibinfo {author} {\bibnamefont {{et al.}}},\ }\href
  {\doibase 10.3847/2041-8213/ab0c0f} {\bibfield  {journal} {\bibinfo
  {journal} {\apjl}\ }\textbf {\bibinfo {volume} {874}},\ \bibinfo {eid} {L2}
  (\bibinfo {year} {2019})}\BibitemShut {NoStop}%
\bibitem [{\citenamefont {{LIGO Scientific Collaboration}}\ and\ \citenamefont
  {{Virgo Collaboration}}(2021)}]{2021ApJ...923...14A}%
  \BibitemOpen
  \bibfield  {author} {\bibinfo {author} {\bibnamefont {{LIGO Scientific
  Collaboration}}}\ and\ \bibinfo {author} {\bibnamefont {{Virgo
  Collaboration}}},\ }\href {\doibase 10.3847/1538-4357/ac23db} {\bibfield
  {journal} {\bibinfo  {journal} {\apj}\ }\textbf {\bibinfo {volume} {923}},\
  \bibinfo {eid} {14} (\bibinfo {year} {2021})}\BibitemShut {NoStop}%
\bibitem [{\citenamefont {{Diego}}\ \emph {et~al.}(2021)\citenamefont
  {{Diego}}, \citenamefont {{Broadhurst}},\ and\ \citenamefont
  {{Smoot}}}]{2021PhRvD.104j3529D}%
  \BibitemOpen
  \bibfield  {author} {\bibinfo {author} {\bibfnamefont {J.~M.}\ \bibnamefont
  {{Diego}}}, \bibinfo {author} {\bibfnamefont {T.}~\bibnamefont
  {{Broadhurst}}}, \ and\ \bibinfo {author} {\bibfnamefont {G.~F.}\
  \bibnamefont {{Smoot}}},\ }\href {\doibase 10.1103/PhysRevD.104.103529}
  {\bibfield  {journal} {\bibinfo  {journal} {\prd}\ }\textbf {\bibinfo
  {volume} {104}},\ \bibinfo {eid} {103529} (\bibinfo {year}
  {2021})}\BibitemShut {NoStop}%
\bibitem [{\citenamefont {{Dai}}\ \emph {et~al.}(2020)\citenamefont {{Dai}},
  \citenamefont {{Zackay}}, \citenamefont {{Venumadhav}}, \citenamefont
  {{Roulet}},\ and\ \citenamefont {{Zaldarriaga}}}]{2020arXiv200712709D}%
  \BibitemOpen
  \bibfield  {author} {\bibinfo {author} {\bibfnamefont {L.}~\bibnamefont
  {{Dai}}}, \bibinfo {author} {\bibfnamefont {B.}~\bibnamefont {{Zackay}}},
  \bibinfo {author} {\bibfnamefont {T.}~\bibnamefont {{Venumadhav}}}, \bibinfo
  {author} {\bibfnamefont {J.}~\bibnamefont {{Roulet}}}, \ and\ \bibinfo
  {author} {\bibfnamefont {M.}~\bibnamefont {{Zaldarriaga}}},\ }\href@noop {}
  {\bibfield  {journal} {\bibinfo  {journal} {arXiv e-prints}\ } (\bibinfo
  {year} {2020})},\ \Eprint {http://arxiv.org/abs/2007.12709} {2007.12709}
  \BibitemShut {NoStop}%
\bibitem [{\citenamefont {{Liu}}\ \emph {et~al.}(2021)\citenamefont {{Liu}},
  \citenamefont {{Maga{\~n}a Hernandez}},\ and\ \citenamefont
  {{Creighton}}}]{2021ApJ...908...97L}%
  \BibitemOpen
  \bibfield  {author} {\bibinfo {author} {\bibfnamefont {X.}~\bibnamefont
  {{Liu}}}, \bibinfo {author} {\bibfnamefont {I.}~\bibnamefont {{Maga{\~n}a
  Hernandez}}}, \ and\ \bibinfo {author} {\bibfnamefont {J.}~\bibnamefont
  {{Creighton}}},\ }\href {\doibase 10.3847/1538-4357/abd7eb} {\bibfield
  {journal} {\bibinfo  {journal} {\apj}\ }\textbf {\bibinfo {volume} {908}},\
  \bibinfo {eid} {97} (\bibinfo {year} {2021})}\BibitemShut {NoStop}%
\bibitem [{\citenamefont {{Kim}}\ \emph {et~al.}(2022)\citenamefont {{Kim}},
  \citenamefont {{Lee}}, \citenamefont {{Hannuksela}},\ and\ \citenamefont
  {{Li}}}]{2022arXiv220608234K}%
  \BibitemOpen
  \bibfield  {author} {\bibinfo {author} {\bibfnamefont {K.}~\bibnamefont
  {{Kim}}}, \bibinfo {author} {\bibfnamefont {J.}~\bibnamefont {{Lee}}},
  \bibinfo {author} {\bibfnamefont {O.~A.}\ \bibnamefont {{Hannuksela}}}, \
  and\ \bibinfo {author} {\bibfnamefont {T.~G.~F.}\ \bibnamefont {{Li}}},\
  }\href@noop {} {\bibfield  {journal} {\bibinfo  {journal} {arXiv e-prints}\ }
  (\bibinfo {year} {2022})},\ \Eprint {http://arxiv.org/abs/2206.08234}
  {2206.08234} \BibitemShut {NoStop}%
\bibitem [{\citenamefont {{Dai}}\ \emph {et~al.}(2018)\citenamefont {{Dai}},
  \citenamefont {{Li}}, \citenamefont {{Zackay}}, \citenamefont {{Mao}},\ and\
  \citenamefont {{Lu}}}]{2018PhRvD..98j4029D}%
  \BibitemOpen
  \bibfield  {author} {\bibinfo {author} {\bibfnamefont {L.}~\bibnamefont
  {{Dai}}}, \bibinfo {author} {\bibfnamefont {S.-S.}\ \bibnamefont {{Li}}},
  \bibinfo {author} {\bibfnamefont {B.}~\bibnamefont {{Zackay}}}, \bibinfo
  {author} {\bibfnamefont {S.}~\bibnamefont {{Mao}}}, \ and\ \bibinfo {author}
  {\bibfnamefont {Y.}~\bibnamefont {{Lu}}},\ }\href {\doibase
  10.1103/PhysRevD.98.104029} {\bibfield  {journal} {\bibinfo  {journal}
  {\prd}\ }\textbf {\bibinfo {volume} {98}},\ \bibinfo {eid} {104029} (\bibinfo
  {year} {2018})}\BibitemShut {NoStop}%
\bibitem [{\citenamefont {{Christian}}\ \emph {et~al.}(2018)\citenamefont
  {{Christian}}, \citenamefont {{Vitale}},\ and\ \citenamefont
  {{Loeb}}}]{2018PhRvD..98j3022C}%
  \BibitemOpen
  \bibfield  {author} {\bibinfo {author} {\bibfnamefont {P.}~\bibnamefont
  {{Christian}}}, \bibinfo {author} {\bibfnamefont {S.}~\bibnamefont
  {{Vitale}}}, \ and\ \bibinfo {author} {\bibfnamefont {A.}~\bibnamefont
  {{Loeb}}},\ }\href {\doibase 10.1103/PhysRevD.98.103022} {\bibfield
  {journal} {\bibinfo  {journal} {\prd}\ }\textbf {\bibinfo {volume} {98}},\
  \bibinfo {eid} {103022} (\bibinfo {year} {2018})}\BibitemShut {NoStop}%
\bibitem [{\citenamefont {{Liao}}\ \emph {et~al.}(2019)\citenamefont {{Liao}},
  \citenamefont {{Biesiada}},\ and\ \citenamefont
  {{Fan}}}]{2019ApJ...875..139L}%
  \BibitemOpen
  \bibfield  {author} {\bibinfo {author} {\bibfnamefont {K.}~\bibnamefont
  {{Liao}}}, \bibinfo {author} {\bibfnamefont {M.}~\bibnamefont {{Biesiada}}},
  \ and\ \bibinfo {author} {\bibfnamefont {X.-L.}\ \bibnamefont {{Fan}}},\
  }\href {\doibase 10.3847/1538-4357/ab1087} {\bibfield  {journal} {\bibinfo
  {journal} {\apj}\ }\textbf {\bibinfo {volume} {875}},\ \bibinfo {eid} {139}
  (\bibinfo {year} {2019})}\BibitemShut {NoStop}%
\bibitem [{\citenamefont {{Paczynski}}(1991)}]{pacz1991}%
  \BibitemOpen
  \bibfield  {author} {\bibinfo {author} {\bibfnamefont {B.}~\bibnamefont
  {{Paczynski}}},\ }\href {\doibase 10.1086/186003} {\bibfield  {journal}
  {\bibinfo  {journal} {\apjl}\ }\textbf {\bibinfo {volume} {371}},\ \bibinfo
  {pages} {L63} (\bibinfo {year} {1991})}\BibitemShut {NoStop}%
\bibitem [{\citenamefont {{Basak}}\ \emph {et~al.}(2022)\citenamefont
  {{Basak}}, \citenamefont {{Sharma}}, \citenamefont {{Kapadia}},\ and\
  \citenamefont {{Ajith}}}]{2022arXiv220500022B}%
  \BibitemOpen
  \bibfield  {author} {\bibinfo {author} {\bibfnamefont {S.}~\bibnamefont
  {{Basak}}}, \bibinfo {author} {\bibfnamefont {A.~K.}\ \bibnamefont
  {{Sharma}}}, \bibinfo {author} {\bibfnamefont {S.~J.}\ \bibnamefont
  {{Kapadia}}}, \ and\ \bibinfo {author} {\bibfnamefont {P.}~\bibnamefont
  {{Ajith}}},\ }\href@noop {} {\bibfield  {journal} {\bibinfo  {journal} {arXiv
  e-prints}\ } (\bibinfo {year} {2022})},\ \Eprint
  {http://arxiv.org/abs/2205.00022} {2205.00022} \BibitemShut {NoStop}%
\bibitem [{\citenamefont {Biesiada}\ and\ \citenamefont
  {Harikumar}(2021)}]{universe7120502}%
  \BibitemOpen
  \bibfield  {author} {\bibinfo {author} {\bibfnamefont {M.}~\bibnamefont
  {Biesiada}}\ and\ \bibinfo {author} {\bibfnamefont {S.}~\bibnamefont
  {Harikumar}},\ }\href {\doibase 10.3390/universe7120502} {\bibfield
  {journal} {\bibinfo  {journal} {Universe}\ }\textbf {\bibinfo {volume} {7}}
  (\bibinfo {year} {2021}),\ 10.3390/universe7120502}\BibitemShut {NoStop}%
\bibitem [{\citenamefont {{Oguri}}\ and\ \citenamefont
  {{Takahashi}}(2022)}]{2022arXiv220400814O}%
  \BibitemOpen
  \bibfield  {author} {\bibinfo {author} {\bibfnamefont {M.}~\bibnamefont
  {{Oguri}}}\ and\ \bibinfo {author} {\bibfnamefont {R.}~\bibnamefont
  {{Takahashi}}},\ }\href@noop {} {\bibfield  {journal} {\bibinfo  {journal}
  {arXiv e-prints}\ } (\bibinfo {year} {2022})},\ \Eprint
  {http://arxiv.org/abs/2204.00814} {2204.00814} \BibitemShut {NoStop}%
\bibitem [{\citenamefont {{Bulashenko}}\ and\ \citenamefont
  {{Ubach}}(2021)}]{2021arXiv211210773B}%
  \BibitemOpen
  \bibfield  {author} {\bibinfo {author} {\bibfnamefont {O.}~\bibnamefont
  {{Bulashenko}}}\ and\ \bibinfo {author} {\bibfnamefont {H.}~\bibnamefont
  {{Ubach}}},\ }\href@noop {} {\bibfield  {journal} {\bibinfo  {journal} {arXiv
  e-prints}\ } (\bibinfo {year} {2021})},\ \Eprint
  {http://arxiv.org/abs/2112.10773} {2112.10773} \BibitemShut {NoStop}%
\bibitem [{\citenamefont {{Itoh}}\ \emph {et~al.}(2009)\citenamefont {{Itoh}},
  \citenamefont {{Futamase}},\ and\ \citenamefont
  {{Hattori}}}]{2009PhRvD..80d4009I}%
  \BibitemOpen
  \bibfield  {author} {\bibinfo {author} {\bibfnamefont {Y.}~\bibnamefont
  {{Itoh}}}, \bibinfo {author} {\bibfnamefont {T.}~\bibnamefont {{Futamase}}},
  \ and\ \bibinfo {author} {\bibfnamefont {M.}~\bibnamefont {{Hattori}}},\
  }\href {\doibase 10.1103/PhysRevD.80.044009} {\bibfield  {journal} {\bibinfo
  {journal} {\prd}\ }\textbf {\bibinfo {volume} {80}},\ \bibinfo {eid} {044009}
  (\bibinfo {year} {2009})}\BibitemShut {NoStop}%
\bibitem [{\citenamefont {{Cremonese}}\ \emph {et~al.}(2021)\citenamefont
  {{Cremonese}}, \citenamefont {{Ezquiaga}},\ and\ \citenamefont
  {{Salzano}}}]{2021PhRvD.104b3503C}%
  \BibitemOpen
  \bibfield  {author} {\bibinfo {author} {\bibfnamefont {P.}~\bibnamefont
  {{Cremonese}}}, \bibinfo {author} {\bibfnamefont {J.~M.}\ \bibnamefont
  {{Ezquiaga}}}, \ and\ \bibinfo {author} {\bibfnamefont {V.}~\bibnamefont
  {{Salzano}}},\ }\href {\doibase 10.1103/PhysRevD.104.023503} {\bibfield
  {journal} {\bibinfo  {journal} {\prd}\ }\textbf {\bibinfo {volume} {104}},\
  \bibinfo {eid} {023503} (\bibinfo {year} {2021})}\BibitemShut {NoStop}%
\bibitem [{\citenamefont {{D'Orazio}}\ and\ \citenamefont
  {{Loeb}}(2020)}]{2020PhRvD.101h3031D}%
  \BibitemOpen
  \bibfield  {author} {\bibinfo {author} {\bibfnamefont {D.~J.}\ \bibnamefont
  {{D'Orazio}}}\ and\ \bibinfo {author} {\bibfnamefont {A.}~\bibnamefont
  {{Loeb}}},\ }\href {\doibase 10.1103/PhysRevD.101.083031} {\bibfield
  {journal} {\bibinfo  {journal} {\prd}\ }\textbf {\bibinfo {volume} {101}},\
  \bibinfo {eid} {083031} (\bibinfo {year} {2020})}\BibitemShut {NoStop}%
\bibitem [{\citenamefont {{Yu}}\ \emph {et~al.}(2021)\citenamefont {{Yu}},
  \citenamefont {{Wang}}, \citenamefont {{Seymour}},\ and\ \citenamefont
  {{Chen}}}]{2021PhRvD.104j3011Y}%
  \BibitemOpen
  \bibfield  {author} {\bibinfo {author} {\bibfnamefont {H.}~\bibnamefont
  {{Yu}}}, \bibinfo {author} {\bibfnamefont {Y.}~\bibnamefont {{Wang}}},
  \bibinfo {author} {\bibfnamefont {B.}~\bibnamefont {{Seymour}}}, \ and\
  \bibinfo {author} {\bibfnamefont {Y.}~\bibnamefont {{Chen}}},\ }\href
  {\doibase 10.1103/PhysRevD.104.103011} {\bibfield  {journal} {\bibinfo
  {journal} {\prd}\ }\textbf {\bibinfo {volume} {104}},\ \bibinfo {eid}
  {103011} (\bibinfo {year} {2021})}\BibitemShut {NoStop}%
\bibitem [{\citenamefont {{Abbott}}\ \emph {et~al.}(2019)\citenamefont
  {{Abbott}}, \citenamefont {{Abbott}}, \citenamefont {{Abbott}},\ and\
  \citenamefont {{et al.}}}]{PhysRevD.100.104036}%
  \BibitemOpen
  \bibfield  {author} {\bibinfo {author} {\bibfnamefont {B.~P.}\ \bibnamefont
  {{Abbott}}}, \bibinfo {author} {\bibfnamefont {R.}~\bibnamefont {{Abbott}}},
  \bibinfo {author} {\bibfnamefont {T.~D.}\ \bibnamefont {{Abbott}}}, \ and\
  \bibinfo {author} {\bibnamefont {{et al.}}} (\bibinfo {collaboration} {The
  LIGO Scientific Collaboration and the Virgo Collaboration}),\ }\href
  {\doibase 10.1103/PhysRevD.100.104036} {\bibfield  {journal} {\bibinfo
  {journal} {Phys. Rev. D}\ }\textbf {\bibinfo {volume} {100}},\ \bibinfo
  {pages} {104036} (\bibinfo {year} {2019})}\BibitemShut {NoStop}%
\bibitem [{\citenamefont {{Collett}}\ and\ \citenamefont
  {{Bacon}}(2017)}]{2017PhRvL.118i1101C}%
  \BibitemOpen
  \bibfield  {author} {\bibinfo {author} {\bibfnamefont {T.~E.}\ \bibnamefont
  {{Collett}}}\ and\ \bibinfo {author} {\bibfnamefont {D.}~\bibnamefont
  {{Bacon}}},\ }\href {\doibase 10.1103/PhysRevLett.118.091101} {\bibfield
  {journal} {\bibinfo  {journal} {\prl}\ }\textbf {\bibinfo {volume} {118}},\
  \bibinfo {eid} {091101} (\bibinfo {year} {2017})}\BibitemShut {NoStop}%
\bibitem [{\citenamefont {{Fan}}\ \emph {et~al.}(2017)\citenamefont {{Fan}},
  \citenamefont {{Liao}}, \citenamefont {{Biesiada}}, \citenamefont
  {{Pi{\'o}rkowska-Kurpas}},\ and\ \citenamefont
  {{Zhu}}}]{2017PhRvL.118i1102F}%
  \BibitemOpen
  \bibfield  {author} {\bibinfo {author} {\bibfnamefont {X.-L.}\ \bibnamefont
  {{Fan}}}, \bibinfo {author} {\bibfnamefont {K.}~\bibnamefont {{Liao}}},
  \bibinfo {author} {\bibfnamefont {M.}~\bibnamefont {{Biesiada}}}, \bibinfo
  {author} {\bibfnamefont {A.}~\bibnamefont {{Pi{\'o}rkowska-Kurpas}}}, \ and\
  \bibinfo {author} {\bibfnamefont {Z.-H.}\ \bibnamefont {{Zhu}}},\ }\href
  {\doibase 10.1103/PhysRevLett.118.091102} {\bibfield  {journal} {\bibinfo
  {journal} {\prl}\ }\textbf {\bibinfo {volume} {118}},\ \bibinfo {eid}
  {091102} (\bibinfo {year} {2017})}\BibitemShut {NoStop}%
\bibitem [{\citenamefont {{Takahashi}}(2017)}]{2017ApJ...835..103T}%
  \BibitemOpen
  \bibfield  {author} {\bibinfo {author} {\bibfnamefont {R.}~\bibnamefont
  {{Takahashi}}},\ }\href {\doibase 10.3847/1538-4357/835/1/103} {\bibfield
  {journal} {\bibinfo  {journal} {\apj}\ }\textbf {\bibinfo {volume} {835}},\
  \bibinfo {eid} {103} (\bibinfo {year} {2017})}\BibitemShut {NoStop}%
\bibitem [{\citenamefont {{Morita}}\ and\ \citenamefont
  {{Soda}}(2019)}]{2019arXiv191107435M}%
  \BibitemOpen
  \bibfield  {author} {\bibinfo {author} {\bibfnamefont {T.}~\bibnamefont
  {{Morita}}}\ and\ \bibinfo {author} {\bibfnamefont {J.}~\bibnamefont
  {{Soda}}},\ }\href@noop {} {\bibfield  {journal} {\bibinfo  {journal} {arXiv
  e-prints}\ } (\bibinfo {year} {2019})},\ \Eprint
  {http://arxiv.org/abs/1911.07435} {1911.07435} \BibitemShut {NoStop}%
\bibitem [{\citenamefont {{Suyama}}(2020)}]{2020ApJ...896...46S}%
  \BibitemOpen
  \bibfield  {author} {\bibinfo {author} {\bibfnamefont {T.}~\bibnamefont
  {{Suyama}}},\ }\href {\doibase 10.3847/1538-4357/ab8d3f} {\bibfield
  {journal} {\bibinfo  {journal} {\apj}\ }\textbf {\bibinfo {volume} {896}},\
  \bibinfo {eid} {46} (\bibinfo {year} {2020})}\BibitemShut {NoStop}%
\bibitem [{\citenamefont {{Ezquiaga}}\ \emph {et~al.}(2020)\citenamefont
  {{Ezquiaga}}, \citenamefont {{Hu}},\ and\ \citenamefont
  {{Lagos}}}]{2020PhRvD.102b3531E}%
  \BibitemOpen
  \bibfield  {author} {\bibinfo {author} {\bibfnamefont {J.~M.}\ \bibnamefont
  {{Ezquiaga}}}, \bibinfo {author} {\bibfnamefont {W.}~\bibnamefont {{Hu}}}, \
  and\ \bibinfo {author} {\bibfnamefont {M.}~\bibnamefont {{Lagos}}},\ }\href
  {\doibase 10.1103/PhysRevD.102.023531} {\bibfield  {journal} {\bibinfo
  {journal} {\prd}\ }\textbf {\bibinfo {volume} {102}},\ \bibinfo {eid}
  {023531} (\bibinfo {year} {2020})}\BibitemShut {NoStop}%
\bibitem [{\citenamefont {{Tasinato}}\ \emph {et~al.}(2021)\citenamefont
  {{Tasinato}}, \citenamefont {{Garoffolo}}, \citenamefont {{Bertacca}},\ and\
  \citenamefont {{Matarrese}}}]{2021JCAP...06..050T}%
  \BibitemOpen
  \bibfield  {author} {\bibinfo {author} {\bibfnamefont {G.}~\bibnamefont
  {{Tasinato}}}, \bibinfo {author} {\bibfnamefont {A.}~\bibnamefont
  {{Garoffolo}}}, \bibinfo {author} {\bibfnamefont {D.}~\bibnamefont
  {{Bertacca}}}, \ and\ \bibinfo {author} {\bibfnamefont {S.}~\bibnamefont
  {{Matarrese}}},\ }\href {\doibase 10.1088/1475-7516/2021/06/050} {\bibfield
  {journal} {\bibinfo  {journal} {\jcap}\ }\textbf {\bibinfo {volume} {2021}},\
  \bibinfo {eid} {050} (\bibinfo {year} {2021})}\BibitemShut {NoStop}%
\bibitem [{\citenamefont {{Chung}}\ and\ \citenamefont
  {{Li}}(2021)}]{2021PhRvD.104l4060C}%
  \BibitemOpen
  \bibfield  {author} {\bibinfo {author} {\bibfnamefont {A.~K.-W.}\
  \bibnamefont {{Chung}}}\ and\ \bibinfo {author} {\bibfnamefont {T.~G.~F.}\
  \bibnamefont {{Li}}},\ }\href {\doibase 10.1103/PhysRevD.104.124060}
  {\bibfield  {journal} {\bibinfo  {journal} {\prd}\ }\textbf {\bibinfo
  {volume} {104}},\ \bibinfo {eid} {124060} (\bibinfo {year}
  {2021})}\BibitemShut {NoStop}%
\bibitem [{\citenamefont {{Finke}}\ \emph {et~al.}(2021)\citenamefont
  {{Finke}}, \citenamefont {{Foffa}}, \citenamefont {{Iacovelli}},
  \citenamefont {{Maggiore}},\ and\ \citenamefont
  {{Mancarella}}}]{2021PhRvD.104h4057F}%
  \BibitemOpen
  \bibfield  {author} {\bibinfo {author} {\bibfnamefont {A.}~\bibnamefont
  {{Finke}}}, \bibinfo {author} {\bibfnamefont {S.}~\bibnamefont {{Foffa}}},
  \bibinfo {author} {\bibfnamefont {F.}~\bibnamefont {{Iacovelli}}}, \bibinfo
  {author} {\bibfnamefont {M.}~\bibnamefont {{Maggiore}}}, \ and\ \bibinfo
  {author} {\bibfnamefont {M.}~\bibnamefont {{Mancarella}}},\ }\href {\doibase
  10.1103/PhysRevD.104.084057} {\bibfield  {journal} {\bibinfo  {journal}
  {\prd}\ }\textbf {\bibinfo {volume} {104}},\ \bibinfo {eid} {084057}
  (\bibinfo {year} {2021})}\BibitemShut {NoStop}%
\bibitem [{\citenamefont {{Albrecht}}\ and\ \citenamefont
  {{Magueijo}}(1999)}]{Albrecht99}%
  \BibitemOpen
  \bibfield  {author} {\bibinfo {author} {\bibfnamefont {A.}~\bibnamefont
  {{Albrecht}}}\ and\ \bibinfo {author} {\bibfnamefont {J.}~\bibnamefont
  {{Magueijo}}},\ }\href {\doibase 10.1103/PhysRevD.59.043516} {\bibfield
  {journal} {\bibinfo  {journal} {\prd}\ }\textbf {\bibinfo {volume} {59}},\
  \bibinfo {eid} {043516} (\bibinfo {year} {1999})}\BibitemShut {NoStop}%
\bibitem [{\citenamefont {{Barrow}}(1999)}]{Barrow99}%
  \BibitemOpen
  \bibfield  {author} {\bibinfo {author} {\bibfnamefont {J.~D.}\ \bibnamefont
  {{Barrow}}},\ }\href {\doibase 10.1103/PhysRevD.59.043515} {\bibfield
  {journal} {\bibinfo  {journal} {\prd}\ }\textbf {\bibinfo {volume} {59}},\
  \bibinfo {eid} {043515} (\bibinfo {year} {1999})}\BibitemShut {NoStop}%
\bibitem [{\citenamefont {Magueijo}(2003)}]{Magueijo03}%
  \BibitemOpen
  \bibfield  {author} {\bibinfo {author} {\bibfnamefont {J.}~\bibnamefont
  {Magueijo}},\ }\href {\doibase 10.1088/0034-4885/66/11/r04} {\bibfield
  {journal} {\bibinfo  {journal} {Reports on Progress in Physics}\ }\textbf
  {\bibinfo {volume} {66}},\ \bibinfo {pages} {2025} (\bibinfo {year}
  {2003})}\BibitemShut {NoStop}%
\bibitem [{\citenamefont {{Cao}}\ \emph {et~al.}(2018)\citenamefont {{Cao}},
  \citenamefont {{Qi}}, \citenamefont {{Biesiada}},\ and\ \citenamefont {{et
  al.}}}]{2018ApJ...867...50C}%
  \BibitemOpen
  \bibfield  {author} {\bibinfo {author} {\bibfnamefont {S.}~\bibnamefont
  {{Cao}}}, \bibinfo {author} {\bibfnamefont {J.}~\bibnamefont {{Qi}}},
  \bibinfo {author} {\bibfnamefont {M.}~\bibnamefont {{Biesiada}}}, \ and\
  \bibinfo {author} {\bibnamefont {{et al.}}},\ }\href {\doibase
  10.3847/1538-4357/aae5f7} {\bibfield  {journal} {\bibinfo  {journal} {\apj}\
  }\textbf {\bibinfo {volume} {867}},\ \bibinfo {eid} {50} (\bibinfo {year}
  {2018})}\BibitemShut {NoStop}%
\bibitem [{\citenamefont {{Hou}}\ \emph {et~al.}(2019)\citenamefont {{Hou}},
  \citenamefont {{Fan}},\ and\ \citenamefont {{Zhu}}}]{2019PhRvD.100f4028H}%
  \BibitemOpen
  \bibfield  {author} {\bibinfo {author} {\bibfnamefont {S.}~\bibnamefont
  {{Hou}}}, \bibinfo {author} {\bibfnamefont {X.-L.}\ \bibnamefont {{Fan}}}, \
  and\ \bibinfo {author} {\bibfnamefont {Z.-H.}\ \bibnamefont {{Zhu}}},\ }\href
  {\doibase 10.1103/PhysRevD.100.064028} {\bibfield  {journal} {\bibinfo
  {journal} {\prd}\ }\textbf {\bibinfo {volume} {100}},\ \bibinfo {eid}
  {064028} (\bibinfo {year} {2019})}\BibitemShut {NoStop}%
\bibitem [{\citenamefont {{Goyal}}\ \emph {et~al.}(2021)\citenamefont
  {{Goyal}}, \citenamefont {{Haris}}, \citenamefont {{Mehta}},\ and\
  \citenamefont {{Ajith}}}]{2021PhRvD.103b4038G}%
  \BibitemOpen
  \bibfield  {author} {\bibinfo {author} {\bibfnamefont {S.}~\bibnamefont
  {{Goyal}}}, \bibinfo {author} {\bibfnamefont {K.}~\bibnamefont {{Haris}}},
  \bibinfo {author} {\bibfnamefont {A.~K.}\ \bibnamefont {{Mehta}}}, \ and\
  \bibinfo {author} {\bibfnamefont {P.}~\bibnamefont {{Ajith}}},\ }\href
  {\doibase 10.1103/PhysRevD.103.024038} {\bibfield  {journal} {\bibinfo
  {journal} {\prd}\ }\textbf {\bibinfo {volume} {103}},\ \bibinfo {eid}
  {024038} (\bibinfo {year} {2021})}\BibitemShut {NoStop}%
\bibitem [{\citenamefont {{Addazi}}\ \emph {et~al.}(2022)\citenamefont
  {{Addazi}}, \citenamefont {{Alvarez-Muniz}}, \citenamefont {{Alves
  Batista}},\ and\ \citenamefont {{et al.}}}]{2022PrPNP.12503948A}%
  \BibitemOpen
  \bibfield  {author} {\bibinfo {author} {\bibfnamefont {A.}~\bibnamefont
  {{Addazi}}}, \bibinfo {author} {\bibfnamefont {J.}~\bibnamefont
  {{Alvarez-Muniz}}}, \bibinfo {author} {\bibfnamefont {R.}~\bibnamefont
  {{Alves Batista}}}, \ and\ \bibinfo {author} {\bibnamefont {{et al.}}},\
  }\href {\doibase 10.1016/j.ppnp.2022.103948} {\bibfield  {journal} {\bibinfo
  {journal} {Progress in Particle and Nuclear Physics}\ }\textbf {\bibinfo
  {volume} {125}},\ \bibinfo {eid} {103948} (\bibinfo {year}
  {2022})}\BibitemShut {NoStop}%
\bibitem [{\citenamefont {{Amelino-Camelia}}(2013)}]{2013LRR....16....5A}%
  \BibitemOpen
  \bibfield  {author} {\bibinfo {author} {\bibfnamefont {G.}~\bibnamefont
  {{Amelino-Camelia}}},\ }\href {\doibase 10.12942/lrr-2013-5} {\bibfield
  {journal} {\bibinfo  {journal} {Living Reviews in Relativity}\ }\textbf
  {\bibinfo {volume} {16}},\ \bibinfo {eid} {5} (\bibinfo {year}
  {2013})}\BibitemShut {NoStop}%
\bibitem [{\citenamefont {{Amelino-Camelia}}\ and\ \citenamefont
  {{Piran}}(2001)}]{2001PhRvD..64c6005A}%
  \BibitemOpen
  \bibfield  {author} {\bibinfo {author} {\bibfnamefont {G.}~\bibnamefont
  {{Amelino-Camelia}}}\ and\ \bibinfo {author} {\bibfnamefont {T.}~\bibnamefont
  {{Piran}}},\ }\href {\doibase 10.1103/PhysRevD.64.036005} {\bibfield
  {journal} {\bibinfo  {journal} {\prd}\ }\textbf {\bibinfo {volume} {64}},\
  \bibinfo {eid} {036005} (\bibinfo {year} {2001})}\BibitemShut {NoStop}%
\bibitem [{\citenamefont {{Amelino-Camelia}}\ \emph {et~al.}(1998)\citenamefont
  {{Amelino-Camelia}}, \citenamefont {{Ellis}}, \citenamefont {{Mavromatos}},
  \citenamefont {{Nanopoulos}},\ and\ \citenamefont
  {{Sarkar}}}]{1998Natur.395Q.525A}%
  \BibitemOpen
  \bibfield  {author} {\bibinfo {author} {\bibfnamefont {G.}~\bibnamefont
  {{Amelino-Camelia}}}, \bibinfo {author} {\bibfnamefont {J.}~\bibnamefont
  {{Ellis}}}, \bibinfo {author} {\bibfnamefont {N.~E.}\ \bibnamefont
  {{Mavromatos}}}, \bibinfo {author} {\bibfnamefont {D.~V.}\ \bibnamefont
  {{Nanopoulos}}}, \ and\ \bibinfo {author} {\bibfnamefont {S.}~\bibnamefont
  {{Sarkar}}},\ }\href {\doibase 10.1038/26793} {\bibfield  {journal} {\bibinfo
   {journal} {\nat}\ }\textbf {\bibinfo {volume} {395}},\ \bibinfo {pages}
  {525} (\bibinfo {year} {1998})}\BibitemShut {NoStop}%
\bibitem [{\citenamefont {{Pan}}\ \emph {et~al.}(2020)\citenamefont {{Pan}},
  \citenamefont {{Qi}}, \citenamefont {{Cao}},\ and\ \citenamefont {{et
  al.}}}]{2020ApJ...890..169P}%
  \BibitemOpen
  \bibfield  {author} {\bibinfo {author} {\bibfnamefont {Y.}~\bibnamefont
  {{Pan}}}, \bibinfo {author} {\bibfnamefont {J.}~\bibnamefont {{Qi}}},
  \bibinfo {author} {\bibfnamefont {S.}~\bibnamefont {{Cao}}}, \ and\ \bibinfo
  {author} {\bibnamefont {{et al.}}},\ }\href {\doibase
  10.3847/1538-4357/ab6ef5} {\bibfield  {journal} {\bibinfo  {journal} {\apj}\
  }\textbf {\bibinfo {volume} {890}},\ \bibinfo {eid} {169} (\bibinfo {year}
  {2020})}\BibitemShut {NoStop}%
\bibitem [{\citenamefont {{Agrawal}}\ \emph {et~al.}(2021)\citenamefont
  {{Agrawal}}, \citenamefont {{Singirikonda}},\ and\ \citenamefont
  {{Desai}}}]{2021JCAP...05..029A}%
  \BibitemOpen
  \bibfield  {author} {\bibinfo {author} {\bibfnamefont {R.}~\bibnamefont
  {{Agrawal}}}, \bibinfo {author} {\bibfnamefont {H.}~\bibnamefont
  {{Singirikonda}}}, \ and\ \bibinfo {author} {\bibfnamefont {S.}~\bibnamefont
  {{Desai}}},\ }\href {\doibase 10.1088/1475-7516/2021/05/029} {\bibfield
  {journal} {\bibinfo  {journal} {\jcap}\ }\textbf {\bibinfo {volume} {2021}},\
  \bibinfo {eid} {029} (\bibinfo {year} {2021})}\BibitemShut {NoStop}%
\bibitem [{\citenamefont {{Biesiada}}\ and\ \citenamefont
  {{Pi{\'o}rkowska}}(2009)}]{2009MNRAS.396..946B}%
  \BibitemOpen
  \bibfield  {author} {\bibinfo {author} {\bibfnamefont {M.}~\bibnamefont
  {{Biesiada}}}\ and\ \bibinfo {author} {\bibfnamefont {A.}~\bibnamefont
  {{Pi{\'o}rkowska}}},\ }\href {\doibase 10.1111/j.1365-2966.2009.14748.x}
  {\bibfield  {journal} {\bibinfo  {journal} {\mnras}\ }\textbf {\bibinfo
  {volume} {396}},\ \bibinfo {pages} {946} (\bibinfo {year}
  {2009})}\BibitemShut {NoStop}%
\bibitem [{\citenamefont {{Pi{\'o}rkowska-Kurpas}}\ and\ \citenamefont
  {{Biesiada}}(2022)}]{2022Univ....8..321P}%
  \BibitemOpen
  \bibfield  {author} {\bibinfo {author} {\bibfnamefont {A.}~\bibnamefont
  {{Pi{\'o}rkowska-Kurpas}}}\ and\ \bibinfo {author} {\bibfnamefont
  {M.}~\bibnamefont {{Biesiada}}},\ }\href {\doibase 10.3390/universe8060321}
  {\bibfield  {journal} {\bibinfo  {journal} {Universe}\ }\textbf {\bibinfo
  {volume} {8}},\ \bibinfo {pages} {321} (\bibinfo {year} {2022})}\BibitemShut
  {NoStop}%
\bibitem [{\citenamefont {{Baker}}\ and\ \citenamefont
  {{Trodden}}(2017)}]{2017PhRvD..95f3512B}%
  \BibitemOpen
  \bibfield  {author} {\bibinfo {author} {\bibfnamefont {T.}~\bibnamefont
  {{Baker}}}\ and\ \bibinfo {author} {\bibfnamefont {M.}~\bibnamefont
  {{Trodden}}},\ }\href {\doibase 10.1103/PhysRevD.95.063512} {\bibfield
  {journal} {\bibinfo  {journal} {\prd}\ }\textbf {\bibinfo {volume} {95}},\
  \bibinfo {eid} {063512} (\bibinfo {year} {2017})}\BibitemShut {NoStop}%
\bibitem [{\citenamefont {{Diemand}}\ \emph {et~al.}(2008)\citenamefont
  {{Diemand}}, \citenamefont {{Kuhlen}}, \citenamefont {{Madau}},\ and\
  \citenamefont {{et al.}}}]{2008Natur.454..735D}%
  \BibitemOpen
  \bibfield  {author} {\bibinfo {author} {\bibfnamefont {J.}~\bibnamefont
  {{Diemand}}}, \bibinfo {author} {\bibfnamefont {M.}~\bibnamefont {{Kuhlen}}},
  \bibinfo {author} {\bibfnamefont {P.}~\bibnamefont {{Madau}}}, \ and\
  \bibinfo {author} {\bibnamefont {{et al.}}},\ }\href {\doibase
  10.1038/nature07153} {\bibfield  {journal} {\bibinfo  {journal} {\nat}\
  }\textbf {\bibinfo {volume} {454}},\ \bibinfo {pages} {735} (\bibinfo {year}
  {2008})}\BibitemShut {NoStop}%
\bibitem [{\citenamefont {{Klypin}}\ \emph {et~al.}(1999)\citenamefont
  {{Klypin}}, \citenamefont {{Kravtsov}}, \citenamefont {{Valenzuela}},\ and\
  \citenamefont {{Prada}}}]{1999ApJ...522...82K}%
  \BibitemOpen
  \bibfield  {author} {\bibinfo {author} {\bibfnamefont {A.}~\bibnamefont
  {{Klypin}}}, \bibinfo {author} {\bibfnamefont {A.~V.}\ \bibnamefont
  {{Kravtsov}}}, \bibinfo {author} {\bibfnamefont {O.}~\bibnamefont
  {{Valenzuela}}}, \ and\ \bibinfo {author} {\bibfnamefont {F.}~\bibnamefont
  {{Prada}}},\ }\href {\doibase 10.1086/307643} {\bibfield  {journal} {\bibinfo
   {journal} {\apj}\ }\textbf {\bibinfo {volume} {522}},\ \bibinfo {pages} {82}
  (\bibinfo {year} {1999})}\BibitemShut {NoStop}%
\bibitem [{\citenamefont {{Drlica-Wagner}}\ \emph {et~al.}(2015)\citenamefont
  {{Drlica-Wagner}}, \citenamefont {{Bechtol}}, \citenamefont {{Rykoff}},\ and\
  \citenamefont {{DES Collaboration}}}]{2015ApJ...813..109D}%
  \BibitemOpen
  \bibfield  {author} {\bibinfo {author} {\bibfnamefont {A.}~\bibnamefont
  {{Drlica-Wagner}}}, \bibinfo {author} {\bibfnamefont {K.}~\bibnamefont
  {{Bechtol}}}, \bibinfo {author} {\bibfnamefont {E.~S.}\ \bibnamefont
  {{Rykoff}}}, \ and\ \bibinfo {author} {\bibnamefont {{DES Collaboration}}},\
  }\href {\doibase 10.1088/0004-637X/813/2/109} {\bibfield  {journal} {\bibinfo
   {journal} {\apj}\ }\textbf {\bibinfo {volume} {813}},\ \bibinfo {eid} {109}
  (\bibinfo {year} {2015})}\BibitemShut {NoStop}%
\bibitem [{\citenamefont {{Cerny}}\ \emph {et~al.}(2022)\citenamefont
  {{Cerny}}, \citenamefont {{Mart{\'\i}nez-V{\'a}zquez}}, \citenamefont
  {{Drlica-Wagner}},\ and\ \citenamefont {{et al.}}}]{2022arXiv220912422C}%
  \BibitemOpen
  \bibfield  {author} {\bibinfo {author} {\bibfnamefont {W.}~\bibnamefont
  {{Cerny}}}, \bibinfo {author} {\bibfnamefont {C.~E.}\ \bibnamefont
  {{Mart{\'\i}nez-V{\'a}zquez}}}, \bibinfo {author} {\bibfnamefont
  {A.}~\bibnamefont {{Drlica-Wagner}}}, \ and\ \bibinfo {author} {\bibnamefont
  {{et al.}}},\ }\href@noop {} {\bibfield  {journal} {\bibinfo  {journal}
  {arXiv e-prints}\ } (\bibinfo {year} {2022})},\ \Eprint
  {http://arxiv.org/abs/2209.12422} {2209.12422} \BibitemShut {NoStop}%
\bibitem [{\citenamefont {{Mao}}\ and\ \citenamefont
  {{Schneider}}(1998)}]{1998MNRAS.295..587M}%
  \BibitemOpen
  \bibfield  {author} {\bibinfo {author} {\bibfnamefont {S.}~\bibnamefont
  {{Mao}}}\ and\ \bibinfo {author} {\bibfnamefont {P.}~\bibnamefont
  {{Schneider}}},\ }\href {\doibase 10.1046/j.1365-8711.1998.01319.x}
  {\bibfield  {journal} {\bibinfo  {journal} {\mnras}\ }\textbf {\bibinfo
  {volume} {295}},\ \bibinfo {pages} {587} (\bibinfo {year}
  {1998})}\BibitemShut {NoStop}%
\bibitem [{\citenamefont {{Congdon}}\ \emph {et~al.}(2008)\citenamefont
  {{Congdon}}, \citenamefont {{Keeton}},\ and\ \citenamefont
  {{Nordgren}}}]{2008MNRAS.389..398C}%
  \BibitemOpen
  \bibfield  {author} {\bibinfo {author} {\bibfnamefont {A.~B.}\ \bibnamefont
  {{Congdon}}}, \bibinfo {author} {\bibfnamefont {C.~R.}\ \bibnamefont
  {{Keeton}}}, \ and\ \bibinfo {author} {\bibfnamefont {C.~E.}\ \bibnamefont
  {{Nordgren}}},\ }\href {\doibase 10.1111/j.1365-2966.2008.13604.x} {\bibfield
   {journal} {\bibinfo  {journal} {\mnras}\ }\textbf {\bibinfo {volume}
  {389}},\ \bibinfo {pages} {398} (\bibinfo {year} {2008})}\BibitemShut
  {NoStop}%
\bibitem [{\citenamefont {{Schechter}}\ and\ \citenamefont
  {{Wambsganss}}(2002)}]{2002ApJ...580..685S}%
  \BibitemOpen
  \bibfield  {author} {\bibinfo {author} {\bibfnamefont {P.~L.}\ \bibnamefont
  {{Schechter}}}\ and\ \bibinfo {author} {\bibfnamefont {J.}~\bibnamefont
  {{Wambsganss}}},\ }\href {\doibase 10.1086/343856} {\bibfield  {journal}
  {\bibinfo  {journal} {\apj}\ }\textbf {\bibinfo {volume} {580}},\ \bibinfo
  {pages} {685} (\bibinfo {year} {2002})}\BibitemShut {NoStop}%
\bibitem [{\citenamefont {{Keeton}}\ and\ \citenamefont
  {{Moustakas}}(2009)}]{2009ApJ...699.1720K}%
  \BibitemOpen
  \bibfield  {author} {\bibinfo {author} {\bibfnamefont {C.~R.}\ \bibnamefont
  {{Keeton}}}\ and\ \bibinfo {author} {\bibfnamefont {L.~A.}\ \bibnamefont
  {{Moustakas}}},\ }\href {\doibase 10.1088/0004-637X/699/2/1720} {\bibfield
  {journal} {\bibinfo  {journal} {\apj}\ }\textbf {\bibinfo {volume} {699}},\
  \bibinfo {pages} {1720} (\bibinfo {year} {2009})}\BibitemShut {NoStop}%
\bibitem [{\citenamefont {{Liao}}\ \emph {et~al.}(2015)\citenamefont {{Liao}},
  \citenamefont {{Treu}}, \citenamefont {{Marshall}},\ and\ \citenamefont {{et
  al.}}}]{2015ApJ...800...11L}%
  \BibitemOpen
  \bibfield  {author} {\bibinfo {author} {\bibfnamefont {K.}~\bibnamefont
  {{Liao}}}, \bibinfo {author} {\bibfnamefont {T.}~\bibnamefont {{Treu}}},
  \bibinfo {author} {\bibfnamefont {P.}~\bibnamefont {{Marshall}}}, \ and\
  \bibinfo {author} {\bibnamefont {{et al.}}},\ }\href {\doibase
  10.1088/0004-637X/800/1/11} {\bibfield  {journal} {\bibinfo  {journal}
  {\apj}\ }\textbf {\bibinfo {volume} {800}},\ \bibinfo {eid} {11} (\bibinfo
  {year} {2015})}\BibitemShut {NoStop}%
\bibitem [{\citenamefont {{Tie}}\ and\ \citenamefont
  {{Kochanek}}(2018)}]{2018MNRAS.473...80T}%
  \BibitemOpen
  \bibfield  {author} {\bibinfo {author} {\bibfnamefont {S.~S.}\ \bibnamefont
  {{Tie}}}\ and\ \bibinfo {author} {\bibfnamefont {C.~S.}\ \bibnamefont
  {{Kochanek}}},\ }\href {\doibase 10.1093/mnras/stx2348} {\bibfield  {journal}
  {\bibinfo  {journal} {\mnras}\ }\textbf {\bibinfo {volume} {473}},\ \bibinfo
  {pages} {80} (\bibinfo {year} {2018})}\BibitemShut {NoStop}%
\bibitem [{\citenamefont {{Liao}}\ \emph {et~al.}(2018)\citenamefont {{Liao}},
  \citenamefont {{Ding}}, \citenamefont {{Biesiada}}, \citenamefont {{Fan}},\
  and\ \citenamefont {{Zhu}}}]{2018ApJ...867...69L}%
  \BibitemOpen
  \bibfield  {author} {\bibinfo {author} {\bibfnamefont {K.}~\bibnamefont
  {{Liao}}}, \bibinfo {author} {\bibfnamefont {X.}~\bibnamefont {{Ding}}},
  \bibinfo {author} {\bibfnamefont {M.}~\bibnamefont {{Biesiada}}}, \bibinfo
  {author} {\bibfnamefont {X.-L.}\ \bibnamefont {{Fan}}}, \ and\ \bibinfo
  {author} {\bibfnamefont {Z.-H.}\ \bibnamefont {{Zhu}}},\ }\href {\doibase
  10.3847/1538-4357/aae30f} {\bibfield  {journal} {\bibinfo  {journal} {\apj}\
  }\textbf {\bibinfo {volume} {867}},\ \bibinfo {eid} {69} (\bibinfo {year}
  {2018})}\BibitemShut {NoStop}%
\bibitem [{\citenamefont {{Pearson}}\ \emph {et~al.}(2021)\citenamefont
  {{Pearson}}, \citenamefont {{Trendafilova}},\ and\ \citenamefont
  {{Meyers}}}]{2021PhRvD.103f3017P}%
  \BibitemOpen
  \bibfield  {author} {\bibinfo {author} {\bibfnamefont {N.}~\bibnamefont
  {{Pearson}}}, \bibinfo {author} {\bibfnamefont {C.}~\bibnamefont
  {{Trendafilova}}}, \ and\ \bibinfo {author} {\bibfnamefont {J.}~\bibnamefont
  {{Meyers}}},\ }\href {\doibase 10.1103/PhysRevD.103.063017} {\bibfield
  {journal} {\bibinfo  {journal} {\prd}\ }\textbf {\bibinfo {volume} {103}},\
  \bibinfo {eid} {063017} (\bibinfo {year} {2021})}\BibitemShut {NoStop}%
\bibitem [{\citenamefont {{Cao}}\ \emph {et~al.}(2022)\citenamefont {{Cao}},
  \citenamefont {{Qi}}, \citenamefont {{Cao}},\ and\ \citenamefont {{et
  al.}}}]{2022A&A...659L...5C}%
  \BibitemOpen
  \bibfield  {author} {\bibinfo {author} {\bibfnamefont {S.}~\bibnamefont
  {{Cao}}}, \bibinfo {author} {\bibfnamefont {J.}~\bibnamefont {{Qi}}},
  \bibinfo {author} {\bibfnamefont {Z.}~\bibnamefont {{Cao}}}, \ and\ \bibinfo
  {author} {\bibnamefont {{et al.}}},\ }\href {\doibase
  10.1051/0004-6361/202142694} {\bibfield  {journal} {\bibinfo  {journal}
  {\aap}\ }\textbf {\bibinfo {volume} {659}},\ \bibinfo {eid} {L5} (\bibinfo
  {year} {2022})}\BibitemShut {NoStop}%
\bibitem [{\citenamefont {{Cao}}\ \emph {et~al.}(2021)\citenamefont {{Cao}},
  \citenamefont {{Qi}}, \citenamefont {{Biesiada}},\ and\ \citenamefont {{et
  al.}}}]{2021MNRAS.502L..16C}%
  \BibitemOpen
  \bibfield  {author} {\bibinfo {author} {\bibfnamefont {S.}~\bibnamefont
  {{Cao}}}, \bibinfo {author} {\bibfnamefont {J.}~\bibnamefont {{Qi}}},
  \bibinfo {author} {\bibfnamefont {M.}~\bibnamefont {{Biesiada}}}, \ and\
  \bibinfo {author} {\bibnamefont {{et al.}}},\ }\href {\doibase
  10.1093/mnrasl/slaa205} {\bibfield  {journal} {\bibinfo  {journal} {\mnras}\
  }\textbf {\bibinfo {volume} {502}},\ \bibinfo {pages} {L16} (\bibinfo {year}
  {2021})}\BibitemShut {NoStop}%
\bibitem [{\citenamefont {{Green}}\ and\ \citenamefont
  {{Kavanagh}}(2021)}]{2021JPhG...48d3001G}%
  \BibitemOpen
  \bibfield  {author} {\bibinfo {author} {\bibfnamefont {A.~M.}\ \bibnamefont
  {{Green}}}\ and\ \bibinfo {author} {\bibfnamefont {B.~J.}\ \bibnamefont
  {{Kavanagh}}},\ }\href {\doibase 10.1088/1361-6471/abc534} {\bibfield
  {journal} {\bibinfo  {journal} {Journal of Physics G Nuclear Physics}\
  }\textbf {\bibinfo {volume} {48}},\ \bibinfo {eid} {043001} (\bibinfo {year}
  {2021})}\BibitemShut {NoStop}%
\bibitem [{\citenamefont {{Metcalf}}\ and\ \citenamefont
  {{Silk}}(2007)}]{2007PhRvL..98g1302M}%
  \BibitemOpen
  \bibfield  {author} {\bibinfo {author} {\bibfnamefont {R.~B.}\ \bibnamefont
  {{Metcalf}}}\ and\ \bibinfo {author} {\bibfnamefont {J.}~\bibnamefont
  {{Silk}}},\ }\href {\doibase 10.1103/PhysRevLett.98.071302} {\bibfield
  {journal} {\bibinfo  {journal} {\prl}\ }\textbf {\bibinfo {volume} {98}},\
  \bibinfo {eid} {071302} (\bibinfo {year} {2007})}\BibitemShut {NoStop}%
\bibitem [{\citenamefont {{Zumalac{\'a}rregui}}\ and\ \citenamefont
  {{Seljak}}(2018)}]{2018PhRvL.121n1101Z}%
  \BibitemOpen
  \bibfield  {author} {\bibinfo {author} {\bibfnamefont {M.}~\bibnamefont
  {{Zumalac{\'a}rregui}}}\ and\ \bibinfo {author} {\bibfnamefont
  {U.}~\bibnamefont {{Seljak}}},\ }\href {\doibase
  10.1103/PhysRevLett.121.141101} {\bibfield  {journal} {\bibinfo  {journal}
  {\prl}\ }\textbf {\bibinfo {volume} {121}},\ \bibinfo {eid} {141101}
  (\bibinfo {year} {2018})}\BibitemShut {NoStop}%
\bibitem [{\citenamefont {{Goldstein}}\ \emph {et~al.}(2018)\citenamefont
  {{Goldstein}}, \citenamefont {{Nugent}}, \citenamefont {{Kasen}},\ and\
  \citenamefont {{Collett}}}]{2018ApJ...855...22G}%
  \BibitemOpen
  \bibfield  {author} {\bibinfo {author} {\bibfnamefont {D.~A.}\ \bibnamefont
  {{Goldstein}}}, \bibinfo {author} {\bibfnamefont {P.~E.}\ \bibnamefont
  {{Nugent}}}, \bibinfo {author} {\bibfnamefont {D.~N.}\ \bibnamefont
  {{Kasen}}}, \ and\ \bibinfo {author} {\bibfnamefont {T.~E.}\ \bibnamefont
  {{Collett}}},\ }\href {\doibase 10.3847/1538-4357/aaa975} {\bibfield
  {journal} {\bibinfo  {journal} {\apj}\ }\textbf {\bibinfo {volume} {855}},\
  \bibinfo {eid} {22} (\bibinfo {year} {2018})}\BibitemShut {NoStop}%
\bibitem [{\citenamefont {{Quinn}}\ \emph {et~al.}(2009)\citenamefont
  {{Quinn}}, \citenamefont {{Wilkinson}}, \citenamefont {{Irwin}},\ and\
  \citenamefont {{et al.}}}]{2009MNRAS.396L..11Q}%
  \BibitemOpen
  \bibfield  {author} {\bibinfo {author} {\bibfnamefont {D.~P.}\ \bibnamefont
  {{Quinn}}}, \bibinfo {author} {\bibfnamefont {M.~I.}\ \bibnamefont
  {{Wilkinson}}}, \bibinfo {author} {\bibfnamefont {M.~J.}\ \bibnamefont
  {{Irwin}}}, \ and\ \bibinfo {author} {\bibnamefont {{et al.}}},\ }\href
  {\doibase 10.1111/j.1745-3933.2009.00652.x} {\bibfield  {journal} {\bibinfo
  {journal} {\mnras}\ }\textbf {\bibinfo {volume} {396}},\ \bibinfo {pages}
  {L11} (\bibinfo {year} {2009})}\BibitemShut {NoStop}%
\bibitem [{\citenamefont {{Laha}}(2020)}]{2020PhRvD.102b3016L}%
  \BibitemOpen
  \bibfield  {author} {\bibinfo {author} {\bibfnamefont {R.}~\bibnamefont
  {{Laha}}},\ }\href {\doibase 10.1103/PhysRevD.102.023016} {\bibfield
  {journal} {\bibinfo  {journal} {\prd}\ }\textbf {\bibinfo {volume} {102}},\
  \bibinfo {eid} {023016} (\bibinfo {year} {2020})}\BibitemShut {NoStop}%
\bibitem [{\citenamefont {{Jung}}\ and\ \citenamefont
  {{Shin}}(2019)}]{2019PhRvL.122d1103J}%
  \BibitemOpen
  \bibfield  {author} {\bibinfo {author} {\bibfnamefont {S.}~\bibnamefont
  {{Jung}}}\ and\ \bibinfo {author} {\bibfnamefont {C.~S.}\ \bibnamefont
  {{Shin}}},\ }\href {\doibase 10.1103/PhysRevLett.122.041103} {\bibfield
  {journal} {\bibinfo  {journal} {\prl}\ }\textbf {\bibinfo {volume} {122}},\
  \bibinfo {eid} {041103} (\bibinfo {year} {2019})}\BibitemShut {NoStop}%
\bibitem [{\citenamefont {{Liao}}\ \emph
  {et~al.}(2020{\natexlab{b}})\citenamefont {{Liao}}, \citenamefont {{Tian}},\
  and\ \citenamefont {{Ding}}}]{2020MNRAS.495.2002L}%
  \BibitemOpen
  \bibfield  {author} {\bibinfo {author} {\bibfnamefont {K.}~\bibnamefont
  {{Liao}}}, \bibinfo {author} {\bibfnamefont {S.}~\bibnamefont {{Tian}}}, \
  and\ \bibinfo {author} {\bibfnamefont {X.}~\bibnamefont {{Ding}}},\ }\href
  {\doibase 10.1093/mnras/staa1388} {\bibfield  {journal} {\bibinfo  {journal}
  {\mnras}\ }\textbf {\bibinfo {volume} {495}},\ \bibinfo {pages} {2002}
  (\bibinfo {year} {2020}{\natexlab{b}})}\BibitemShut {NoStop}%
\bibitem [{\citenamefont {{Urrutia}}\ and\ \citenamefont
  {{Vaskonen}}(2022)}]{2022MNRAS.509.1358U}%
  \BibitemOpen
  \bibfield  {author} {\bibinfo {author} {\bibfnamefont {J.}~\bibnamefont
  {{Urrutia}}}\ and\ \bibinfo {author} {\bibfnamefont {V.}~\bibnamefont
  {{Vaskonen}}},\ }\href {\doibase 10.1093/mnras/stab3118} {\bibfield
  {journal} {\bibinfo  {journal} {\mnras}\ }\textbf {\bibinfo {volume} {509}},\
  \bibinfo {pages} {1358} (\bibinfo {year} {2022})}\BibitemShut {NoStop}%
\bibitem [{\citenamefont {{Guo}}\ and\ \citenamefont
  {{Lu}}(2022)}]{2022PhRvD.106b3018G}%
  \BibitemOpen
  \bibfield  {author} {\bibinfo {author} {\bibfnamefont {X.}~\bibnamefont
  {{Guo}}}\ and\ \bibinfo {author} {\bibfnamefont {Y.}~\bibnamefont {{Lu}}},\
  }\href {\doibase 10.1103/PhysRevD.106.023018} {\bibfield  {journal} {\bibinfo
   {journal} {\prd}\ }\textbf {\bibinfo {volume} {106}},\ \bibinfo {eid}
  {023018} (\bibinfo {year} {2022})}\BibitemShut {NoStop}%
\bibitem [{\citenamefont {{Oguri}}\ \emph {et~al.}(2003)\citenamefont
  {{Oguri}}, \citenamefont {{Suto}},\ and\ \citenamefont
  {{Turner}}}]{2003ApJ...583..584O}%
  \BibitemOpen
  \bibfield  {author} {\bibinfo {author} {\bibfnamefont {M.}~\bibnamefont
  {{Oguri}}}, \bibinfo {author} {\bibfnamefont {Y.}~\bibnamefont {{Suto}}}, \
  and\ \bibinfo {author} {\bibfnamefont {E.~L.}\ \bibnamefont {{Turner}}},\
  }\href {\doibase 10.1086/345431} {\bibfield  {journal} {\bibinfo  {journal}
  {\apj}\ }\textbf {\bibinfo {volume} {583}},\ \bibinfo {pages} {584} (\bibinfo
  {year} {2003})}\BibitemShut {NoStop}%
\bibitem [{\citenamefont {{Suwa}}(2018)}]{2018MNRAS.474.2612S}%
  \BibitemOpen
  \bibfield  {author} {\bibinfo {author} {\bibfnamefont {Y.}~\bibnamefont
  {{Suwa}}},\ }\href {\doibase 10.1093/mnras/stx2953} {\bibfield  {journal}
  {\bibinfo  {journal} {\mnras}\ }\textbf {\bibinfo {volume} {474}},\ \bibinfo
  {pages} {2612} (\bibinfo {year} {2018})}\BibitemShut {NoStop}%
\bibitem [{\citenamefont {{Foxley-Marrable}}\ \emph {et~al.}(2020)\citenamefont
  {{Foxley-Marrable}}, \citenamefont {{Collett}}, \citenamefont {{Frohmaier}},\
  and\ \citenamefont {{et al.}}}]{2020MNRAS.495.4622F}%
  \BibitemOpen
  \bibfield  {author} {\bibinfo {author} {\bibfnamefont {M.}~\bibnamefont
  {{Foxley-Marrable}}}, \bibinfo {author} {\bibfnamefont {T.~E.}\ \bibnamefont
  {{Collett}}}, \bibinfo {author} {\bibfnamefont {C.}~\bibnamefont
  {{Frohmaier}}}, \ and\ \bibinfo {author} {\bibnamefont {{et al.}}},\ }\href
  {\doibase 10.1093/mnras/staa1289} {\bibfield  {journal} {\bibinfo  {journal}
  {\mnras}\ }\textbf {\bibinfo {volume} {495}},\ \bibinfo {pages} {4622}
  (\bibinfo {year} {2020})}\BibitemShut {NoStop}%
\bibitem [{\citenamefont {{Huber}}\ \emph {et~al.}(2021)\citenamefont
  {{Huber}}, \citenamefont {{Suyu}}, \citenamefont {{Noebauer}},\ and\
  \citenamefont {{et al.}}}]{2021A&A...646A.110H}%
  \BibitemOpen
  \bibfield  {author} {\bibinfo {author} {\bibfnamefont {S.}~\bibnamefont
  {{Huber}}}, \bibinfo {author} {\bibfnamefont {S.~H.}\ \bibnamefont {{Suyu}}},
  \bibinfo {author} {\bibfnamefont {U.~M.}\ \bibnamefont {{Noebauer}}}, \ and\
  \bibinfo {author} {\bibnamefont {{et al.}}},\ }\href {\doibase
  10.1051/0004-6361/202039218} {\bibfield  {journal} {\bibinfo  {journal}
  {\aap}\ }\textbf {\bibinfo {volume} {646}},\ \bibinfo {eid} {A110} (\bibinfo
  {year} {2021})}\BibitemShut {NoStop}%
\bibitem [{\citenamefont {{Rydberg}}\ \emph {et~al.}(2020)\citenamefont
  {{Rydberg}}, \citenamefont {{Whalen}}, \citenamefont {{Maturi}},\ and\
  \citenamefont {{et al.}}}]{2020MNRAS.491.2447R}%
  \BibitemOpen
  \bibfield  {author} {\bibinfo {author} {\bibfnamefont {C.-E.}\ \bibnamefont
  {{Rydberg}}}, \bibinfo {author} {\bibfnamefont {D.~J.}\ \bibnamefont
  {{Whalen}}}, \bibinfo {author} {\bibfnamefont {M.}~\bibnamefont {{Maturi}}},
  \ and\ \bibinfo {author} {\bibnamefont {{et al.}}},\ }\href {\doibase
  10.1093/mnras/stz3203} {\bibfield  {journal} {\bibinfo  {journal} {\mnras}\
  }\textbf {\bibinfo {volume} {491}},\ \bibinfo {pages} {2447} (\bibinfo {year}
  {2020})}\BibitemShut {NoStop}%
\bibitem [{\citenamefont {{Pan}}\ and\ \citenamefont
  {{Loeb}}(2013)}]{2013MNRAS.435L..33P}%
  \BibitemOpen
  \bibfield  {author} {\bibinfo {author} {\bibfnamefont {T.}~\bibnamefont
  {{Pan}}}\ and\ \bibinfo {author} {\bibfnamefont {A.}~\bibnamefont {{Loeb}}},\
  }\href {\doibase 10.1093/mnrasl/slt089} {\bibfield  {journal} {\bibinfo
  {journal} {\mnras}\ }\textbf {\bibinfo {volume} {435}},\ \bibinfo {pages}
  {L33} (\bibinfo {year} {2013})}\BibitemShut {NoStop}%
\bibitem [{\citenamefont {{Petrushevska}}\ \emph {et~al.}(2016)\citenamefont
  {{Petrushevska}}, \citenamefont {{Amanullah}}, \citenamefont {{Goobar}},\
  and\ \citenamefont {{et al.}}}]{2016A&A...594A..54P}%
  \BibitemOpen
  \bibfield  {author} {\bibinfo {author} {\bibfnamefont {T.}~\bibnamefont
  {{Petrushevska}}}, \bibinfo {author} {\bibfnamefont {R.}~\bibnamefont
  {{Amanullah}}}, \bibinfo {author} {\bibfnamefont {A.}~\bibnamefont
  {{Goobar}}}, \ and\ \bibinfo {author} {\bibnamefont {{et al.}}},\ }\href
  {\doibase 10.1051/0004-6361/201628925} {\bibfield  {journal} {\bibinfo
  {journal} {\aap}\ }\textbf {\bibinfo {volume} {594}},\ \bibinfo {eid} {A54}
  (\bibinfo {year} {2016})}\BibitemShut {NoStop}%
\bibitem [{\citenamefont {{Petrushevska}}\ \emph
  {et~al.}(2018{\natexlab{b}})\citenamefont {{Petrushevska}}, \citenamefont
  {{Okamura}}, \citenamefont {{Kawamata}},\ and\ \citenamefont {{et
  al.}}}]{2018ARep...62..917P}%
  \BibitemOpen
  \bibfield  {author} {\bibinfo {author} {\bibfnamefont {T.}~\bibnamefont
  {{Petrushevska}}}, \bibinfo {author} {\bibfnamefont {T.}~\bibnamefont
  {{Okamura}}}, \bibinfo {author} {\bibfnamefont {R.}~\bibnamefont
  {{Kawamata}}}, \ and\ \bibinfo {author} {\bibnamefont {{et al.}}},\ }\href
  {\doibase 10.1134/S1063772918120272} {\bibfield  {journal} {\bibinfo
  {journal} {Astronomy Reports}\ }\textbf {\bibinfo {volume} {62}},\ \bibinfo
  {pages} {917} (\bibinfo {year} {2018}{\natexlab{b}})}\BibitemShut {NoStop}%
\bibitem [{\citenamefont {{Wong}}\ \emph {et~al.}(2019)\citenamefont {{Wong}},
  \citenamefont {{Moriya}}, \citenamefont {{Oguri}},\ and\ \citenamefont {{et
  al.}}}]{2019PASJ...71...60W}%
  \BibitemOpen
  \bibfield  {author} {\bibinfo {author} {\bibfnamefont {K.~C.}\ \bibnamefont
  {{Wong}}}, \bibinfo {author} {\bibfnamefont {T.~J.}\ \bibnamefont
  {{Moriya}}}, \bibinfo {author} {\bibfnamefont {M.}~\bibnamefont {{Oguri}}}, \
  and\ \bibinfo {author} {\bibnamefont {{et al.}}},\ }\href {\doibase
  10.1093/pasj/psz037} {\bibfield  {journal} {\bibinfo  {journal} {\pasj}\
  }\textbf {\bibinfo {volume} {71}},\ \bibinfo {eid} {60} (\bibinfo {year}
  {2019})}\BibitemShut {NoStop}%
\bibitem [{\citenamefont {{Boco}}\ \emph {et~al.}(2019)\citenamefont {{Boco}},
  \citenamefont {{Lapi}}, \citenamefont {{Goswami}},\ and\ \citenamefont {{et
  al.}}}]{2019ApJ...881..157B}%
  \BibitemOpen
  \bibfield  {author} {\bibinfo {author} {\bibfnamefont {L.}~\bibnamefont
  {{Boco}}}, \bibinfo {author} {\bibfnamefont {A.}~\bibnamefont {{Lapi}}},
  \bibinfo {author} {\bibfnamefont {S.}~\bibnamefont {{Goswami}}}, \ and\
  \bibinfo {author} {\bibnamefont {{et al.}}},\ }\href {\doibase
  10.3847/1538-4357/ab328e} {\bibfield  {journal} {\bibinfo  {journal} {\apj}\
  }\textbf {\bibinfo {volume} {881}},\ \bibinfo {eid} {157} (\bibinfo {year}
  {2019})}\BibitemShut {NoStop}%
\bibitem [{\citenamefont {{Oguri}}(2018)}]{2018MNRAS.480.3842O}%
  \BibitemOpen
  \bibfield  {author} {\bibinfo {author} {\bibfnamefont {M.}~\bibnamefont
  {{Oguri}}},\ }\href {\doibase 10.1093/mnras/sty2145} {\bibfield  {journal}
  {\bibinfo  {journal} {\mnras}\ }\textbf {\bibinfo {volume} {480}},\ \bibinfo
  {pages} {3842} (\bibinfo {year} {2018})}\BibitemShut {NoStop}%
\bibitem [{\citenamefont {{He}}\ \emph {et~al.}(2022)\citenamefont {{He}},
  \citenamefont {{Liao}}, \citenamefont {{Ding}},\ and\ \citenamefont {{et
  al.}}}]{2022arXiv220515515H}%
  \BibitemOpen
  \bibfield  {author} {\bibinfo {author} {\bibfnamefont {X.}~\bibnamefont
  {{He}}}, \bibinfo {author} {\bibfnamefont {K.}~\bibnamefont {{Liao}}},
  \bibinfo {author} {\bibfnamefont {X.}~\bibnamefont {{Ding}}}, \ and\ \bibinfo
  {author} {\bibnamefont {{et al.}}},\ }\href@noop {} {\bibfield  {journal}
  {\bibinfo  {journal} {arXiv e-prints}\ } (\bibinfo {year} {2022})},\ \Eprint
  {http://arxiv.org/abs/2205.15515} {2205.15515} \BibitemShut {NoStop}%
\bibitem [{\citenamefont {{Diego}}(2019)}]{2019A&A...625A..84D}%
  \BibitemOpen
  \bibfield  {author} {\bibinfo {author} {\bibfnamefont {J.~M.}\ \bibnamefont
  {{Diego}}},\ }\href {\doibase 10.1051/0004-6361/201833670} {\bibfield
  {journal} {\bibinfo  {journal} {\aap}\ }\textbf {\bibinfo {volume} {625}},\
  \bibinfo {eid} {A84} (\bibinfo {year} {2019})}\BibitemShut {NoStop}%
\bibitem [{\citenamefont {{Hannuksela}}\ \emph {et~al.}(2020)\citenamefont
  {{Hannuksela}}, \citenamefont {{Collett}}, \citenamefont
  {{{\c{C}}al{\i}{\c{s}}kan}},\ and\ \citenamefont
  {{Li}}}]{2020MNRAS.498.3395H}%
  \BibitemOpen
  \bibfield  {author} {\bibinfo {author} {\bibfnamefont {O.~A.}\ \bibnamefont
  {{Hannuksela}}}, \bibinfo {author} {\bibfnamefont {T.~E.}\ \bibnamefont
  {{Collett}}}, \bibinfo {author} {\bibfnamefont {M.}~\bibnamefont
  {{{\c{C}}al{\i}{\c{s}}kan}}}, \ and\ \bibinfo {author} {\bibfnamefont
  {T.~G.~F.}\ \bibnamefont {{Li}}},\ }\href {\doibase 10.1093/mnras/staa2577}
  {\bibfield  {journal} {\bibinfo  {journal} {\mnras}\ }\textbf {\bibinfo
  {volume} {498}},\ \bibinfo {pages} {3395} (\bibinfo {year}
  {2020})}\BibitemShut {NoStop}%
\bibitem [{\citenamefont {{Planck Collaboration}}(2020)}]{2020A&A...641A...6P}%
  \BibitemOpen
  \bibfield  {author} {\bibinfo {author} {\bibnamefont {{Planck
  Collaboration}}},\ }\href {\doibase 10.1051/0004-6361/201833910} {\bibfield
  {journal} {\bibinfo  {journal} {\aap}\ }\textbf {\bibinfo {volume} {641}},\
  \bibinfo {eid} {A6} (\bibinfo {year} {2020})}\BibitemShut {NoStop}%
\bibitem [{\citenamefont {{Aiola}}\ \emph {et~al.}(2020)\citenamefont
  {{Aiola}}, \citenamefont {{Calabrese}}, \citenamefont {{Maurin}},\ and\
  \citenamefont {{et al.}}}]{2020JCAP...12..047A}%
  \BibitemOpen
  \bibfield  {author} {\bibinfo {author} {\bibfnamefont {S.}~\bibnamefont
  {{Aiola}}}, \bibinfo {author} {\bibfnamefont {E.}~\bibnamefont
  {{Calabrese}}}, \bibinfo {author} {\bibfnamefont {L.}~\bibnamefont
  {{Maurin}}}, \ and\ \bibinfo {author} {\bibnamefont {{et al.}}},\ }\href
  {\doibase 10.1088/1475-7516/2020/12/047} {\bibfield  {journal} {\bibinfo
  {journal} {\jcap}\ }\textbf {\bibinfo {volume} {2020}},\ \bibinfo {eid} {047}
  (\bibinfo {year} {2020})}\BibitemShut {NoStop}%
\bibitem [{\citenamefont {{Riess}}\ \emph {et~al.}(2019)\citenamefont
  {{Riess}}, \citenamefont {{Casertano}}, \citenamefont {{Yuan}}, \citenamefont
  {{Macri}},\ and\ \citenamefont {{Scolnic}}}]{2019ApJ...876...85R}%
  \BibitemOpen
  \bibfield  {author} {\bibinfo {author} {\bibfnamefont {A.~G.}\ \bibnamefont
  {{Riess}}}, \bibinfo {author} {\bibfnamefont {S.}~\bibnamefont
  {{Casertano}}}, \bibinfo {author} {\bibfnamefont {W.}~\bibnamefont {{Yuan}}},
  \bibinfo {author} {\bibfnamefont {L.~M.}\ \bibnamefont {{Macri}}}, \ and\
  \bibinfo {author} {\bibfnamefont {D.}~\bibnamefont {{Scolnic}}},\ }\href
  {\doibase 10.3847/1538-4357/ab1422} {\bibfield  {journal} {\bibinfo
  {journal} {\apj}\ }\textbf {\bibinfo {volume} {876}},\ \bibinfo {eid} {85}
  (\bibinfo {year} {2019})}\BibitemShut {NoStop}%
\bibitem [{\citenamefont {{Riess}}\ \emph {et~al.}(2021)\citenamefont
  {{Riess}}, \citenamefont {{Yuan}}, \citenamefont {{Macri}},\ and\
  \citenamefont {{et al.}}}]{2021arXiv211204510R}%
  \BibitemOpen
  \bibfield  {author} {\bibinfo {author} {\bibfnamefont {A.~G.}\ \bibnamefont
  {{Riess}}}, \bibinfo {author} {\bibfnamefont {W.}~\bibnamefont {{Yuan}}},
  \bibinfo {author} {\bibfnamefont {L.~M.}\ \bibnamefont {{Macri}}}, \ and\
  \bibinfo {author} {\bibnamefont {{et al.}}},\ }\href@noop {} {\bibfield
  {journal} {\bibinfo  {journal} {arXiv e-prints}\ } (\bibinfo {year}
  {2021})},\ \Eprint {http://arxiv.org/abs/2112.04510} {2112.04510}
  \BibitemShut {NoStop}%
\bibitem [{\citenamefont {{Paraficz}}\ and\ \citenamefont
  {{Hjorth}}(2009)}]{2009A&A...507L..49P}%
  \BibitemOpen
  \bibfield  {author} {\bibinfo {author} {\bibfnamefont {D.}~\bibnamefont
  {{Paraficz}}}\ and\ \bibinfo {author} {\bibfnamefont {J.}~\bibnamefont
  {{Hjorth}}},\ }\href {\doibase 10.1051/0004-6361/200913307} {\bibfield
  {journal} {\bibinfo  {journal} {\aap}\ }\textbf {\bibinfo {volume} {507}},\
  \bibinfo {pages} {L49} (\bibinfo {year} {2009})}\BibitemShut {NoStop}%
\bibitem [{\citenamefont {{Jee}}\ \emph {et~al.}(2015)\citenamefont {{Jee}},
  \citenamefont {{Komatsu}},\ and\ \citenamefont
  {{Suyu}}}]{2015JCAP...11..033J}%
  \BibitemOpen
  \bibfield  {author} {\bibinfo {author} {\bibfnamefont {I.}~\bibnamefont
  {{Jee}}}, \bibinfo {author} {\bibfnamefont {E.}~\bibnamefont {{Komatsu}}}, \
  and\ \bibinfo {author} {\bibfnamefont {S.~H.}\ \bibnamefont {{Suyu}}},\
  }\href {\doibase 10.1088/1475-7516/2015/11/033} {\bibfield  {journal}
  {\bibinfo  {journal} {\jcap}\ }\textbf {\bibinfo {volume} {2015}},\ \bibinfo
  {eid} {033} (\bibinfo {year} {2015})}\BibitemShut {NoStop}%
\bibitem [{\citenamefont {{Blum}}\ \emph {et~al.}(2020)\citenamefont {{Blum}},
  \citenamefont {{Castorina}},\ and\ \citenamefont
  {{Simonovi{\'c}}}}]{2020ApJ...892L..27B}%
  \BibitemOpen
  \bibfield  {author} {\bibinfo {author} {\bibfnamefont {K.}~\bibnamefont
  {{Blum}}}, \bibinfo {author} {\bibfnamefont {E.}~\bibnamefont {{Castorina}}},
  \ and\ \bibinfo {author} {\bibfnamefont {M.}~\bibnamefont
  {{Simonovi{\'c}}}},\ }\href {\doibase 10.3847/2041-8213/ab8012} {\bibfield
  {journal} {\bibinfo  {journal} {\apjl}\ }\textbf {\bibinfo {volume} {892}},\
  \bibinfo {eid} {L27} (\bibinfo {year} {2020})}\BibitemShut {NoStop}%
\bibitem [{\citenamefont {{Suyu}}\ \emph {et~al.}(2017)\citenamefont {{Suyu}},
  \citenamefont {{Bonvin}}, \citenamefont {{Courbin}},\ and\ \citenamefont {{et
  al.}}}]{2017MNRAS.468.2590S}%
  \BibitemOpen
  \bibfield  {author} {\bibinfo {author} {\bibfnamefont {S.~H.}\ \bibnamefont
  {{Suyu}}}, \bibinfo {author} {\bibfnamefont {V.}~\bibnamefont {{Bonvin}}},
  \bibinfo {author} {\bibfnamefont {F.}~\bibnamefont {{Courbin}}}, \ and\
  \bibinfo {author} {\bibnamefont {{et al.}}},\ }\href {\doibase
  10.1093/mnras/stx483} {\bibfield  {journal} {\bibinfo  {journal} {\mnras}\
  }\textbf {\bibinfo {volume} {468}},\ \bibinfo {pages} {2590} (\bibinfo {year}
  {2017})}\BibitemShut {NoStop}%
\bibitem [{\citenamefont {{Eigenbrod}}\ \emph {et~al.}(2005)\citenamefont
  {{Eigenbrod}}, \citenamefont {{Courbin}}, \citenamefont {{Vuissoz}},
  \citenamefont {{Meylan}}, \citenamefont {{Saha}},\ and\ \citenamefont
  {{Dye}}}]{2005A&A...436...25E}%
  \BibitemOpen
  \bibfield  {author} {\bibinfo {author} {\bibfnamefont {A.}~\bibnamefont
  {{Eigenbrod}}}, \bibinfo {author} {\bibfnamefont {F.}~\bibnamefont
  {{Courbin}}}, \bibinfo {author} {\bibfnamefont {C.}~\bibnamefont
  {{Vuissoz}}}, \bibinfo {author} {\bibfnamefont {G.}~\bibnamefont {{Meylan}}},
  \bibinfo {author} {\bibfnamefont {P.}~\bibnamefont {{Saha}}}, \ and\ \bibinfo
  {author} {\bibfnamefont {S.}~\bibnamefont {{Dye}}},\ }\href {\doibase
  10.1051/0004-6361:20042422} {\bibfield  {journal} {\bibinfo  {journal}
  {\aap}\ }\textbf {\bibinfo {volume} {436}},\ \bibinfo {pages} {25} (\bibinfo
  {year} {2005})}\BibitemShut {NoStop}%
\bibitem [{\citenamefont {{Treu}}\ \emph {et~al.}(2018)\citenamefont {{Treu}},
  \citenamefont {{Agnello}}, \citenamefont {{Baumer}},\ and\ \citenamefont {{et
  al.}}}]{2018MNRAS.481.1041T}%
  \BibitemOpen
  \bibfield  {author} {\bibinfo {author} {\bibfnamefont {T.}~\bibnamefont
  {{Treu}}}, \bibinfo {author} {\bibfnamefont {A.}~\bibnamefont {{Agnello}}},
  \bibinfo {author} {\bibfnamefont {M.~A.}\ \bibnamefont {{Baumer}}}, \ and\
  \bibinfo {author} {\bibnamefont {{et al.}}},\ }\href {\doibase
  10.1093/mnras/sty2329} {\bibfield  {journal} {\bibinfo  {journal} {\mnras}\
  }\textbf {\bibinfo {volume} {481}},\ \bibinfo {pages} {1041} (\bibinfo {year}
  {2018})}\BibitemShut {NoStop}%
\bibitem [{\citenamefont {{Millon}}\ \emph {et~al.}(2020)\citenamefont
  {{Millon}}, \citenamefont {{Galan}}, \citenamefont {{Courbin}},\ and\
  \citenamefont {{et al.}}}]{2020A&A...639A.101M}%
  \BibitemOpen
  \bibfield  {author} {\bibinfo {author} {\bibfnamefont {M.}~\bibnamefont
  {{Millon}}}, \bibinfo {author} {\bibfnamefont {A.}~\bibnamefont {{Galan}}},
  \bibinfo {author} {\bibfnamefont {F.}~\bibnamefont {{Courbin}}}, \ and\
  \bibinfo {author} {\bibnamefont {{et al.}}},\ }\href {\doibase
  10.1051/0004-6361/201937351} {\bibfield  {journal} {\bibinfo  {journal}
  {\aap}\ }\textbf {\bibinfo {volume} {639}},\ \bibinfo {eid} {A101} (\bibinfo
  {year} {2020})}\BibitemShut {NoStop}%
\bibitem [{\citenamefont {{Wong}}\ \emph {et~al.}(2020)\citenamefont {{Wong}},
  \citenamefont {{Suyu}}, \citenamefont {{Chen}},\ and\ \citenamefont {{et
  al.}}}]{2020MNRAS.498.1420W}%
  \BibitemOpen
  \bibfield  {author} {\bibinfo {author} {\bibfnamefont {K.~C.}\ \bibnamefont
  {{Wong}}}, \bibinfo {author} {\bibfnamefont {S.~H.}\ \bibnamefont {{Suyu}}},
  \bibinfo {author} {\bibfnamefont {G.~C.~F.}\ \bibnamefont {{Chen}}}, \ and\
  \bibinfo {author} {\bibnamefont {{et al.}}},\ }\href {\doibase
  10.1093/mnras/stz3094} {\bibfield  {journal} {\bibinfo  {journal} {\mnras}\
  }\textbf {\bibinfo {volume} {498}},\ \bibinfo {pages} {1420} (\bibinfo {year}
  {2020})}\BibitemShut {NoStop}%
\bibitem [{\citenamefont {{Birrer}}\ \emph {et~al.}(2020)\citenamefont
  {{Birrer}}, \citenamefont {{Shajib}}, \citenamefont {{Galan}},\ and\
  \citenamefont {{et al.}}}]{2020A&A...643A.165B}%
  \BibitemOpen
  \bibfield  {author} {\bibinfo {author} {\bibfnamefont {S.}~\bibnamefont
  {{Birrer}}}, \bibinfo {author} {\bibfnamefont {A.~J.}\ \bibnamefont
  {{Shajib}}}, \bibinfo {author} {\bibfnamefont {A.}~\bibnamefont {{Galan}}}, \
  and\ \bibinfo {author} {\bibnamefont {{et al.}}},\ }\href {\doibase
  10.1051/0004-6361/202038861} {\bibfield  {journal} {\bibinfo  {journal}
  {\aap}\ }\textbf {\bibinfo {volume} {643}},\ \bibinfo {eid} {A165} (\bibinfo
  {year} {2020})}\BibitemShut {NoStop}%
\bibitem [{\citenamefont {{Liao}}\ \emph {et~al.}(2017)\citenamefont {{Liao}},
  \citenamefont {{Fan}}, \citenamefont {{Ding}}, \citenamefont {{Biesiada}},\
  and\ \citenamefont {{Zhu}}}]{2017NatCo...8.1148L}%
  \BibitemOpen
  \bibfield  {author} {\bibinfo {author} {\bibfnamefont {K.}~\bibnamefont
  {{Liao}}}, \bibinfo {author} {\bibfnamefont {X.-L.}\ \bibnamefont {{Fan}}},
  \bibinfo {author} {\bibfnamefont {X.}~\bibnamefont {{Ding}}}, \bibinfo
  {author} {\bibfnamefont {M.}~\bibnamefont {{Biesiada}}}, \ and\ \bibinfo
  {author} {\bibfnamefont {Z.-H.}\ \bibnamefont {{Zhu}}},\ }\href {\doibase
  10.1038/s41467-017-01152-9} {\bibfield  {journal} {\bibinfo  {journal}
  {Nature Communications}\ }\textbf {\bibinfo {volume} {8}},\ \bibinfo {eid}
  {1148} (\bibinfo {year} {2017})}\BibitemShut {NoStop}%
\bibitem [{\citenamefont {{Oguri}}\ and\ \citenamefont
  {{Kawano}}(2003)}]{2003MNRAS.338L..25O}%
  \BibitemOpen
  \bibfield  {author} {\bibinfo {author} {\bibfnamefont {M.}~\bibnamefont
  {{Oguri}}}\ and\ \bibinfo {author} {\bibfnamefont {Y.}~\bibnamefont
  {{Kawano}}},\ }\href {\doibase 10.1046/j.1365-8711.2003.06290.x} {\bibfield
  {journal} {\bibinfo  {journal} {\mnras}\ }\textbf {\bibinfo {volume} {338}},\
  \bibinfo {pages} {L25} (\bibinfo {year} {2003})}\BibitemShut {NoStop}%
\bibitem [{\citenamefont {{Sereno}}\ \emph {et~al.}(2011)\citenamefont
  {{Sereno}}, \citenamefont {{Jetzer}}, \citenamefont {{Sesana}},\ and\
  \citenamefont {{Volonteri}}}]{2011MNRAS.415.2773S}%
  \BibitemOpen
  \bibfield  {author} {\bibinfo {author} {\bibfnamefont {M.}~\bibnamefont
  {{Sereno}}}, \bibinfo {author} {\bibfnamefont {P.}~\bibnamefont {{Jetzer}}},
  \bibinfo {author} {\bibfnamefont {A.}~\bibnamefont {{Sesana}}}, \ and\
  \bibinfo {author} {\bibfnamefont {M.}~\bibnamefont {{Volonteri}}},\ }\href
  {\doibase 10.1111/j.1365-2966.2011.18895.x} {\bibfield  {journal} {\bibinfo
  {journal} {\mnras}\ }\textbf {\bibinfo {volume} {415}},\ \bibinfo {pages}
  {2773} (\bibinfo {year} {2011})}\BibitemShut {NoStop}%
\bibitem [{\citenamefont {{Hou}}\ \emph
  {et~al.}(2021{\natexlab{b}})\citenamefont {{Hou}}, \citenamefont {{Fan}},\
  and\ \citenamefont {{Zhu}}}]{2021MNRAS.507..761H}%
  \BibitemOpen
  \bibfield  {author} {\bibinfo {author} {\bibfnamefont {S.}~\bibnamefont
  {{Hou}}}, \bibinfo {author} {\bibfnamefont {X.-L.}\ \bibnamefont {{Fan}}}, \
  and\ \bibinfo {author} {\bibfnamefont {Z.-H.}\ \bibnamefont {{Zhu}}},\ }\href
  {\doibase 10.1093/mnras/stab2221} {\bibfield  {journal} {\bibinfo  {journal}
  {\mnras}\ }\textbf {\bibinfo {volume} {507}},\ \bibinfo {pages} {761}
  (\bibinfo {year} {2021}{\natexlab{b}})}\BibitemShut {NoStop}%
\bibitem [{\citenamefont {{Li}}\ \emph
  {et~al.}(2018{\natexlab{b}})\citenamefont {{Li}}, \citenamefont {{Gao}},
  \citenamefont {{Ding}}, \citenamefont {{Wang}},\ and\ \citenamefont
  {{Zhang}}}]{2018NatCo...9.3833L}%
  \BibitemOpen
  \bibfield  {author} {\bibinfo {author} {\bibfnamefont {Z.-X.}\ \bibnamefont
  {{Li}}}, \bibinfo {author} {\bibfnamefont {H.}~\bibnamefont {{Gao}}},
  \bibinfo {author} {\bibfnamefont {X.-H.}\ \bibnamefont {{Ding}}}, \bibinfo
  {author} {\bibfnamefont {G.-J.}\ \bibnamefont {{Wang}}}, \ and\ \bibinfo
  {author} {\bibfnamefont {B.}~\bibnamefont {{Zhang}}},\ }\href {\doibase
  10.1038/s41467-018-06303-0} {\bibfield  {journal} {\bibinfo  {journal}
  {Nature Communications}\ }\textbf {\bibinfo {volume} {9}},\ \bibinfo {eid}
  {3833} (\bibinfo {year} {2018}{\natexlab{b}})}\BibitemShut {NoStop}%
\bibitem [{\citenamefont {{Gao}}\ \emph {et~al.}(2022)\citenamefont {{Gao}},
  \citenamefont {{Li}},\ and\ \citenamefont {{Gao}}}]{2022MNRAS.516.1977G}%
  \BibitemOpen
  \bibfield  {author} {\bibinfo {author} {\bibfnamefont {R.}~\bibnamefont
  {{Gao}}}, \bibinfo {author} {\bibfnamefont {Z.}~\bibnamefont {{Li}}}, \ and\
  \bibinfo {author} {\bibfnamefont {H.}~\bibnamefont {{Gao}}},\ }\href
  {\doibase 10.1093/mnras/stac2270} {\bibfield  {journal} {\bibinfo  {journal}
  {\mnras}\ }\textbf {\bibinfo {volume} {516}},\ \bibinfo {pages} {1977}
  (\bibinfo {year} {2022})}\BibitemShut {NoStop}%
\bibitem [{\citenamefont {{Suyu}}\ \emph {et~al.}(2020)\citenamefont {{Suyu}},
  \citenamefont {{Huber}}, \citenamefont {{Ca{\~n}ameras}},\ and\ \citenamefont
  {{et al.}}}]{2020A&A...644A.162S}%
  \BibitemOpen
  \bibfield  {author} {\bibinfo {author} {\bibfnamefont {S.~H.}\ \bibnamefont
  {{Suyu}}}, \bibinfo {author} {\bibfnamefont {S.}~\bibnamefont {{Huber}}},
  \bibinfo {author} {\bibfnamefont {R.}~\bibnamefont {{Ca{\~n}ameras}}}, \ and\
  \bibinfo {author} {\bibnamefont {{et al.}}},\ }\href {\doibase
  10.1051/0004-6361/202037757} {\bibfield  {journal} {\bibinfo  {journal}
  {\aap}\ }\textbf {\bibinfo {volume} {644}},\ \bibinfo {eid} {A162} (\bibinfo
  {year} {2020})}\BibitemShut {NoStop}%
\bibitem [{\citenamefont {{Huber}}\ \emph {et~al.}(2022)\citenamefont
  {{Huber}}, \citenamefont {{Suyu}}, \citenamefont {{Ghoshdastidar}},\ and\
  \citenamefont {{et al.}}}]{2022A&A...658A.157H}%
  \BibitemOpen
  \bibfield  {author} {\bibinfo {author} {\bibfnamefont {S.}~\bibnamefont
  {{Huber}}}, \bibinfo {author} {\bibfnamefont {S.~H.}\ \bibnamefont {{Suyu}}},
  \bibinfo {author} {\bibfnamefont {D.}~\bibnamefont {{Ghoshdastidar}}}, \ and\
  \bibinfo {author} {\bibnamefont {{et al.}}},\ }\href {\doibase
  10.1051/0004-6361/202141956} {\bibfield  {journal} {\bibinfo  {journal}
  {\aap}\ }\textbf {\bibinfo {volume} {658}},\ \bibinfo {eid} {A157} (\bibinfo
  {year} {2022})}\BibitemShut {NoStop}%
\bibitem [{\citenamefont {{Bayer}}\ \emph {et~al.}(2021)\citenamefont
  {{Bayer}}, \citenamefont {{Huber}}, \citenamefont {{Vogl}}, \citenamefont
  {{Suyu}},\ and\ \citenamefont {{et al.}}}]{2021A&A...653A..29B}%
  \BibitemOpen
  \bibfield  {author} {\bibinfo {author} {\bibfnamefont {J.}~\bibnamefont
  {{Bayer}}}, \bibinfo {author} {\bibfnamefont {S.}~\bibnamefont {{Huber}}},
  \bibinfo {author} {\bibfnamefont {C.}~\bibnamefont {{Vogl}}}, \bibinfo
  {author} {\bibfnamefont {S.~H.}\ \bibnamefont {{Suyu}}}, \ and\ \bibinfo
  {author} {\bibnamefont {{et al.}}},\ }\href {\doibase
  10.1051/0004-6361/202040169} {\bibfield  {journal} {\bibinfo  {journal}
  {\aap}\ }\textbf {\bibinfo {volume} {653}},\ \bibinfo {eid} {A29} (\bibinfo
  {year} {2021})}\BibitemShut {NoStop}%
\bibitem [{\citenamefont {{Ding}}\ \emph {et~al.}(2021)\citenamefont {{Ding}},
  \citenamefont {{Liao}}, \citenamefont {{Birrer}},\ and\ \citenamefont {{et
  al.}}}]{2021MNRAS.504.5621D}%
  \BibitemOpen
  \bibfield  {author} {\bibinfo {author} {\bibfnamefont {X.}~\bibnamefont
  {{Ding}}}, \bibinfo {author} {\bibfnamefont {K.}~\bibnamefont {{Liao}}},
  \bibinfo {author} {\bibfnamefont {S.}~\bibnamefont {{Birrer}}}, \ and\
  \bibinfo {author} {\bibnamefont {{et al.}}},\ }\href {\doibase
  10.1093/mnras/stab1240} {\bibfield  {journal} {\bibinfo  {journal} {\mnras}\
  }\textbf {\bibinfo {volume} {504}},\ \bibinfo {pages} {5621} (\bibinfo {year}
  {2021})}\BibitemShut {NoStop}%
\bibitem [{\citenamefont {{Birrer}}\ \emph {et~al.}(2022)\citenamefont
  {{Birrer}}, \citenamefont {{Dhawan}},\ and\ \citenamefont
  {{Shajib}}}]{2022ApJ...924....2B}%
  \BibitemOpen
  \bibfield  {author} {\bibinfo {author} {\bibfnamefont {S.}~\bibnamefont
  {{Birrer}}}, \bibinfo {author} {\bibfnamefont {S.}~\bibnamefont {{Dhawan}}},
  \ and\ \bibinfo {author} {\bibfnamefont {A.~J.}\ \bibnamefont {{Shajib}}},\
  }\href {\doibase 10.3847/1538-4357/ac323a} {\bibfield  {journal} {\bibinfo
  {journal} {\apj}\ }\textbf {\bibinfo {volume} {924}},\ \bibinfo {eid} {2}
  (\bibinfo {year} {2022})}\BibitemShut {NoStop}%
\bibitem [{\citenamefont {{Qi}}\ \emph {et~al.}(2022)\citenamefont {{Qi}},
  \citenamefont {{Cui}}, \citenamefont {{Hu}},\ and\ \citenamefont {{et
  al.}}}]{2022arXiv220201396Q}%
  \BibitemOpen
  \bibfield  {author} {\bibinfo {author} {\bibfnamefont {J.-Z.}\ \bibnamefont
  {{Qi}}}, \bibinfo {author} {\bibfnamefont {Y.}~\bibnamefont {{Cui}}},
  \bibinfo {author} {\bibfnamefont {W.-H.}\ \bibnamefont {{Hu}}}, \ and\
  \bibinfo {author} {\bibnamefont {{et al.}}},\ }\href@noop {} {\bibfield
  {journal} {\bibinfo  {journal} {arXiv e-prints}\ } (\bibinfo {year}
  {2022})},\ \Eprint {http://arxiv.org/abs/2202.01396} {2202.01396}
  \BibitemShut {NoStop}%
\bibitem [{\citenamefont {{Qi}}\ \emph {et~al.}(2019)\citenamefont {{Qi}},
  \citenamefont {{Cao}}, \citenamefont {{Biesiada}},\ and\ \citenamefont {{et
  al.}}}]{2019PhRvD.100b3530Q}%
  \BibitemOpen
  \bibfield  {author} {\bibinfo {author} {\bibfnamefont {J.}~\bibnamefont
  {{Qi}}}, \bibinfo {author} {\bibfnamefont {S.}~\bibnamefont {{Cao}}},
  \bibinfo {author} {\bibfnamefont {M.}~\bibnamefont {{Biesiada}}}, \ and\
  \bibinfo {author} {\bibnamefont {{et al.}}},\ }\href {\doibase
  10.1103/PhysRevD.100.023530} {\bibfield  {journal} {\bibinfo  {journal}
  {\prd}\ }\textbf {\bibinfo {volume} {100}},\ \bibinfo {eid} {023530}
  (\bibinfo {year} {2019})}\BibitemShut {NoStop}%
\end{thebibliography}%

\end{document}